\documentclass[amssymb,nobibnotes,nofootinbib,aps,prd]{revtex4}

 \usepackage{here}

\usepackage[dvipdfmx]{graphicx}
\usepackage{amssymb,amsfonts,amsmath,bm,color,multirow,cases,empheq,hyperref}
\usepackage{mathrsfs}
\definecolor{darkergreen}{rgb}{0,0.5,0}

\usepackage{enumerate}
\usepackage{ulem}
\usepackage{afterpage}

\hypersetup{%
 setpagesize=false,
 bookmarksnumbered=true,%
 bookmarksopen=true,%
 colorlinks=true,%
 linkcolor=blue,%
 citecolor=red}

\usepackage{tikz}
\usetikzlibrary{positioning,fit}
\usetikzlibrary{calc,intersections,through,backgrounds}
\usetikzlibrary{calc,arrows,decorations.markings,shapes.arrows,patterns, decorations.pathmorphing}
\usetikzlibrary{positioning,matrix,arrows,decorations.pathmorphing,decorations.pathreplacing}
\usetikzlibrary{arrows,matrix,positioning,shapes,arrows}
\usetikzlibrary{shapes.geometric}
\usetikzlibrary{decorations.text}
\usetikzlibrary{arrows.meta}

\tikzset{
  ->-/.style={decoration={markings, mark=at position 0.5 with {\arrow{to}}},
              postaction={decorate}},}
\tikzset{
  -<-/.style={decoration={markings, mark=at position 0.5 with {\arrow{to reversed}}},
              postaction={decorate}},}
              
\tikzset{
  pics/torus/.style n args={3}{
    code = {
      \providecolor{pgffillcolor}{rgb}{1,1,1}
      \begin{scope}[
          yscale=cos(#3),
          outer torus/.style = {draw,line width/.expanded={\the\dimexpr2\pgflinewidth+#2*2},line join=round},
          inner torus/.style = {draw=pgffillcolor,line width={#2*2}}
        ]
        \draw[outer torus] circle(#1);\draw[inner torus] circle(#1);
        \draw[outer torus] (180:#1) arc (180:360:#1);\draw[inner torus,line cap=round] (180:#1) arc (180:360:#1);
      \end{scope}
    }
  }
}

\newcommand{\tikznode}[2]{\relax
    \ifmmode%
    \tikz[remember picture,baseline=(#1.base),inner sep=0pt] \node (#1) {$#2$};
    \else
    \tikz[remember picture,baseline=(#1.base),inner sep=0pt] \node (#1) {#2};%
    \fi
}

\newcommand{\no}{\nonumber}

\newcommand{\cA}{\mathcal A}
\newcommand{\cC}{\mathcal C}
\newcommand{\cF}{\mathcal F}

\newcommand{\cK}{\mathcal K}\newcommand{\cL}{\mathcal L}
\newcommand{\cM}{\mathcal M}\newcommand{\cN}{\mathcal N}
\newcommand{\cO}{\mathcal O}

\newcommand{\cV}{\mathcal V}

\newcommand{\cW}{\mathcal W}

\newcommand{\la}{\lambda}
\newcommand{\Tr}{{\rm Tr}}

\newcommand{\ads}{$AdS_3\times S^3\times T^4$ }

\allowdisplaybreaks

\begin{document}

\begin{flushright}
TTI-MATHPHYS-34
\end{flushright}
\vspace*{0.5cm}

\title{
Building multi-BTZ black holes through Riemann-Hilbert problem
}

\author{Jun-ichi Sakamoto}
\email{jsakamoto@toyota-ti.ac.jp}
\author{Shinya Tomizawa}
\email{tomizawa@toyota-ti.ac.jp}
\affiliation{\vspace{3mm}Mathematical Physics Laboratory, Toyota Technological Institute\vspace{2mm}\\Hisakata 2-12-1, Tempaku-ku, Nagoya, Japan 468-8511\vspace{3mm}}

\begin{abstract}

We construct a recently found class of non-BPS black hole solutions with asymptotically \ads in type IIB supergravity, consisting of multiple BTZ black holes localized on an $S^3$, within the group theoretical framework of Breitenlohner and Maison (BM). 
Starting with the multi-neutral black string solution as a seed, we solve the associated Riemann-Hilbert problem for the BM linear system. 
First, we determine the monodromy matrix corresponding to this seed solution by generalizing the early work of Katsimpouri et al. on the four-charged black hole of STU supergravity, where some assumptions must be relaxed for the solutions with multiple horizons.
By applying the Harrison transformation, a charge-generating transformation in the $SO(4,4)$ group, to the monodromy matrix, we obtain the multi-charged black string solution.
Furthermore, through a ``subtraction'' procedure--- an $SO(4,4)$ transformation that changes the asymptotic structure from $R^{1,4}\times S^1\times  T^4$ to \ads spacetime--- we derive the multi-BTZ black hole solution. 
This is the first example in which the subtraction procedure is applied to multiple black holes, and it may also have potential applications to other cases.
\end{abstract}

\date{\today}

\maketitle

\section{Introduction}

Exact solutions may lead to the prediction of new physical phenomena or provide new perspectives on known phenomena. For example, exact solutions for black hole models or field theory solutions may provide insights into physical situations that had not been previously considered, advancing our understanding of the universe. These solutions may also guide experimental predictions and provide new avenues for verification through observation or experimentation.
The mathematical techniques developed for finding exact solutions, such as inverse scattering methods, Riemann-Hilbert problems, and Harrison transformations, can be applied to other areas of mathematics as well. By developing exact solutions, new approaches are often introduced that can benefit other mathematical fields. Techniques in integrable systems can bridge gaps with number theory, group theory, and analysis, expanding the toolkit of mathematical problem-solving.
In the study of integrable systems, obtaining exact solutions is key to revealing the symmetries, conserved quantities, and the diversity of solutions these systems possess. For instance, in integrable systems such as the Korteweg-de Vries (KdV) equation or the nonlinear Schr\"odinger equation, exact solutions can reveal the behavior of solitary waves, periodic solutions, and other physical or mathematical features. These characteristics are essential for understanding the deeper structure of integrable systems. 

In Einstein and supergravity theories, the nonlinearity of the equations of motion makes obtaining exact solutions a highly challenging task. However, in $D$-dimensional gravity theories, the assumption of the existence of $D-2$ commuting Killing vectors reduces the Einstein equations to the equations of motion for an integrable coset sigma model \cite{Maison:1979kx} on a 2D conformally flat space $ds_2^2=e^{2\nu}(d\rho^2+dz^2)$, whose action is written in the form $S=\int d\rho dz \rho{\rm Tr}(M^{-1}\partial_m M M^{-1}\partial^m M)$ with a certain coset matrix $M$. 
This remarkable connection to integrable systems allows a broad class of exact axisymmetric solutions to be interpreted as soliton solutions, which can be systematically constructed using the inverse scattering method (ISM).
The ISM is one of the most valuable tools for obtaining exact solutions to the vacuum Einstein equations with $D-2$ Killing isometries. This method enables the systematic generation of new solutions with the same isometries through soliton transformations, starting from a known simple solution. While the original formulation of the ISM by Belinski and Zakharov \cite{Belinsky:1971nt, Belinsky:1979mh,Belinski:2001ph} typically yields singular solutions when applied directly to higher dimensions, Pomeransky modified the ISM to produce regular solutions, even in higher dimensions \cite{Pomeransky:2005sj}. Notably, when combined with the rod structure \cite{Emparan:2001wk, Harmark:2004rm}, this modified ISM has proven highly effective, especially in the context of five-dimensional vacuum black hole solutions~\cite{
Mishima:2005id,
Tomizawa:2005wv,Tomizawa:2006jz,Tomizawa:2006vp,
Iguchi:2006rd,
Elvang:2007rd,
Tomizawa:2007mz,Iguchi:2007xs,
Pomeransky:2006bd,
Iguchi:2007is,Evslin:2007fv,
Elvang:2007hs,Izumi:2007qx,
Chen:2011jb,
Chen:2008fa,
Chen:2012kd,
Rocha:2011vv,
Rocha:2012vs,
Chen:2015iex,
Chen:2012zb,
Lucietti:2020ltw,
Lucietti:2020phh,
Morisawa:2007di,Evslin:2008gx,
Feldman:2012vd,
Chen:2010ih,
Iguchi:2011qi,
Tomizawa:2019acu,
Tomizawa:2022qyd,
Suzuki:2023nqf,
Figueras:2009mc}.

While significant progress has been made in constructing exact solutions for asymptotically flat black holes and Kaluza-Klein black holes, much less is known about exact non-BPS black hole solutions with asymptotically AdS spacetime, despite their relevance to applications in the AdS/CFT correspondence. 
One major obstacle is that when higher-dimensional AdS black hole solutions are dimensionally reduced, additional scalar potentials often arise, which can break classical integrability. 
As a result, most known black hole solutions in AdS spacetime are restricted to BPS solutions (except solutions with spherical horizon  topology), leaving the construction of exact non-BPS solutions largely unexplored.
Recently, a novel class of non-BPS multi-black hole solutions with asymptotically \ads spacetime has been constructed in \cite{Bah:2022pdn}. 
These solutions describe smooth and regular configurations in type IIB supergravity, consisting of multiple BTZ black holes localized on $S^3$, connected by bubbles stabilized by RR flux. 
We refer to this class of solutions as multi-BTZ black holes.
This configuration exhibits several intriguing features, as discussed in \cite{Bena:2024gmp}. 
For example, in the microcanonical ensemble, the multi-BTZ black holes are found to dominate in a certain range of positive energies, suggesting they could be a candidate for the final state of BTZ black hole instability in the low-energy regime. These solutions also exist in the negative energy region, where the BTZ black hole is not present. 
In particular, it has been shown that they contribute to a portion of the entropy of the D1-D5 symmetric orbifold CFT in this negative energy regime.
In this context, the construction of more general non-BPS black hole solutions with asymptotically AdS$_3$ geometry is expected to not only extend existing solution-generating techniques based on integrable structures of the Einstein equations but also provide new insights into the AdS$_3$/CFT$_2$ correspondence.

The multi-BTZ black hole was initially constructed by exploiting the fact that, under a static ansatz for the D1-D5-KKm system, the type IIB supergravity equations reduce to a set of well-known Ernst equations~\cite{Ernst:1967wx,Ernst:1967by} for the Einstein-Maxwell system in the absence of both twist and  magnetic potentials, as desribed below in Eqs.~(\ref{gen-eom1})--(\ref{gen-eom3})~\cite{Bah:2022pdn}.
However, when attempting to construct more general solutions, such as stationary solutions, the set of Ernst equations is no longer independent. In such cases, one must work directly with the equations of motion of the corresponding 2D integrable coset sigma model \footnote{In this work, we do not consider solutions described by the most general D1-D5-KKm ansatz. For a discussion on the connection between such solutions and a 2D integrable coset sigma model, we refer the reader to, for example, \cite{Chakraborty:2025ger}.}.
Therefore, this work aims to develop solution-generating techniques based on this integrable structure to construct non-BPS black holes with asymptotically AdS geometry. As a first step, we reconstruct the multi-BTZ black hole solution by utilizing the Breitenlohner-Maison (BM) linear system \cite{Breitenlohner:1986um}, which describes the integrable structure of the underlying coset sigma model.
A key advantage of the BM linear system is that it explicitly reveals the connection to the Geroch group, an infinite-dimensional symmetry group that governs the classical integrability of the underlying gravitational system. This formulation clarifies the mathematical structure of the solution-generating technique more effectively than the BZ method, thereby facilitating a natural and systematic extension to multi-black hole solutions, such as those explored in this work.

Similar to the construction of asymptotically flat black hole solutions, we use the BM approach to construct soliton solutions corresponding to asymptotically \ads black holes. 
Following the procedure developed in \cite{Breitenlohner:1986um, Katsimpouri:2012ky, Chakrabarty:2014ora, Katsimpouri:2013wka, Katsimpouri:2014ara} (see also \cite{Camara:2017hez}), this is achieved by solving the Riemann-Hilbert problem. 
Given the monodromy matrix $ \cM(w)$, which is meromorphic for the constant complex spectral parameter $w$,  one factorizes it as $\cM(w)=\cV(\lambda)^{\natural}\cV(\lambda)$ for another coordinate-dependent spectral parameter $\lambda$. 
The coset matrix $M=V^{\natural}V$ is then obtained, where $V=\cV(0)$,  as the solution to the equation of motion for the sigma model. 
This factorization process is referred to as the Riemann-Hilbert problem.
In general, solving the Riemann-Hilbert problem is a highly nontrivial task due to the difficulty of factorizing a given monodromy matrix. 
However, fortunately, this becomes possible when we restrict ourselves to soliton solutions, which are characterized by monodromy matrices with poles in the $w$-plane.
Similar to many known asymptotically flat non-extremal black holes, the monodromy matrix for these solutions has only simple poles. 
Moreover, as shown in~Ref.\cite{Camara:2017hez}, the monodromy matrices with double poles describe extremal black hole solutions. 
In these cases, the factorization problem simplifies considerably and reduces to solving a set of algebraic equations.

Despite these simplifications, directly factorizing the monodromy matrix corresponding to the multi-BTZ black hole is not computationally efficient due to the presence of RR flux.
To manage this, we first construct the monodromy matrix $\cM_{\rm mString}$ associated with a non-extremal multi-black string solution without RR-flux and carry out its factorization. 
As shown for the four-charged black hole solution of Cveti\v{c}-Youm in the earlier work \cite{Katsimpouri:2013wka}, we find that this monodromy matrix corresponding to the black string solution consists of only simple poles with rank-2 residue matrices. 
This allows us to employ the techniques developed in \cite{Katsimpouri:2012ky, Katsimpouri:2013wka, Katsimpouri:2014ara}.
However, to apply the procedure to the multi-black string case, we must relax one of the assumptions made in \cite{Katsimpouri:2013wka}. 
By considering a more general setting, we succeed in explicitly factorizing the monodromy matrix for the multi-black string with an arbitrary number of bubbles. 
In particular, when computing the conformal factor $e^{2\nu}$, we cannot use the approach in Refs.~\cite{Katsimpouri:2012ky, Katsimpouri:2013wka} due to the existence of multiple bubbles; instead, we adapt the procedure in \cite{Korotkin:1994au, Korotkin:1996vi}, which relies on the analytic structure of the monodromy matrix. 
We extend this procedure to the current setting and derive a more general formula for the conformal factor, which successfully reproduces the known result for the multi-BTZ black hole.
After doing that, applying the $SO(4,4)$-transformation to the vacuum multi-black string solution as $\cM_{\rm mString}\to \cM_{\rm cmString}=g_{\rm c}^{\natural}\cM g_{\rm c}\ (g_{\rm c}\in SO(4,4))$,  we thereby reproduce the non-extremal charged multi-black string solutions constructed in \cite{Bah:2021owp}. 
Finally, by applying another $SO(4,4)$-transformation as $\cM_{\rm cmString}\to \cM_{\rm mBTZ}=g_{\rm sub}^{\natural}\cM_{\rm cmString} g_{\rm sub}\ (g_{\rm sub}\in SO(4,4))$, which changes the asymptotic structure from $\rm R^{1,4}\times \rm S^1\times  \rm T^4$ to \ads spacetime, we can obtain the multi-BTZ black hole solutions.
This procedure is a natural extension of the earlier works~\cite{Cvetic:2011hp, Cvetic:2013cja}, which showed that a 6D BTZ black hole solution with asymptotically AdS$_3 \times$ S$^3$ geometry can be obtained from a 6D black string solution through the $SO(4,4)$-transformation.
In the terminology used in \cite{Cvetic:2011hp}, the multi-BTZ black hole can be regarded as the “subtracted geometry” of a multi-black string.
Our result is the first successful factorization of monodromy matrices describing black holes with an arbitrary number of bubbles, not only in asymptotically AdS$_3$ spacetime but also in asymptotically flat spacetime. This demonstrates that the monodromy matrix approach based on the BM linear system is a powerful method for constructing higher-dimensional non-BPS black hole solutions.

The rest of this paper is organized as follows: In Sec.\ref{sec2}, we provide an overview of the basic properties of the multi-BTZ black holes and multi-black strings, demonstrating that they can be described by a static D1-D5-KKm system. In Sec.\ref{sec:reduction}, we perform a dimensional reduction of the 10D type IIB supergravity with the static D1-D5-KKm system down to two dimensions, showing that it reduces to an integrable 2D symmetric coset sigma model coupled to dilaton gravity. Section~\ref{sec:mnblack} is devoted to the construction of a multi-neutral black string solution by solving the associated Riemann-Hilbert problem, and we present a closed formula for the conformal factor. In Sec.~\ref{sec:harrison}, we apply a sequence of 
$SO(4,4)$ Harrison transformations to the asymptotically flat solution to construct multi-BTZ black hole configurations and derive the corresponding monodromy matrices.

\section{multi-BTZ black holes and multi-black strings}\label{sec2}

In this section, we provide an overview of the key properties of two regular black hole solutions with distinct asymptotic behaviors  considered in this paper\,: 
\begin{itemize}
    \item Multi-BTZ black hole solutions which are asymptotically \ads \cite{Bah:2022pdn,Bena:2024gmp},
    \item Multi-neutral black string solutions which are asymptotically $R^{1,4}\times S^{1}\times T^4$ \cite{Bah:2021owp}.
\end{itemize}
These solutions, which are related by $SO(4,4)$ transformations as seen in Sec.~\ref{sec:harrison}, satisfy the equations of motion derived from the type IIB supergravity framework describing the D1-D5-KKm system. The solutions feature multiple horizons that are interconnected by bubbles in an alternating fashion. While each individual bubble is generally unstable, the introduction of a RR 3-form flux can stabilize them. 
By appropriately adjusting the sizes of the bubbles, the resulting gravitational solutions can be made regular.

\subsection{D1-D5-KKm system}

We first consider ten-dimensional type IIB supergravity solutions with eight commuting Killing vectors characterized by the D1-D5-KKm system.
The brane configuration we consider consists of D1-branes wrapped along the $S^1(y)$ direction, D5-branes are wrapped along the $S^1(y)\times T^4(y^a)$ directions, and Kaluza-Klein monopoles\,(KKm) are supported along the $S^1(\psi)$ direction. This configuration is summarized in the table below:
\begin{center}
\begin{tabular}{lccccccccccc}
\hline
   branes & $t$ & $\rho$ & $z$ & $\phi$ & $\psi$ & $y$ & $y^1$ & $y^2$ & $y^3$ & $y^4$  \\ \hline
   D1 & - &  &    & &  & - &  & &  &  \\
   D5 & - &  &    & &  & - & - & - & - & -  \\
   KKm & - &  &    & & $\psi$ & - & - & - & - & -  \\
 \hline
\end{tabular}
\end{center}
We work on the string frame and take a specific ansatz for axisymmetric solutions corresponding to this brane configuration as \cite{Bah:2022pdn}
\begin{align}\label{d1d5-bg}
\begin{split}
    ds^2_{10}&=ds_{6}^2+\sqrt{\frac{Z_1}{Z_5}}\sum_{a=1}^{4}(dy^a)^2\,,\\
    ds_{6}^2&=\frac{1}{\sqrt{Z_1Z_5}}\bigg[-\frac{dt^2}{W}+W\,dy^2\biggr]\\
    &\qquad+\sqrt{Z_1Z_5}\biggl[\frac{1}{Z_0}(d\psi+H_0 d\phi)^2+Z_0\left(e^{2\nu}(d\rho^2+dz^2)+\rho^2 d\phi^2\right)\biggr]\,,\\
     e^{2\Phi_{10}}&=\frac{Z_1}{Z_5}\,,\\
    C_2&=-T_1\,dt\wedge dy+H_5\,d\phi\wedge d\psi\,,\qquad
    C_0=C_4=B_2=0\,.
\end{split}
\end{align}
The solution is described by the potential functions $W, Z_I (I=0,1,5), T_1, H_0, H_5$ and the conformal factor $e^{2\nu}$, which depend only on the Weyl coordinates $\rho$ and $z$ ($\rho \geq 0, z\in \mathbb{R}$).
The functions $Z_0,Z_1$, and $Z_5$ are the warp factors generated by the KKm, D1-brane, and D5-brane, respectively.
Similarly, the electric potentials $T_I$ and magnetic potentials $H_I$, associated with KKm, D1-brane, and D5-brane, are introduced. Each pair of potentials is related by the Hodge dual $\star_2$ in the two dimensional Euclidean space $(\rho,z)$:
\begin{align}\label{TH-dual}
    dT_I=-\frac{1}{\rho Z^2_I}\star_2dH_I\,.
\end{align}
We also define the $8\times 8$ Killing metric $g_{\mu\nu}^{\rm Killing}$, which represents the remaining part of the 10D metric $g_{mn}^{E}$ in the Einstein frame after excluding the components involving with $(\rho,z)$\footnote{The 10D metric $g_{mn}^{E}$ in the Einstein frame is related to the metric $g_{mn}^{\rm st}$ in the string frame by $g_{mn}^{\rm st}=e^{\frac{1}{2}\Phi_{10}}g_{mn}^{E}$. The determinant of the Killing metric satisfies ${\rm det}(g_{\mu\nu}^{\rm Killing})=-\rho^2$.}.

The type IIB supergravity equations of motion (\ref{iib-eom}) are significantly simplified under the ansatz (\ref{d1d5-bg}) and decomposed into four independent Ernst equations \cite{Bah:2022pdn} for the potential functions $W, Z_I, T_I$ \footnote{The equations of motion for the magnetic potentials $H_I$ can be obtained from (\ref{gen-eom2}) and (\ref{gen-eom3}) via the duality relation (\ref{TH-dual}).}:
\begin{align}
    &\nabla^2\log W=0\,,\label{gen-eom1}\\
    &\nabla^2\log Z_I=-Z_I^2 \left[(\partial_{\rho}T_I)^2+(\partial_z T_I)^2\right]\,,\label{gen-eom2}\\
    &\partial_{\rho}\left(\rho Z_I^2\partial_{\rho}T_I\right)+\partial_z(\rho Z_I^2\partial_{z}T_I)=0\,,\label{gen-eom3}
\end{align}
where $\nabla^2$ is the Laplacian with respect to the abstract 3D Euclidean space metric $d\rho^2+dz^2+\rho^2d\varphi$. When acting on axisymmetric functions, this operator takes the form
\begin{align}\label{Laplacian}
    \nabla^2=\frac{1}{\rho}\partial_{\rho}(\rho \partial_{\rho})+\partial_z^2\,.
\end{align}
It should be noted that Eq.~(\ref{gen-eom1}) corresponds to the vacuum Ernst equation  in the absence of a twist potential, while Eqs.~(\ref{gen-eom2}) and (\ref{gen-eom3}) represent the Ernst eqautions for the Einstein-Maxwell  system in the absence of both twist and magnetic potentials.
The conformal factor $e^{2\nu}$ is determined by
\begin{align}
    \frac{2}{\rho}\partial_{z}\nu&=\partial_{\rho}(\log W)\partial_{z}(\log W)+\partial_{\rho}(\log Z_0)\partial_{z}(\log Z_0)+\frac{1}{\rho^2Z_0^2}\partial_{\rho}H_0\partial_{z}H_0\no\\
    &\quad +\partial_{\rho}(\log Z_1)\partial_{z}(\log Z_1)-Z_1^2\partial_{\rho}T_1\partial_{z}T_1\no\\
    &\quad +\partial_{\rho}(\log Z_5)\partial_{z}(\log Z_5)+\frac{1}{\rho^2Z_5^2}\partial_{\rho}H_5\partial_{z}H_5\,,\\
  \frac{4}{\rho}\partial_{\rho}\nu&=(\partial_{\rho}\log W)^2-(\partial_{z}\log W)^2+(\partial_{\rho}\log Z_0)^2-(\partial_{z}\log Z_0)^2+\frac{1}{\rho^2Z_0^2}((\partial_{\rho}H_0)^2-(\partial_{z}H_0)^2)\no\\
  &\quad +(\partial_{\rho}\log Z_1)^2-(\partial_{z}\log Z_1)^2-Z_1^2((\partial_{\rho}T_1)^2-(\partial_{z}T_1)^2)\no\\
 &\quad +(\partial_{\rho}\log Z_5)^2-(\partial_{z}\log Z_5)^2+\frac{1}{\rho^2Z_5^2}((\partial_{\rho}H_5)^2-(\partial_{z}H_5)^2)\,.
\end{align}
The integrality $\nu_{,\rho z}=\nu_{,z \rho}$ is assured by Eqs.~(\ref{gen-eom1})-(\ref{gen-eom3}).
Thus,  once the potential functions $W, Z_I, T_I$ are given, one can completely determine $e^{2\nu}$.
As shown in Sec.\,\ref{sec:reduction},  Eqs.~(\ref{gen-eom1})-(\ref{gen-eom3}) can be combined into the equations of motion describing a classically integrable 2D coset sigma model.

\subsection{multi-BTZ black holes}

The main focus of this paper is on regular bound states of non-extremal BTZ black hole solutions with asymptotically \ads geometry, as discussed in Ref.~\cite{Bah:2022pdn}. 
The gravitational solution belongs to a specific class of type IIB supergravity solutions, which are described by the ansatz (\ref{d1d5-bg}). 
Below, we provide a brief overview of these black hole solutions. 
For further details, refer to \cite{Bah:2022pdn, Bena:2024gmp}.

\subsubsection*{Rod structure}

The above Weyl-Papapetrou coordinate system $(\rho,z)$ allows for the introduction of the rod structure and rod diagram,  which were first introduced by Harmark~\cite{Harmark:2004rm} for stationary solutions based on earlier work for static solutions by Emparan-Reall~\cite{Emparan:2001wk} in Einstein gravity without a cosmological constant.  
The rods are specified by the values of $z$ where the determinant of the Killing metric $g^{\text{Killing}}$  vanishes at $\rho=0$ and their degenerate directions are characterized by the corresponding eigenvectors for the zero eigenvalue, known as rod vectors.
It should be note that although the multi-BTZ black hole spacetime in this paper is asymptotically AdS geometry, the concept of the rod structure can still be applied, as the 10D IIB supergravity does not include a cosmological constant. 

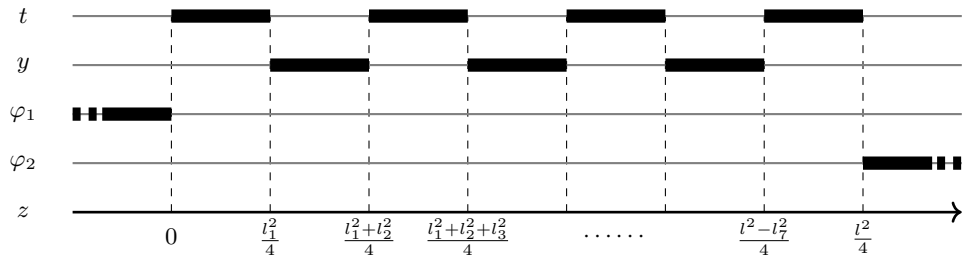
\begin{figure}
\begin{center}
\begin{tikzpicture}[scale=0.65]
\node[font=\small ] at (-10,4) {$t$};
\node[font=\small ] at (-10,3) {$y$};
\node[font=\small ] at (-10,2) {$\varphi_1$};
\node[font=\small ] at (-10,1) {$\varphi_2$};
\node[font=\small ] at (-10,0) {$z$};
\node[font=\small ] at (-7,-0.5) {$0$};
\node[font=\small ] at (-5,-0.5) {$\frac{l^2_1}{4}$};
\node[font=\small ] at (-3,-0.5) {$\frac{l^2_1+l_2^2}{4}$};
\node[font=\small ] at (-1,-0.5) {$\frac{l^2_1+l_2^2+l_3^2}{4}$};
\node[font=\small ] at (2,-0.5) {$\cdots\cdots$};
\node[font=\small ] at (5,-0.5) {$\frac{l^2-l_7^2}{4}$};
\node[font=\small ] at (7,-0.5) {$\frac{l^2}{4}$};
\draw[black,dashed ] (-7,0) -- (-7,4);
\draw[black,dashed ] (-5,0) -- (-5,4);
\draw[black,dashed ] (-3,0) -- (-3,4);
\draw[black,dashed ] (-1,0) -- (-1,4);
\draw[black,dashed ] (1,0) -- (1,4);
\draw[black,dashed ] (3,0) -- (3,4);
\draw[black,dashed ] (5,0) -- (5,4);
\draw[black,dashed ] (7,0) -- (7,4);
\draw[gray,line width = 0.8] (-9,4) -- (9,4);
\draw[black,line width = 5] (-7,4) -- (-5,4);
\draw[black,line width = 5] (-3,4) -- (-1,4);
\draw[black,line width = 5] (1,4) -- (3,4);
\draw[black,line width = 5] (5,4) -- (7,4);
\draw[gray,line width = 0.8] (-9,3) -- (9,3);
\draw[black,line width = 5] (-5,3) -- (-3,3);
\draw[black,line width = 5] (-1,3) -- (1,3);
\draw[black,line width = 5] (3,3) -- (5,3);
\draw[gray,line width = 0.8] (-9,2) -- (9,2);
\draw[black,line width = 5,dashed ] (-9,2) -- (-8.5,2);
\draw[black,line width = 5] (-8.4,2) -- (-7,2);
\draw[gray,line width = 0.8] (-9,1) -- (9,1);
\draw[black,line width = 5] (7,1) -- (8.4,1);
\draw[black,line width = 5,dashed ] (8.5,1) -- (9,1);
\draw[->,black,line width = 1] (-9,0) -- (9,0);
\end{tikzpicture}
\caption{The rod structure of multi-BTZ black hole solution with $n=7$. The sets of $U_t$ and $U_y$ for this case are $U_t=\{1,3,5,7\}$ and $U_y=\{2,4,6\}$.}\label{rod-mbtz-able}
\end{center}
\end{figure}

The multi-BTZ black hole solutions constructed in \cite{Bah:2022pdn} are multi-black hole solutions with asymptotically \ads spacetime in type IIB supergravity, where $n_h$ BTZ black holes are localized on $S^3$ and connected by $n_b$ bubbles that smoothly shrink along the compactified $y$-direction. 
The rod structure for the multi-BTZ black hole solutions is a chain of horizon of non-extremal black holes and bubbles, as illustrated in Fig.\,\ref{rod-mbtz-able}, where the horizons and bubbles are represented by the timelike and spacelike rods, corresponding to $g^{\rm Killing}_{tt}=0$ and $g^{\rm Killing}_{yy}=0$, respectively.
The size of the $i$-th rod with a finite length is characterized by a real parameter $l_i\,(i=1,2,\dots,n=n_h+n_b)$, and  the size of the entire rod is given by  a parameter $l$, which are related by
\begin{align}
    l^2=\sum_{i=1}^{n}l_i^2\,.
\end{align}
This means that the $i$-th rod is sitting at 
\begin{align}
    \frac{1}{4}\sum_{j=1}^{i-1}l_j^2<z<\frac{1}{4}\sum_{j=1}^{i}l_j^2\,,\qquad \rho=0\,.
\end{align}
The two semi-infinite rods correspond to the axis of rotations $\varphi_1$ and $\varphi_2$ \cite{Emparan:2001wk} which are defined in (\ref{u1-fib-coord}).
For later convenience, we introduce the sets $U_t$ and $U_y$ of indices $i$ that specify the finite rods for horizons and bubbles, respectively.

\subsubsection*{multi-BTZ solutions}

Here, we explicitly write down the multi-BTZ black hole solutions, which fit the ansatz (\ref{d1d5-bg}), characterized by the rod structure shown in Fig.\,\ref{rod-mbtz-able}. 
In the spherical coordinates introduced later, this gravitaitonal solution takes the form \cite{Bah:2022pdn,Bena:2024gmp}
\begin{align}
\begin{split}\label{btzs-sol}
    ds_{10}^2&=\frac{1}{\sqrt{Q_1Q_5}}\biggl[-\frac{r^2dt^2}{W_b(r,\theta)}+ (r^2+l^2)W_b(r,\theta)\,dy^2\biggr]+\sqrt{\frac{Q_1}{Q_5}}\sum_{a=1}^{4}(dy^a)^2\\
    &\quad+\sqrt{Q_1Q_5}\biggl[G(r,\theta)\left(\frac{dr^2}{r^2+l^2}+d\theta^2\right)+\cos^2\theta\,d\varphi_1^2+\sin^2\theta\,d\varphi_2^2\biggr]\,,\\
    C_2&=-\frac{r^2+l^2}{Q_1}dt\wedge dy+\frac{Q_5}{2}\,\cos2\theta\,d\varphi_2\wedge d\varphi_1\,,\qquad e^{2\Phi_{10}}=\frac{Q_1}{Q_5}\,,
\end{split}
\end{align}
where the scalar functions $G(r,\theta)$ and $W_b(r,\theta)$ are
\begin{align}
    G&=\prod_{\substack{i\in U_y\\j\in U_t}}\frac{\left((r_i^2+l_i^2)\cos^2\theta_i+r^2_j\sin^2\theta_j\right)
    \left(r_i^2\cos^2\theta_i+(r_j^2+l_j^2)\sin^2\theta_j\right)}{\left((r_i^2+l_i^2)\cos^2\theta_i+(r_j^2+l_j^2)\sin^2\theta_j\right)\left(r_i^2\cos^2\theta_i+r_j^2\sin^2\theta_j\right)}\,,\label{conf-part}\\
    W_b&=\prod_{\substack{i\in U_y}}\left(1+\frac{l_i^2}{r_i^2}\right)^{-1}\,.\label{bubble-part}
\end{align}
This solution satisfies the equations of motion (\ref{iib-eom}) of type IIB supergravity and describes a bound system of D1-branes, D5-branes, and a KKm, carrying a total D1 charge $Q_1$, D5 charge $Q_5$, and $k=1$ KKm charge. To introduce spherical coordinates $(r, \theta)$, we make a change from the Weyl-Papapetrou coordinates as
\begin{align}\label{weyl-def}
    \rho=\frac{r\sqrt{r^2+l^2}}{4}\sin2\theta\,,\qquad z=\frac{2r^2+l^2}{8}\cos2\theta+\frac{l^2}{8}\,,
\end{align}
where the region of the angle variable $\theta$ is
\begin{align}
      0\leq \theta \leq \frac{\pi}{2}\,.
\end{align}
The angular coordinates $(\varphi_1,\varphi_2)$ are related to $(\phi,\psi)$ by the transformation
\begin{align}\label{u1-fib-coord}
    \varphi_1=\frac{1}{2}\left(\phi+\psi\right)\,,\qquad  \varphi_2=\frac{1}{2}\left(\phi-\psi\right)\,,
\end{align}
and the angle variables $(\theta,\varphi_1,\varphi_2)$ parametrizes the warped $S^3$ in Hopf coordinates for fixed $r$.
The coordinates $(y,\varphi_1,\varphi_2, y^a)$ describe compact directions with the periodicity
\begin{align}
  y\sim y+2\pi R_y\,,\qquad \varphi_1\sim \varphi_1+2\pi\,,\qquad \varphi_2\sim \varphi_2+2\pi\,,\qquad y^a\sim y^a+2\pi\,.
\end{align}
The solution (\ref{btzs-sol}) is characterized by a total of $(n+3)$ real parameters: $Q_1,Q_5,R_y$ and $l_i\,(i=1,2,\dots,n=2n_b+1)$. 
As we will show below, for the multi-BTZ solution to describe a smooth geometry, regularity conditions must be imposed on each rod corresponding to the $n_b$ bubbles. These constraints reduce the number of independent parameters to $n_b+4$.

The scalar functions $G(r,\theta)$ and $W_b(r,\theta)$ can be compactly expressed using the local spherical coordinates $(r_i,\theta_i)$ around the $i$-th rod.
The local coordinates are defined as\footnote{
The inverse of (\ref{ri-cosi}) is 
\begin{align}
    \rho=\frac{1}{4}r_i\sqrt{r_i^2+l_i^2}\sin2\theta_i\,,\qquad 
    z=\frac{2r_i^2+l_i^2}{8}\cos2\theta_i+\frac{1}{4}\sum_{j=1}^{i}l_j^2
    -\frac{l_i^2}{8}\,.
\end{align}
}
\begin{align}\label{ri-cosi}
    r_i^2=\rho_i+\rho_{i-1}-\frac{l_i^2}{2}\,,\qquad
    l_i^2\cos^2\theta_i&=\rho_{i-1}-\rho_i+\frac{l_i^2}{2}\,,
\end{align}
where $\rho_0$ and $\rho_i (i=1,\dots,n)$ are defined as
\begin{align}
\rho_0=2\sqrt{\rho^2+z^2}\,,\qquad
    \rho_i&=2\sqrt{\rho^2+\left(z-\frac{1}{4}\sum_{j=1}^{i}l_j^2\right)^2}\,.
\end{align}
Using this local spherical coordinate system $(r_i,\theta_i)$, the region of the $i$-th rod is described by
\begin{align}
    r_i=0\,,\qquad  0\leq \theta_i \leq \frac{\pi}{2}\,,
\end{align}
and for the other local spherical coordinates $(r_j,\theta_j)$ with $j\neq i$, we have
\begin{align}
    r_j>0\,,\qquad \theta_j=
    \begin{cases}
        0\qquad\qquad &j<i\\
        \frac{\pi}{2}\qquad &j>i
    \end{cases}\,.
\end{align}
In the global spherical coordinate $(r,\theta)$, all rods are localized at $r=0$, and the segment of $i$-th rod is determined by the range of $\theta$\,:
\begin{align}
   \frac{1}{l^2}\sum_{j=1}^{i-1}l_j^2<\cos^2\theta< \frac{1}{l^2}\sum_{j=1}^{i}l_j^2\,.
\end{align}

\subsubsection*{Asymptotic structure}

 The black hole solution (\ref{btzs-sol}) asymptotically approaches a \ads spacetime with equal radii $(\sqrt{Q_1Q_5})^{1/2}$ of AdS$_3$ and S$^3$ at large $r$ :
\begin{align}
\begin{split}\label{btzs-asy-sol}
    ds_{10}^2&\sim \sqrt{Q_1Q_5}\biggl[-\frac{r^2}{Q_1Q_5}(-dt^2+dy^2)+\frac{dr^2}{r^2}+d\Omega_3^2 \biggr]+\sqrt{\frac{Q_1}{Q_5}}\sum_{a=1}^{4}(dy^a)^2\,,\\
    C_2&\sim -\frac{r^2}{Q_1}dt\wedge dy+\frac{Q_5}{2}\,\cos2\theta\,d\varphi_2\wedge d\varphi_1\,,\qquad e^{2\Phi_{10}}\sim\frac{Q_1}{Q_5}\,,
\end{split}
\end{align}
where the metric $d\Omega_3^2$ of  three-dimensional sphere with unit radius is
\begin{align}\label{s3-metric}
    d\Omega_3^2 &=\cos^2\theta d\varphi_1^2+ \sin^2\theta d\varphi_2^2+d\theta^2\no\\
    &=\frac{1}{4}(d\psi+\cos2\theta\,d\phi)^2+\sin^2\theta\cos^2\theta\,d\phi^2+d\theta^2\,.
\end{align}
The asymptotic behavior is used to determine the values of $Q_1$ and $Q_5$.
The total RR charges $Q_1$ and $Q_5$ are normalized as
\begin{align}
\begin{split}\label{charge-def}
        Q_1&=\frac{1}{2\text{Vol}(S^3\times T^4)}\int_{S^3_{\infty}\times T^4_{\infty}} \star_{10}F_{3}\,,\\
    Q_5&=\frac{1}{2\text{Vol}(S^3)}\int_{S^3_{\infty}} F_3\,,
\end{split}
\end{align}
where $S^3_{\infty}$ denotes the three-sphere with the metric (\ref{s3-metric}) at large $r$, enclosing all the horizons.

\subsubsection*{Regularity and thermal equilibrium conditions}

For given RR charges $Q_1, Q_5$, the multi-BTZ black hole with arbitrary $l_i$ and $R_y$ does not necessarily yield regular bubbles. To ensure regularity, it is necessary to appropriately impose constraints on these parameters. In addition, we also impose the thermodynamic equilibrium conditions at each horizon.
After imposing all these constraints, the remaining parameters of the multi-BTZ black hole are reduced to the radius $R_y$ in the $y$-direction and the black hole temperature $T$.

The regularity near each rod can be conveniently analyzed based on the associated local spherical coordinates $(r_i,\theta_i)$. 
In the vicinity of the $(2s-1)$-th and $2s$-th rods, which describe the horizon and the bubble respectively, the multi-BTZ black holes can be rewritten in terms of the local spherical coordinates $(r_{2s-1},\theta_{2s-1})$ and $(r_{2s},\theta_{2s})$ as follows:
\begin{align}\label{2s-1-metric}
       ds_{10}^2
        &\propto dr_{2s-1}^2-\frac{1}{Q_1Q_5} \frac{r^2(r_{2s-1}^2+l_{2s-1}^2)}{W_bGF_{2s-1}}dt^2+ds^2_{\text{horizon}_{2s-1}}\,,\\
 ds_{10}^2\bigl\lvert_{dt=0}
    &\propto dr_{2s}^2+\frac{1}{Q_1Q_5}\frac{(r^2+l^2)(r_{2s}^2+l_{2s}^2)W_b}{GF_{2s}}\,dy^2+ds_{\text{bubble}_{2s}}^{2}\,,\label{2s-metric}
\end{align}
where we used the relation
\begin{align}
\begin{split}
    &\frac{dr^2}{r^2+l^2}+d\theta^2
    =F_{i}(r,\theta)\left(\frac{dr_i^2}{r_i^2+l_i^2}+d\theta_i^2\right)\,,\\
       &F_{i}(r,\theta)=\frac{(r_i^2+l_i^2\cos^2\theta_i)(r_i^2+l_i^2\sin^2\theta_i)}{(r^2+l^2\cos^2\theta)(r^2+l^2\sin^2\theta)}\,,
\end{split}
\end{align}
and the line elements $ds^2_{\text{horizon}_{2s-1}}, ds_{\text{bubble}_{2s}}^{2}$ describe the geometries of the $(2s-1)$-th horizon and the $(2s)$-th bubble, respectively.
In the $\rho \to 0^+$ limit in the region where each rod exists,  (\ref{2s-1-metric}) and (\ref{2s-metric}) reduce to
\begin{align}
   ds_{10}^2
        &\propto dr_{2s-1}^2-\frac{r_{2s-1}^2}{C^2_{2s-1}}dt^2+ds^2_{\text{horizon}_{2s-1}}\,,\label{eq:btz_horizon}\\
  ds_{10}^2\bigl\lvert_{dt=0}&\propto dr_{2s}^2+\frac{r_{2s}^2}{C^2_{2s}}dy^2+ds_{\text{bubble}_{2s}}^{2}\,,\label{eq:btz_bubble}
\end{align}
where the constants $C^2_{i}\,(i=1,\dots,n)$ are given by 
\begin{align}
C_{i}^2=\frac{kQ_1Q_5l_{i}^2}{l^4}d_{i}^2\,.
\end{align}
Here, the quantity $d_i^2$ is given by \cite{Bah:2022pdn}
\begin{align}\label{d-formula}
    d_i^2=\prod_{p=1}^{i-1}\prod_{q=i+1}^{n}\left[\frac{1+\frac{l_q^2}{\sum_{k=p}^{q-1}l_k^2}}{1+\frac{l_q^2}{\sum_{k=p+1}^{q-1}l_k^2}}\right]^{\alpha_{pq}}\prod_{p=1}^{i-1}\left(1+\frac{l_p^2}{\sum_{k=p+1}^{i}l_k^2}\right)^{\alpha_{ip}}
    \prod_{p=i+1}^{n}\left(1+\frac{l_p^2}{\sum_{k=i}^{p-1}l_k^2}\right)^{\alpha_{ip}}\,,
\end{align}
where $\alpha_{pq}=1$ for $p,q \in U_t$ or $p,q \in U_y$, and $\alpha_{pq}=0$ for $p\in U_t, q\in U_y$ or $p\in U_y, q\in U_t$.
Recalling that the $y$-direction is compactified with radius $R_y$ ($y\simeq y+2\pi R_y$), the thermal equilibrium and the regularity conditions for each horizon and bubble impose on the following constraints:
\begin{align}
    T&=\frac{l^2}{2\pi \sqrt{Q_1Q_5}l_id_i}\qquad \text{for}\quad i\in U_t\,,\label{horizon-reg}\\
    R_y&=\frac{\sqrt{Q_1Q_5}l_id_i}{l^2}\qquad\quad \text{for}\quad i\in U_y\,.\label{bubble-reg}
\end{align}
The condition fixes the length parameter $l_i^2$ for each rod in terms of $Q_1,Q_5,k,R_y$ and $T$.
Finding a solution to this balance condition for arbitrary $n$ is challenging, and explicit solutions are available only for a few small values of $n$ and for very large $n$.

\subsubsection*{Potential function}

Finally, we provide an expression for the multi-BTZ black holes in terms of the potential functions given in (\ref{d1d5-bg}). From the multi-BTZ solution (\ref{btzs-sol}) combined with (\ref{d1d5-bg}), we can read off the corresponding scalar functions as follows:
\begin{align}
\begin{split}\label{mbtz-pot}
  Z_0&= \frac{4}{r^2\sqrt{1+\frac{l^2}{r^2}}}\,,\quad
    Z_1=\frac{Q_1}{r^2\sqrt{1+\frac{l^2}{r^2}}}\,,\quad
    Z_5=\frac{Q_5}{r^2\sqrt{1+\frac{l^2}{r^2}}}\,,\\
   W&=\sqrt{1+\frac{l^2}{r^2}}\prod_{i\in U_y}\left(1+\frac{l_i^2}{r_i^2}\right)^{-1}\,,\\
   H_0&=\cos2\theta\,,\qquad  T_1=\frac{r^2+\frac{l^2}{2}}{Q_1}\,,\qquad H_5=\frac{Q_5}{4}\cos 2\theta\,,\\
    T_0&=\frac{r^2+\frac{l^2}{2}}{4k}\,,\qquad  H_1=\frac{Q_1}{4}\cos 2\theta\,,\qquad T_5=\frac{r^2+\frac{l^2}{2}}{Q_5}\,.
\end{split}
\end{align}
The scalar function $G$ is related to the conformal factor $e^{2\nu}$ through the equation
\begin{align}\label{conf-mbtz}
    e^{2\nu}=G(r,\theta)\frac{r^2(r^2+l^2)}{(r^2+l^2\cos^2\theta)(r^2+l^2\sin^2\theta)}\,.
\end{align}

\subsection{Concrete examples of multi-BTZ solutions}

Here, we present two simple examples of the multi-BTZ black hole solution (\ref{btzs-sol}): (i) BTZ black hole and (ii) black bipole solution.

\subsubsection{BTZ black hole}

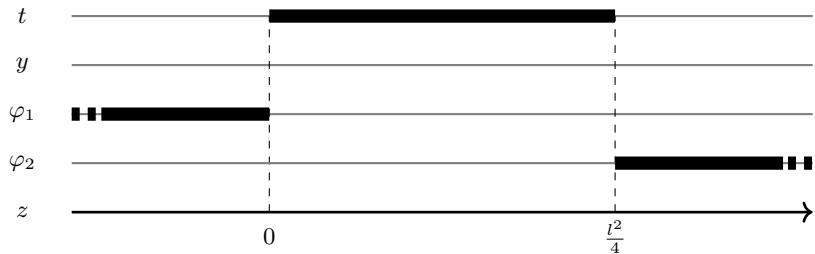
\begin{figure}
\begin{center}
\begin{tikzpicture}[scale=0.65]
\node[font=\small ] at (-10,4) {$t$};
\node[font=\small ] at (-10,3) {$y$};
\node[font=\small ] at (-10,2) {$\varphi_1$};
\node[font=\small ] at (-10,1) {$\varphi_2$};
\node[font=\small ] at (-10,0) {$z$};
\node[font=\small ] at (-5,-0.5) {$0$};
\node[font=\small ] at (2,-0.5) {$\frac{l^2}{4}$};
\draw[gray,line width = 0.8] (-9,4) -- (6,4);
\draw[black,line width = 5] (-5,4) -- (2,4);
\draw[gray,line width = 0.8] (-9,3) -- (6,3);
\draw[gray,line width = 0.8] (-9,2) -- (6,2);
\draw[black,line width = 5,dashed ] (-9,2) -- (-8.5,2);
\draw[black,line width = 5] (-8.4,2) -- (-5,2);
\draw[black,dashed ] (-5,0) -- (-5,4);
\draw[black,dashed ] (2,0) -- (2,4);
\draw[gray,line width = 0.8] (-9,1) -- (6,1);
\draw[black,line width = 5] (2,1) -- (5.4,1);
\draw[black,line width = 5,dashed ] (5.5,1) -- (6,1);
\draw[->,black,line width = 1] (-9,0) -- (6,0);
\end{tikzpicture}
\caption{The rod structure of BTZ black hole solution }\label{rod-btz-able}
\end{center}
\end{figure}
The multi-BTZ black hole solutions (\ref{btzs-sol}) coincides with a single BTZ black hole solution,  which corresponds to $U_{t}=\{1\}$ and $U_y=\emptyset$.
In this case, the functions $G$ and $W_b$ become $G=1$ and $W_b=1$, and then the solution (\ref{btzs-sol}) simplifies to
\begin{align}
\begin{split}\label{10dBTZ-sol}
    ds_{10}^2&=\frac{1}{\sqrt{Q_1Q_2}}\biggl[-r^2dt^2+(r^2+l^2) dy^2\biggr]+\sqrt{Q_1Q_5}\biggl[\frac{dr^2}{r^2+l^2}+d\Omega_3^2\biggr]+\sqrt{\frac{Q_1}{Q_5}}ds_{T^4}^2\,,\\
    C_2&=-\frac{r^2+l^2}{Q_1}dt\wedge dy+\frac{Q_5}{2}\,\cos2\theta\,d\varphi_2\wedge d\varphi_1\,,\qquad e^{\Phi}=\sqrt{\frac{Q_1}{Q_5}}\,,
\end{split}
\end{align}
where we replaced $r_1$ with $r$.
Then, the scalar functions $(Z_{0,1,5},W,e^{2\nu})$ take the forms
\begin{align}
\begin{split}
    Z_0&=\frac{4}{r^2\sqrt{1+\frac{l^2}{r^2}}}\,,\qquad Z_1=\frac{Q_1}{r^2\sqrt{1+\frac{l^2}{r^2}}}\,,\qquad Z_5=\frac{Q_5}{r^2\sqrt{1+\frac{l^2}{r^2}}}\,,\\
   W&=\sqrt{1+\frac{l^2}{r^2}}\,,\qquad e^{2\nu}=\frac{r^2(r^2+l^2)}{(r^2+l^2\cos^2\theta)(r^2+l^2\sin^2\theta)}\,.
\end{split}
\end{align}
If we perform a reduction over $S^3\times T^4$, the solution (\ref{10dBTZ-sol}) describes the static BTZ black hole.
The temperature from (\ref{horizon-reg}) is given by
\begin{align}
    T=\frac{l}{2\pi \sqrt{Q_1Q_5}}\,.
\end{align}
The extremal limit corresponds to taking the limit $l\to 0$.

\subsubsection{Black bi-pole case}

Next, we consider the black bi-pole solution,  which corresponds to the case $n=3$ with $ U_t=\{1,3\}, U_y=\{2\}$. The solution can be written as
\begin{align}
\begin{split}\label{btz2-sol}
    ds_{10}^2&=\frac{1}{\sqrt{Q_1Q_5}}\biggl[-r^2\left(1+\frac{l_2^2}{r_2^2}\right)dt^2+ \frac{r^2+l^2}{\left(1+\frac{l_2^2}{r_2^2}\right)}dy^2\biggr]+\sqrt{\frac{Q_1}{Q_5}}ds_{T^4}^2\\
    &\quad+\sqrt{Q_1Q_5}\biggl[G(r,\theta)\left(\frac{dr^2}{r^2+l^2}+d\theta^2\right)+\cos^2\theta\,d\varphi_1^2+\sin^2\theta\,d\varphi_2^2\biggr]\,,\\
    C_2&=-\frac{r^2+l^2}{Q_1}dt\wedge dy+\frac{Q_5}{2}\,\cos2\theta\,d\varphi_2\wedge d\varphi_1\,,\qquad e^{2\Phi_{10}}=\frac{Q_1}{Q_5}\,,
\end{split}
\end{align}
where the scalar function $G(r,\theta)$ is given by
\begin{align}
    G&=\prod_{\substack{1,3\in U_t}}\left(\frac{\left(r^2_2\cos^2\theta_2+(r_j^2+l_j^2)\sin^2\theta_j\right)
    \left((r_2^2+l_2^2)\cos^2\theta_2+r_j^2\sin^2\theta_j\right)}{\left((r_2^2+l_2^2)\cos^2\theta_2+(r_j^2+l_j^2)\sin^2\theta_j\right)\left(r_2^2\cos^2\theta_2+r_j^2\sin^2\theta_j\right)}\right)\,.
\end{align}
The associated scalar functions are
\begin{align}
\begin{split}
    Z_0&= \frac{4}{r^2\sqrt{1+\frac{l^2}{r^2}}}\,,\quad
    Z_1=\frac{Q_1}{r^2\sqrt{1+\frac{l^2}{r^2}}}\,,\quad
    Z_5=\frac{Q_5}{r^2\sqrt{1+\frac{l^2}{r^2}}}\,,\\
   W&=\sqrt{1+\frac{l^2}{r^2}}\left(1+\frac{l_2^2}{r_2^2}\right)^{-1}\,.
\end{split}
\end{align}
\begin{figure}
\begin{center}
\begin{tikzpicture}[scale=0.65]
\node[font=\small ] at (-10,4) {$t$};
\node[font=\small ] at (-10,3) {$y$};
\node[font=\small ] at (-10,2) {$\varphi_1$};
\node[font=\small ] at (-10,1) {$\varphi_2$};
\node[font=\small ] at (-10,0) {$z$};
\node[font=\small ] at (-6,-0.5) {$0$};
\node[font=\small ] at (-2,-0.5) {$\frac{l^2_1}{4}$};
\node[font=\small ] at (2,-0.5) {$\frac{l^2_1+l_2^2}{4}$};
\node[font=\small ] at (6,-0.5) {$\frac{l^2}{4}$};
\draw[gray,line width = 0.8] (-9,4) -- (9,4);
\draw[black,line width = 5] (-6,4) -- (-2,4);
\draw[black,line width = 5] (2,4) -- (6,4);
\draw[gray,line width = 0.8] (-9,3) -- (9,3);
\draw[black,line width = 5] (-2,3) -- (2,3);
\draw[gray,line width = 0.8] (-9,2) -- (9,2);
\draw[black,line width = 5,dashed ] (-9,2) -- (-8.5,2);
\draw[black,line width = 5] (-8.4,2) -- (-6,2);
\draw[black,dashed ] (-6,0) -- (-6,4);
\draw[black,dashed ] (-2,0) -- (-2,4);
\draw[black,dashed ] (2,0) -- (2,4);
\draw[black,dashed ] (6,0) -- (6,4);
\draw[gray,line width = 0.8] (-9,1) -- (9,1);
\draw[black,line width = 5] (6,1) -- (8.4,1);
\draw[black,line width = 5,dashed ] (8.5,1) -- (9,1);
\draw[->,black,line width = 1] (-9,0) -- (9,0);
\end{tikzpicture}
\caption{The rod structure of black bi-pole solution }\label{rod-m2btz-able}
\end{center}
\end{figure}
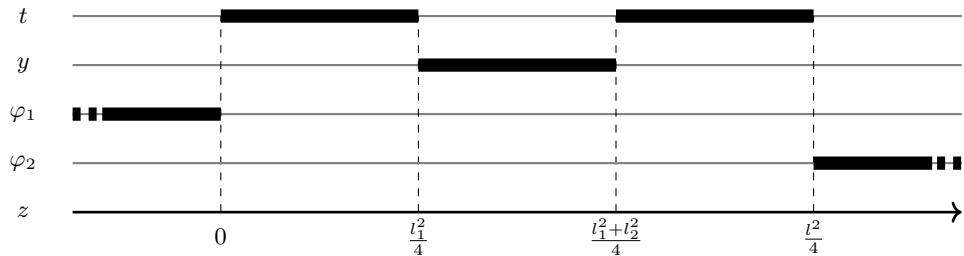
This solution describes two  black holes with an $S^3\times S^1\times T^4$-topology, localized at both poles $\theta=0$ and $\theta=\pi/2$ of the $S^3$, where the two three-spheres are spanned by $(\theta,\varphi_1,y)$ and $(\theta,\varphi_2,y)$, respectively.
In this case, the thermal equilibrium conditions (\ref{horizon-reg}) on horizons and the regularity conditions (\ref{bubble-reg}) on bubbles  can be solved explicitly in terms of $l_i$ \cite{Bena:2024gmp}
\footnote{The condition (\ref{horizon-reg}) of thermal equilibrium has been solved under the assumption that all horizons have the same temperature $T$. It is also possible to solve the constraints in which each horizon has a different temperature, following a similar procedure. In such cases, $l_1$ and $l_3$ are no longer equal, and one can consider limits of temperature in which the bi-pole black hole reduces to, for example, a black pole \cite{Bena:2024gmp} with a single horizon located near the north pole of the S$^3$.}
\begin{align}
\begin{split}\label{birod-sol}
    l_1&=l_3=\frac{l^2}{2}\left(1-\frac{\pi T}{\sqrt{1+\pi^2 T^2}}\right)\,,\quad 
l_2=\frac{\pi l^2 T}{\sqrt{1+\pi^2 T^2}}\,,
\end{split}
\end{align}
and the total length parameter $l$ becomes
\begin{align}
        l^2&=\frac{4\pi^2Q_1Q_5 T^2(\sqrt{1+\pi^2 T^2}-\pi T)^2}{R_y^2}\,.
\end{align}
From the expression (\ref{birod-sol}) with fixed $l^2$, in the low-temperature limit $T \to 0$, the black bi-pole reduces to the BTZ black hole, while in the high-temperature limit $T \to \infty$, it approaches global AdS$_3\times$S$^3\times$T$^4$.

\subsection{Multi-black string solution}\label{sec:mbst}

Next, we consider the multi-black string solutions with an asymptotic structure of $R^{1,4}\times S^{1}\times T^4$ in type IIB supergravity. 
The explicit expression of the solution is given in~Ref.\cite{Bah:2021owp} as follows: 
\begin{align}
\begin{split}\label{mbstring}
    ds_{10}^2&=ds_6^2+\sqrt{\frac{\hat{Z}_1}{\hat{Z}_5}}\sum_{a=1}^{4}dy^{a}dy^{a}\,,
    \qquad e^{2\Phi_{10}}=\frac{\hat{Z}_1}{\hat{Z}_5}\,,\\
    C_2&=-\frac{2ms_1c_1}{\hat{r}^2\hat{Z}_1}dt\wedge dy-2m\,s_5c_5\cos^2\theta\,d\varphi_1\wedge d\varphi_2\,,
\end{split}
\end{align}
where the six dimensional metric is
\begin{align}\label{6dmst-metric}
    ds_6^2&=\frac{1}{\sqrt{\hat{Z}_1\hat{Z}_5}}\biggl[-\left(1-\frac{2m}{\hat{r}^2}\right)\frac{dt^2}{W_{b}(r,\theta)}+W_{b}(r,\theta)\,dy^2\biggr]\no\\
    &\quad+\hat{r}^2\sqrt{\hat{Z}_1\hat{Z}_5}\biggl[G(r,\theta)\left(\frac{d\hat{r}^2}{\hat{r}^2-2m}+d\theta^2\right)+ \cos^2\theta d\varphi_1^2+\sin^2\theta d\varphi_2^2\biggr]\,.
\end{align}
The scalar functions $G(r,\theta)$ and $W_b(r,\theta)$ are common to both the multi-BTZ black hole solution and the multi-black string solution, as given by Eqs.~(\ref{conf-part}) and (\ref{bubble-part}), respectively. 
The functions $\hat{Z}_I\,(I=1,5)$ are characterized by real parameters $\delta_I$ and take the form
\begin{align}
        \hat{Z}_I&=1+\frac{2m\,s_I^2}{\hat{r}^2}\,,
\end{align}
where we introduce an abbreviated notation for hyperbolic functions:
\begin{align}
    s_I=\sinh\delta_I\,,\qquad c_I=\cosh\delta_I\,.
\end{align}
The real parameters $\delta_I$ parametrize the electric charges of the black string solution (\ref{d1d5p-e}).
Here, we assume that the multi-black string has the same D1 and D5 charges as the multi-BTZ black hole, in which case using the definition (\ref{charge-def}) the real parameters $\delta_I$ satisfy 
\begin{align}\label{flat-q1q5}
    Q_1&=2m  s_1c_1\,,\qquad
    Q_5=2m  s_5c_5\,.
\end{align}
Additionally, we have introduced a new radial coordinate $\hat{r}$ such that the simplest example ($n=1$) of Eq.~(\ref{mbstring}), the non-extremal static black string \cite{Cvetic:1998xh}, takes a standard form:
\begin{align}\label{d1d5p-e}
    ds_6^2&=\frac{1}{\sqrt{\hat{Z}_1\hat{Z}_5}}\biggl[-\left(1-\frac{2m}{\hat{r}^2}\right)dt^2+dy^2\biggr]+\hat{r}^2\sqrt{\hat{Z}_1\hat{Z}_5}\biggl[\frac{d\hat{r}^2}{\hat{r}^2-2m}+ d\Omega_3^2\biggr]\,.
\end{align}
This solution is an uplift to six dimensions of the non-rotating two-charge subclass of five dimensional Cvetic-Youm black holes \cite{Cvetic:1996xz,Cvetic:1998xh}.
The relation between $\hat{r}$ and $r$ is given by
\begin{align}
    \hat{r}^2=r^2+2m\,.
\end{align}
Their geometric structure is characterized by the same rod structure as that of the multi-BTZ black holes, as illustrated in Fig.\,\ref{rod-mbtz-able}.

\subsubsection*{Regularity and thermal equilibrium conditions}

Let us examine the regularity of the multi-black string solutions. To do so, we work on the radial coordinate $r$ instead of $\hat{r}$, and then the 6D black string part of (\ref{mbstring}) takes the form 
\begin{align}
        ds_6^2&=\frac{1}{(r^2+l^2)\sqrt{\hat{Z}_1\hat{Z}_5}}\biggl[-\frac{r^2\,dt^2}{W_{b}}+(r^2+l^2)W_{b}\,dy^2\biggr]\no\\
    &\quad+(r^2+l^2)\sqrt{\hat{Z}_1\hat{Z}_5}\biggl[G(r,\theta)\left(\frac{dr^2}{r^2+l^2}+d\theta^2\right)+ \cos^2\theta d\varphi_1^2+\sin^2\theta d\varphi_2^2\biggr]\,,
\end{align}
where the scalar functions $\hat{Z}_I$ become
\begin{align}
    \hat{Z}_I=1+\frac{l^2s_I^2}{r^2+l^2}\,.
\end{align}
Similar to Eqs.~(\ref{eq:btz_horizon}) and (\ref{eq:btz_bubble}) for the multi-BTZ black hole, we rewrite the multi-black string solution in terms of the local spherical coordinates around the $(2s-1)$-th and $2s$-th rods, as described in
\begin{align}
       ds_{10}^2
        &\propto dr_{2s-1}^2-\frac{1}{(r^2+l^2)^2\hat{Z}_1\hat{Z}_5} \frac{r^2(r_{2s-1}^2+l_{2s-1}^2)}{W_bGF_{2s-1}}dt^2+ds(\cC_{\text{horizon}}^{(2s-1)})^2\,,\\
 ds_{10}^2\bigl\lvert_{dt=0}
    &\propto dr_{2s}^2+\frac{1}{(r^2+l^2)^2\hat{Z}_1\hat{Z}_5}\frac{(r^2+l^2)(r_{2s}^2+l_{2s}^2)W_b}{GF_{2s}}\,dy^2+ds(\cC_{\text{bubble}}^{(2s)})^2\,.
\end{align}
In the limit approaching each rod, the product of the scalar functions $\hat{Z}_I$ and $(r^2+l^2)$ becomes constants
\begin{align}
    \lim_{\rho\to 0^+}(r^2+l^2)\hat{Z}_I=l^2c_I^2\,.
\end{align}
Thus, we obtain the same thermal equilibrium and balance conditions as in the case of the multi-BTZ black hole, except that $Q_I$ are replaced with $l^2c_I^2$. The regular multi-black string can be realized by solving these constraints

\subsubsection*{Potential function}

The multi-black string solutions (\ref{mbstring}) are also expressed by using the ansatz~(\ref{d1d5-bg}).
By comparing Eq.~(\ref{mbstring}) with Eq.~(\ref{d1d5-bg}), the potential functions $(Z_{0,1,5},W,T_{1,5})$ can be written as
\begin{align}
\begin{split}\label{bs-scalar-z}
    Z_0&=\frac{4}{\hat{r}^2\sqrt{1-\frac{2m}{\hat{r}^2}}} \,,\qquad Z_1=\frac{\hat{Z}_1}{ \sqrt{1-\frac{2m}{\hat{r}^2}}}\,,\qquad Z_5=\frac{\hat{Z}_5}{ \sqrt{1-\frac{2m}{\hat{r}^2}}}\,,\\
    W&=\frac{1}{\sqrt{1-\frac{2m}{\hat{r}^2}}}\,,\qquad T_1=\frac{2ms_1c_1}{\hat{r}^2\hat{Z}_1}\,,\qquad T_5=\frac{2ms_5c_5}{\hat{r}^2\hat{Z}_5}\,.
\end{split}
\end{align} 
As we will see in the next section, the potential functions characterize the scalar moduli in the gravitational solutions, which form a coset space with isometry group $SO(4,4)$. The $SO(4,4)$ transformations act transitively on this coset space and, interestingly, realize an interpolation between the multi-BTZ black hole and the multi-black string solutions. This fact significantly simplifies the construction of the multi-BTZ black hole via the Riemann-Hilbert problem. While the details of the relation between these two solutions will be discussed in Sec.\,\ref{sec:harrison},  we briefly outline here, as preparation, which potential functions are subject to the action of the $SO(4,4)$ transformations.
To make this connection more transparent, it is convenient to use the $r$ coordinate and set $l^2=2m$. Then the potential functions (\ref{bs-scalar-z}) take the form
\begin{align}
\begin{split}\label{warp-mst}
    Z_0&=\frac{4}{r^2\sqrt{1+\frac{l^2}{r^2}}} \,,\quad Z_1=\left(1+\frac{l^2(1+s_1^2)}{r^2}\right)\frac{1}{ \sqrt{1+\frac{l^2}{r^2}}}\,,\quad Z_5=\left(1+\frac{l^2(1+s_5^2)}{r^2}\right)\frac{1}{ \sqrt{1+\frac{l^2}{r^2}}}\,,\\
    W&=\sqrt{1+\frac{l^2}{r^2}}\prod_{i\in U_y}\left(1+\frac{l_i^2}{r_i^2}\right)^{-1}\,,\qquad T_1=\frac{2ms_1c_1}{r^2+l^2(1+s_1^2)}\,,\qquad T_5=\frac{2ms_5c_5}{r^2+l^2(1+s_5^2)}\,.
\end{split}
\end{align}
By comparing (\ref{mbtz-pot}) and (\ref{warp-mst}), we find that the two scalar functions $Z_0$ and $W$ are identical in both solutions
\begin{align}
    Z_0\lvert_{\text{black string}}&=Z_0\lvert_{\text{BTZ}}\,,\\
  W\lvert_{\text{black string}}&=W\lvert_{\text{BTZ}}\,.
\end{align}
Thus, the $SO(4,4)$ transformations connecting two solutions act on the scalar functions $(Z_{1,5}, T_{1,5})$.
This aspect will be further elaborated in Sec.\,\ref{sec:harrison}.

\section{Coset description of D1-D5-KKm system and its classical integrability}\label{sec:reduction}

This section outlines how, through an appropriate dimensional reduction, the type IIB supergravity action for the D1-D5-KKm systems (with the ansatz~(\ref{d1d5-bg})) can be reformulated as a 2D integrable coset sigma model coupled to dilaton gravity. 
Subsequently, we provide a brief discussion on the classical integrability of the resulting 2D coset sigma model, viewed from the perspective of the Breitenlohner-Maison linear system.

\subsection{Summary of dimensional reduction from 10D to 3D}

\begin{figure}
\begin{center}
\begin{tikzpicture}[scale=0.65]
\node[font=\small ] at (-6,1) {10D type IIB supergravity (\ref{IIBaction})};
\node[font=\small ] at (5,1) {6D supergravity (\ref{6dsugra-action})};
\node[font=\small ] at (-6,-4) {5D $U(1)^3$ supergravity (\ref{5d_sugra_action})};
\node[font=\small ] at (5,-4) {4D Euclidean STU model (\ref{4dstu})};
\node[font=\small ] at (-6,-7) {3D theory (\ref{3daction-iib})};
\node[font=\small ] at (-1,2) {$T^4$ reduction};
\node[font=\small ] at (-1,-0.4) {$S^1(y)$ reduction};
\node[font=\small ] at (-1,-3) {$S^1(t)$ reduction};
\node[font=\small ] at (-1,-6) {$S^1(\psi)$ reduction};
\draw[->,black,line width = 1] (-2,1) -- (1,1);
\draw[-,black,line width = 1] (5,0.5) -- (5,-1);
\draw[-,black,line width = 1] (5,-1) -- (-6,-1);
\draw[->,black,line width = 1] (-6,-1) -- (-6,-3);
\draw[->,black,line width = 1] (-2,-4) -- (1,-4);
\draw[-,black,line width = 1] (5,-4.5) -- (5,-7);
\draw[->,black,line width = 1] (5,-7) -- (-3,-7);
\end{tikzpicture}\caption{A sequence illustrating the order of dimensional reductions and the corresponding reduced supergravity theories.}\label{sugra-map}
\end{center}
\end{figure}
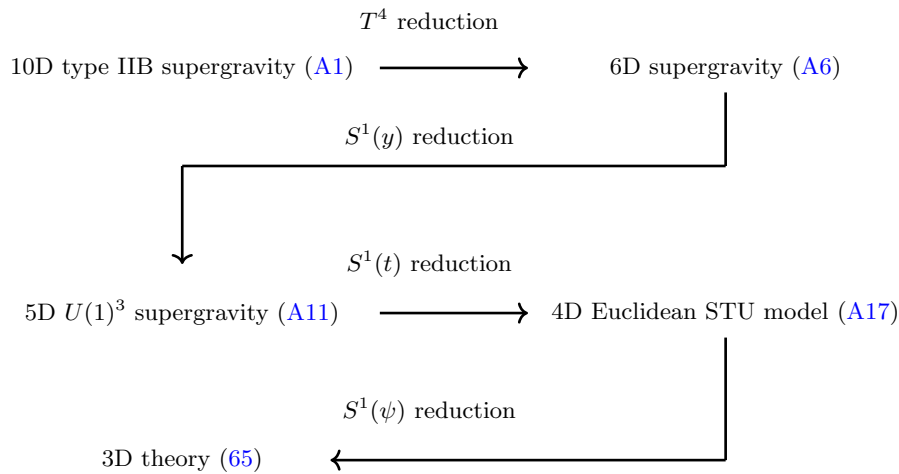

For our purposes, we consider a dimensional reduction of type IIB supergravity from 10D to 2D.
We begin by reducing the 10D theory to a 3D theory by compactifying along the Killing directions $S^1(t) \times S^1(y) \times S^1(\psi) \times T^4$.
The order of the dimensional reductions does not affect the discussion of classical integrability.
In this work, we proceed with the following sequence of the reductions:
\begin{align}
    T^4\to S^1(y) \to S^1(t) \to S^1(\psi)\,.
\end{align}
The supergravity theories that appear at each step of the dimensional reduction are summarized in Fig.\,\ref{sugra-map}.
To elucidate the underlying classical integrability, it is useful to appropriately dualize the RR fields and gauge fields at each stage of the dimensional reductions. By following, for example, \cite{Figueras:2009mc,Virmani:2012kw,Sahay:2013xda,Roy:2018ptt} (see also \cite{Chakraborty:2025ger}), we implement the following two dualization procedures:
\begin{itemize}
    \item In five dimensions, we dualize the RR 2-form potential into a gauge field (e.g. \cite{Virmani:2012kw})
    \item In three dimensions, we dualize the vector fields into scalar fields (e.g. \cite{Sahay:2013xda}) 
\end{itemize}
The latter procedure is often referred to as an Ehlers reduction.
For further details on the full reduction procedure, see appendix \ref{sec:dim_red} and \cite{Figueras:2009mc,Virmani:2012kw,Sahay:2013xda,Roy:2018ptt}.

Through the above sequence of the dimensional reductions including the dualization procedure, the compactified space of the original 10D gravitational solution can be fully characterized by 16 scalar fields,
\begin{align}
    \{\varphi^a\}=\{U,x^I,y^I, \tilde{\zeta}_\Lambda,\zeta^\Lambda,\sigma\}\,,
\end{align}
where $I=1,2,3$ and $\Lambda=0,1,2,3$.
The bosonic part of type IIB supergravity reduces to a coset sigma model coupled to 3D gravity with the metric,
\begin{align}\label{3d-metric}
\begin{split}
    ds_3^2&=g_{3,mn}dx^mdx^n=e^{2\nu}(d\rho^2+dz^2)+\rho^2 d\phi^2\,.
\end{split}
\end{align}
and the resulting 3D action is given by
\begin{align}\label{3daction-iib}
    S_{\rm 3D}=\int d^3x\sqrt{{\rm det}g_{3}}\left(R_3-\frac{1}{2}G_{ab}(\varphi)\partial^{\mu}\varphi^a\partial_{\mu}\varphi^b\right)\,,
\end{align}
where $G_{ab}$ is the metric of the moduli space of the scalar fields $\varphi^a$ described by a symmetric coset
\begin{align}\label{so44-coset-iib}
    \frac{SO(4,4)}{SO(2,2)\times SO(2,2)}\,.
\end{align}
Each Lie group has a $8\times 8$ matrix realization defined by
\begin{align}
    G&=SO(4,4)=\{~g \in GL(8,\mathbb{R}) ~\lvert~ g^{T}\eta g=\eta\,,~ {\rm det}g=1~\}\,,\\
    H&=SO(2,2)\times SO(2,2)=\{~g \in SO(4,4)~ \lvert ~g^{T}\eta' g=\eta'~\}\,,
\end{align}
where the invariant metrics $\eta$ and $\eta'$ for $G$ and $H$ are
\begin{align}
    \eta=
    \begin{pmatrix}
        0_4&1_4\\
        1_4&0_4
    \end{pmatrix}\,,\qquad
    \eta'=\text{diag}(-1,-1,1,1,-1,-1,1,1)\,.
\end{align}
The semisimple Lie algebra $\mathfrak{g}=\mathfrak{so}(4,4)$ is spanned by the 28 generators $\{H_{\Lambda}, E_{\Lambda},E_{q_{\Lambda}}, E_{p^{\Lambda}}\}$, and their matrix representations are presented in appendix \ref{sec:so44_rep}.

The symmetric coset (\ref{so44-coset-iib}) has the $\mathbb{Z}_2$ grading structure which is a crucial property for the construction of the Lax pair of the underlying two-dimensional coset sigma model.
To see this, we decompose $\mathfrak{g}$ as a vector space like 
\begin{align}
    \mathfrak{g}=\mathfrak{h}\oplus \mathfrak{p}\,,
\end{align}
where $\mathfrak{h}=\mathfrak{so}(2,2)\times \mathfrak{so}(2,2)$.
The $\mathbb{Z}_2$ grading structure of the coset (\ref{so44-coset-iib}) follows from the fact that the decomposition satisfies the commutation relations 
\begin{align}
    [\mathfrak{h},\mathfrak{h}]\subset \mathfrak{h}\,,\quad 
 [\mathfrak{h},\mathfrak{p}]\subset \mathfrak{p}\,,\quad 
 [\mathfrak{p},\mathfrak{p}]\subset \mathfrak{h}\,,
\end{align}
which are preserved under the $\mathbb{Z}_2$ involution $\tau:\mathfrak{g}\to \mathfrak{g}$ defined by
\begin{align}
     \tau(x)=
     \begin{cases}
         x \qquad &x \in \mathfrak{h}\\
         -x \qquad &x \in \mathfrak{p}
     \end{cases}
     \,.
\end{align}
When a matrix realization of $\mathfrak{so}(4,4)$ is given, the automorphism $\tau$ can be explicitly defined by
\begin{align}\label{tau-rep}
    \tau(x)=-\eta' \cdot x^{T} \cdot \eta'\,.
\end{align}
Such an automorphism $\tau$ defines the $\mathbb{Z}_2$ grading structure for the symmetric coset (\ref{so44-coset-iib}).

\subsection{Classical integrability of 2D coset sigma model}

Finally, we perform the dimensional reduction along the angular direction $\phi$, and then the 3D action (\ref{3d-metric}) reduces to 2D symmetric coset sigma model coupled with a dilaton gravity. 
To facilitate access to the integrable structure, it is useful to specify a representative element of the coset (\ref{so44-coset-iib}) and rewrite the symmetric coset sigma model action in terms of it.

The coset representative $V(z,\rho)\in G$ in the coset space $H\backslash G$ is subject to a gauge transformation from the left by an element $h(z,\rho) \in H$, whereas a group element $g \in G$ acts transitively from the right i.e. 
\begin{align}
    V(z,\rho)\mapsto h(z,\rho)V(z,\rho)g\,.
\end{align}
By performing a gauge transformation, we can fix the coset representative in the Iwasawa gauge and parametrize it in terms of 16 scalar fields $\{\varphi^a\}$ as \cite{Bossard:2009we}
\begin{align}\label{iwasawa-rep}
    V&=e^{-U\,H_0}\cdot\left(\prod_{I=1}^{3}e^{-\frac{1}{2}(\log y^I)H_I}\cdot e^{-x^IE_{I}}\right)\cdot e^{-\zeta^{\Lambda}E_{q_{\Lambda}}-\tilde{\zeta}_{\Lambda}E_{p^{\Lambda}}}\cdot e^{-\frac{1}{2}\sigma E_0}\,.
\end{align}
When $g\in G$ acts on (\ref{iwasawa-rep}), $V(z,\rho)$ deviates from the Iwasawa gauge, making it difficult to directly extract the corresponding 16 scalar fields $\{\varphi^a\}$. Hence, it is convenient to introduce a gauge invariant element
\begin{align}\label{ginv-M}
    M(z,\rho)=V^{\natural}(z,\rho)V(z,\rho) \in G\,,
\end{align}
where $\natural:G\to G$ is an anti-involutive automorphism 
\begin{align}
    x^{\natural}=\eta' x^{T}\eta'\qquad \text{for}\quad x\in G\,.
\end{align}
In the matrix representation of $\mathfrak{g}=\mathfrak{so}(4,4)$ that we are using, $\natural$ satisfies $h^{\natural}=h^{-1}$ for $h\in H$ by comparing with the definition (\ref{tau-rep}) of $\tau$.
Since $(g_1g_2)^{\natural}=g_2^{\natural}g_1^{\natural}$ for $g_1,g_2\in G$, $M$ satisfies $M^{\natural}=M$ and transforms under the action of $g\in G$ as 
\begin{align}
    M'(z,\rho)=g^{\natural}M(z,\rho)g\,.
\end{align}

By employing the gauge invariant element $M(z,\rho)$, we can write down the resulting 2D action as
\begin{align}\label{2d-gravity-sigma}
    S_{\rm 2D}=\int d\rho dz\,\sqrt{g_2}\,\rho\Bigl[R_2-2g_2^{mn}\Tr(\partial_{m}MM^{-1}\,\partial_{n}MM^{-1})\Bigr]\,.
\end{align}
The 2D nonlinear sigma model is defined on the conformally flat space:
\begin{align}
    ds_2^2=e^{2\nu}(d\rho^2+dz^2)\,.
\end{align}
If we interpret $\rho$ as a dilaton $\Phi$, this model can be regarded as a 2D sigma model coupled to 2D dilaton gravity.
The equations of motion for the sigma model are
\begin{align}\label{sigma-eom}
    &\partial_{\rho}(\rho\,\partial_{\rho}MM^{-1})+\partial_{z}(\rho\,\partial_{z}MM^{-1})=0\,,\\
    &\partial_{\pm}\left(\ln e^{2\nu}\right)=\mp i\frac{\rho}{8}\Tr(\partial_{\pm}MM^{-1}\partial_{\pm}MM^{-1})\,,\label{conf-eom}
\end{align}
where we introduced
\begin{align}
    x^{\pm}=\frac{1}{2}(z\mp i \rho)\,,\qquad \partial_{\pm}=\partial_z\pm i \partial_{\rho}\,.
\end{align}
Equations~(\ref{conf-eom}) are so-called Virasoro constraints in the context of string sigma models.
We see that Eq.~(\ref{sigma-eom}) are equivalent to the set of equations of motion (\ref{gen-eom1})--(\ref{gen-eom3}) for the scalar functions $(Z_I, T_I, W)$.

\subsubsection*{Classical integrability}

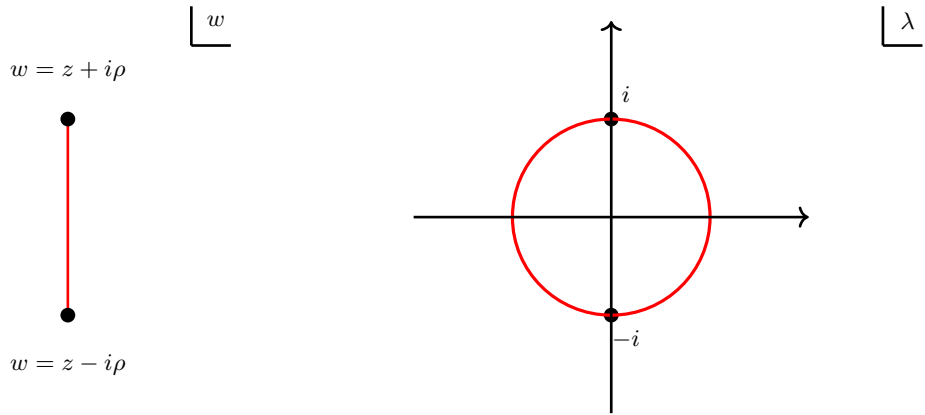
\begin{figure}
\begin{center}
\begin{tikzpicture}[scale=0.65]
\node[font=\small ] at (-10,3) {$w=z+i\rho$};
\node[font=\small ] at (-10,-3) {$w=z-i \rho$};
\node[font=\small ] at (-7,4) {$w$};
\node[font=\small ] at (7,4) {$\la$};
\node[font=\small ] at (1.3,2.5) {$i$};
\node[font=\small ] at (1.3,-2.5) {$-i$};
\fill[black] (1,2) circle (0.15);
\fill[black] (1,-2) circle (0.15);
\draw[-,red,line width = 1] (-10,-2) -- (-10,2);
\fill[black] (-10,2) circle (0.15);
\fill[black] (-10,-2) circle (0.15);
\draw[-,black,line width = 1] (-7.5,4.3) -- (-7.5,3.5);
\draw[-,black,line width = 1] (-7.5,3.5) -- (-6.7,3.5);
\draw[-,black,line width = 1] (6.5,4.3) -- (6.5,3.5);
\draw[-,black,line width = 1] (6.5,3.5) -- (7.3,3.5);
\draw [red,very thick](1,0)circle[radius=2];
\draw[->,black,line width = 1] (1,-4) -- (1,4);
\draw[->,black,line width = 1] (-3,0) -- (5,0);
\end{tikzpicture}
\caption{$w$- and $\la$-planes. The red line in the $w$-plane represents the branch cut connecting $w=z\pm i\rho$ and is mapped to the unit circle in the $\la$-plane. The possible values of $\la_{(\pm)}$ are determined by the rational curve (\ref{r-alg}).}
\label{bm-surface}
\end{center}
\end{figure}

The sigma model given in (\ref{2d-gravity-sigma}) is known to be classically integrable, even in the presence of the overall prefactor $\rho$ multiplying the usual coset sigma model Lagrangian. The (weak) classical integrability of a 2D field theory is characterized by the existence of a Lax pair—a connection on the two-dimensional spacetime that depends meromorphically on an auxiliary complex variable called the spectral parameter. Crucially, the flatness condition of this Lax pair is equivalent to the equations of motion of the theory.
To construct the Lax pair for the sigma model in (\ref{2d-gravity-sigma}), we begin by decomposing the right-invariant current into two components:
\begin{align}
    dVV^{-1}=Q+P\,,
\end{align}
where $P$ and $Q$ are defined as
\begin{align}
    P&=P_+dx^++P_-dx^-=\frac{1}{2}\left(dVV^{-1}+(dVV^{-1})^{\natural}\right)\in \mathfrak{p}\,,\\
    Q&=Q_+dx^++Q_-dx^-=\frac{1}{2}\left(dVV^{-1}-(dVV^{-1})^{\natural}\right)\in \mathfrak{h}\,.
\end{align}
In terms of $P$ and $Q$, the Lax pair is given by
\begin{align}\label{BM-lax}
    \cL_+&=Q_{+}+\frac{1-i \la}{1+i\la}P_{+}\,,\qquad
    \cL_{-}=Q_{-}+\frac{1+i \la}{1-i\la}P_{-}\,.
\end{align}
This is the standard form of the Lax pair for the symmetric coset sigma model. However due to the coupling with a dilaton $\Phi=\rho$, the spectral parameter $\la$ becomes a function of the coordinates $\rho$ and $z$.
In order for the flatness condition
\begin{align}\label{flat-lax}
    \partial_+\cL_{-}-\partial_{-}\cL_{+}-[\cL_{+},\cL_{-}]=0
\end{align}
to be equivalent to the equations of motion (\ref{sigma-eom})\footnote{This can be easily seen by using the relation $P=\frac{1}{2}(V^{\natural})^{-1}(dMM^{-1})V^{\natural}$.}, the spectral parameter $\la$ has to be 
\begin{align}\label{r-alg}
    \frac{1}{\la}-\la=\frac{2}{\rho}(w-z)
\end{align}
or
\begin{align}\label{rpm-alg}
    \la_{(\pm)}=\frac{1}{\rho}\left[(z-w)\pm \sqrt{(z-w)^2+\rho^2}\right]=-\frac{1}{\la_{(\mp)}}\,,
\end{align}
where $w\in\mathbb{C}$ is a constant spectral parameter. The algebraic equation (\ref{r-alg}) describes a genus zero Riemann surface, which is a two-sheeted cover of the $w$-plane with two branch points at $w=z\pm i\rho$ illustrated in Fig. \ref{bm-surface}.
The Lax pair (\ref{BM-lax}) defines the Breitenlohner-Maison (BM) linear system\footnote{For the relation between BZ system and BM system, see for example \cite{Figueras:2009mc}.
}
\begin{align}\label{bm-eq}
    \partial_{\pm}\Psi=\cL_{\pm}\Psi\,,
\end{align}
where $\Psi$ is an $SO(4,4)$ matrix valued function of $w,z,\rho$.
The flatness condition (\ref{flat-lax}) of the Lax pair is realized as the compatibility condition of the BM linear system.
The monodromy matrix associated with the Lax pair generates an infinite number of conserved quantities and is often considered to form an infinite-dimensional algebra known as the Geroch group. For further discussions on this point, see, for example, \cite{Geroch:1970nt,Geroch:1972yt,Breitenlohner:1986um}. Here, we assume the existence of such an infinite-dimensional symmetry and denote the Geroch group as $G^{(\infty)}$, while representing the subgroup of the Geroch group corresponding to $H$ as $H^{(\infty)}$.

\section{multi-neutral black string from solving Riemann-Hilbert problem}\label{sec:mnblack}

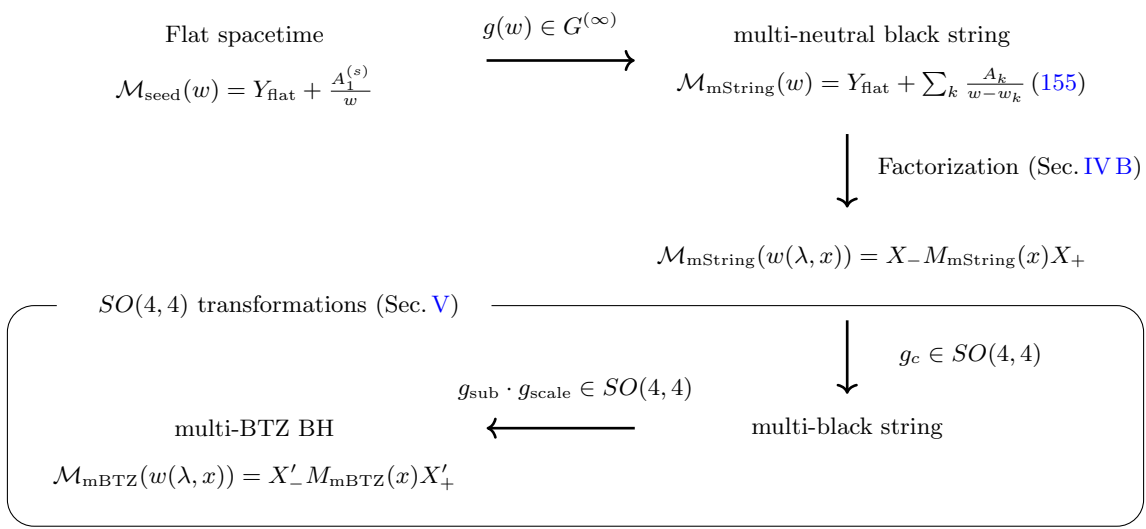
\begin{figure}
\begin{center}
\begin{tikzpicture}[scale=0.65]
\node[font=\small ] at (-12.2,3) {Flat spacetime};
\node[font=\small ] at (-12.2,2) {$\cM_{\text{seed}}(w)=Y_{\text{flat}}+\frac{A_1^{(s)}}{w}$};
\node[font=\small ] at (-6.0,3.2) {$g(w)\in G^{(\infty)}$};
\node[font=\small ] at (0.5,3) {multi-neutral black string};
\node[font=\small ] at (0.7,2) {$\cM_{\rm mString}(w)=Y_{\text{flat}}+\sum_{k}\frac{A_k}{w-w_k}$\,(\ref{mnst-mono})};
\node[font=\small ] at (0.5,-1.5) {$\cM_{\rm mString}(w(\la,x))=X_-M_{\rm mString}(x)X_+$};
\node[font=\small ] at (3.3,0.3) {Factorization (Sec.\,\ref{sec:mnst-fac})};
\node[font=\small ] at (-12,-5) {multi-BTZ BH};
\node[font=\small ] at (2.5,-3.5) {$g_c\in SO(4,4)$};
\node[font=\small ] at (-5.5,-4.2) {$g_{\text{sub}}\cdot g_{\text{scale}}\in SO(4,4)$};
\node[font=\small ] at (0,-5) {multi-black string};
\node[font=\small ] at (-12,-6) {$\cM_{\rm mBTZ}(w(\la,x))
=X_{-}'M_{\rm mBTZ}(x)X_{+}'$};
\node[font=\small ] at (-11.5,-2.5) {$SO(4,4)$ transformations (Sec.\,\ref{sec:harrison})};
\draw[->,black,line width = 1] (0,1) -- (0,-0.5);
\draw[->,black,line width = 1] (-7.3,2.5) -- (-4.3,2.5);
\draw[->,black,line width = 1] (0,-2.8) -- (0,-4.3);
\draw[->,black,line width = 1] (-4.3,-5) -- (-7.3,-5);
\draw[rounded corners=10pt] (-7.2,-2.5) --(6,-2.5) --(6,-7)
--(-17,-7) --(-17,-2.5) --(-16,-2.5);
\end{tikzpicture}
\caption{This figure describes a schematic overview of the whole process for constructing the multi-BTZ black hole solution, starting from the multi-neutral black string.
In Sec.\,\ref{sec:mnst-fac}, we show the multi-neutral black string can be obtained by factorizing the monodromy matrix $\cM_{\rm mString}(w)$ given in (\ref{mnst-mono}), and in Sec.\,\ref{sec:harrison}, we construct the monodromy matrix $\cM_{\rm mBTZ}(w)$ describing the multi-BTZ black hole by applying $SO(4,4)$ transformations. Since the factorization of the monodromy matrix is preserved under $SO(4,4)$ transformations, the multi-BTZ black hole solution can also be constructed via the factorization procedure. The symbols used in this figure will be explained in detail in later sections.}
\label{flow}
\end{center}
\end{figure}

As shown in previous works~\cite{Breitenlohner:1986um, Katsimpouri:2012ky, Chakrabarty:2014ora, Katsimpouri:2013wka, Katsimpouri:2014ara}, static and axisymmetric black hole solutions in type IIB supergravity under the ansatz~ (\ref{d1d5-bg}) can be constructed as soliton solutions of a 2D integrable coset sigma model. 
To obtain such soliton solutions, we follow the procedure developed in \cite{Breitenlohner:1986um, Katsimpouri:2012ky, Chakrabarty:2014ora, Katsimpouri:2013wka, Katsimpouri:2014ara} (see also \cite{Camara:2017hez} for a related discussion), where soliton solutions are derived by solving a Riemann-Hilbert problem for the monodromy matrix associated with the BM linear system. 
As mentioned in the introduction, solving the Riemann-Hilbert problem for the monodromy matrix corresponding to the multi-BTZ black hole (\ref{btzs-sol}) is not computationally efficient due to the presence of RR flux.
Instead of this, we first consider a simpler case, namely the multi-neutral black string, which can be obtained by turning off the RR 3-form flux in the multi-black string~(\ref{mbstring}).
The multi-BTZ black hole solution is then constructed by applying three consecutive $SO(4,4)$ transformations to the multi-neutral black string. 
A diagrammatic overview of this construction is presented in Fig. \ref{flow}.

\subsection{Riemann-Hilbert problem}

We begin by outlining the procedure for solving the Riemann-Hilbert problem for a given monodromy matrix in the BM linear system, following the approach developed in \cite{Breitenlohner:1986um, Katsimpouri:2012ky, Chakrabarty:2014ora, Katsimpouri:2013wka, Katsimpouri:2014ara}.
First, we introduce the monodromy matrix from the Lax pair~(\ref{BM-lax}).
Next, we describe the procedure for factorizing monodromy matrices composed of simple poles with rank-2 residue matrices by basically following Ref.~\cite{Katsimpouri:2013wka}.
Finally, we present a formula for the conformal factor based on the discussion in \cite{Korotkin:1994au,Korotkin:1996vi}.

\subsubsection{Dressing transformation}

We are interested in axisymmetric gravitational solutions described by soliton solutions of the sigma model in (\ref{2d-gravity-sigma}), which can be effectively constructed using the Lax pair (\ref{BM-lax}).
Since the Lax pair satisfies the on-shell flatness condition (\ref{flat-lax}), it can be written as a pure gauge at the on-shell level: 
\begin{align}\label{lax-pure}
    \cL=d\cV(\la,z,\rho)\cV(\la,z,\rho)^{-1}\,.
\end{align}
The information of classical solutions is encoded in the analytic matrix function $\cV(\la,z,\rho)$ \footnote{The matrix function $\cV$ is identified with a solution $\Psi(\la,z,\rho)$ to the BM linear system (\ref{bm-eq}) up to a constant matrix.}.
In this work, we consider classical solutions for which $\cV(\la,z,\rho)$ satisfies the following analytic properties with respect to the spectral parameter:
The matrix valued function $\cV(\la,z,\rho)\in G^{(\infty)}$ is assumed to be analytic at $\lambda=0$ with $\cV(\la,z,0)=V(z,\rho)$ and containing only positive powers of $\la$.
    Since the Lax pair (\ref{BM-lax}) at $\la=0$ reduces to $dVV^{-1}$,  $\cV(\la,z,\rho)$ can be expanded as
    \begin{align}\label{cV-expansion}
        \cV(\la,z,\rho)=\sum_{\la=0}^{\infty}\la^{n}\,\cV_{n}(z,\rho)\,,\qquad \cV_{0}(z,\rho)=V(z,\rho)\,.
    \end{align}
In particular, we require that such matrix valued function $\cV$ does not have poles at $\la=\pm i$, which correspond to the branch points $w=z\pm i\rho$. 
This is utilized to simplify a derivation of the formula of the conformal factor $e^{2\nu}$ later on.

A procedure to generate a new gravitational solution from a given seed solution (e.g., flat spacetime) is to apply an element $g(w)\in G^{(\infty)}$ from the right to $\cV_{\text{seed}}$ corresponding to the seed solution.
The transformed seed matrix $\cV_{\text{seed}}\,g(w)$ generally no longer satisfies the triangular gauge and the boundary condition (\ref{cV-expansion}). Therefore, to preserve these properties, it is necessary to simultaneously perform a compensating gauge transformation generated by $h(\la,z,\rho) \in H^{(\infty)}$:
\begin{align} \label{dressing-v}
    \cV_{\text{seed}}(\la,z,\rho)\quad \mapsto \quad \cV(\la,z,\rho)=h(\la,z,\rho)\cV_{\text{seed}}(\la,z,\rho)g(w)\,.
\end{align}
Since the transformation (\ref{dressing-v}) preserves the flatness condition of the Lax pair, the transformed $\cV(\la,z,\rho)$ also describes a classical solution.
This solution generating technique is commonly referred to as the dressing transformation.

\subsubsection{Monodromy matrix $\cM(w)$ and Riemann-Hilbert problem}

To extract a gravitational solution from $\cV(\la,z,\rho)$, it is necessary to always take the Iwasawa gauge (\ref{iwasawa-rep}). However applying the dressing transformation generally causes a deviation from the triangular gauge, requiring a compensating gauge transformation.
To avoid this cumbersomeness, we introduce a matrix valued function $\cM(\la,z,\rho)$, constructed from $\cV(\la,z,\rho)$, which is an extension of $M(z,\rho)$ by incorporating dependence on the spectral parameter $\la$.

For this purpose, we define the anti-involution $\natural^{(\infty)}:G^{(\infty)}\to G^{(\infty)}$ as
\begin{align}
    (\cV(\la,z,\rho))^{\natural^{(\infty)}}=\cV^{\natural}\left(-\frac{1}{\la}, z,\rho\right)\qquad \cV(\la,z,\rho)\in G^{(\infty)}\,.
\end{align}
This map acts as an automorphism of the Geroch group $G^{(\infty)}$ \cite{Breitenlohner:1986um}, which is a combination of an anti-involution in $G=SO(4,4)$ and mapping $\la_{(+)}$ in the first Riemann sheet to $\la_{(-)}=-1/\la_{(+)}$ in the second.
The matrix valued function $\cM(\la,z,\rho)\in G^{(\infty)}$ is then defined as
\begin{align}\label{m-fac}
  \cM(\la,z,\rho)=(\cV(\la,z,\rho))^{\natural^{(\infty)}}\cV(\la,z,\rho)\,,
\end{align}
where by definition satisfies
\begin{align}\label{m-con}
    \cM^{-1}=\eta \cM^{T}\eta\,,\qquad   \cM^{\natural^{(\infty)}}=\cM\,.
\end{align}
A key property of $\cM(\la,z,\rho)$ is that it depends only on the spectral parameter $w$, and not explicitly on the Weyl-Papapetrou coordinates $\rho$ and $z$.
This can be verified as follows
\begin{align}
    d\cM=d\cV^{\natural^{(\infty)}}\cV+\cV^{\natural^{(\infty)}}d\cV=\cV^{\natural^{(\infty)}}\left(\cL^{\natural^{(\infty)}}+\cL\right)\cV=0\,,
\end{align}
where we used $\cL^{\natural^{(\infty)}}=-\cL$. Hence, we denote $\cM(\la,\rho,z)$ simply as $\cM(w)$, and it is referred to as the monodromy matrix.
To make the expression (\ref{m-fac}) more practical, we introduce a matrix valued function
\begin{align}\label{xp-def}
   X_+(\la,z,\rho)=V(z,\rho)^{-1} \cV(\la,z,\rho)\in G^{(\infty)}\,,
\end{align}
which is defined such that for a fixed $(z,\rho)$. The matrix valued functions $X_+(\la,z,\rho)$ and $X_-(\la,z,\rho)=X_+^{\natural}(-1/\la,z,\rho)$ are normalized to satisfy the boundary conditions
\begin{align}\label{Xpm-bc}
    X_+(0,z,\rho)=1_{8\times 8}=X_-(\infty,z,\rho)\,.
\end{align}
Then, the monodromy matrix admits the factorized form:
\begin{align}\label{fac-m}
    \cM(w(\la,z,\rho))=X_-^{}(\la,z,\rho)M(z,\rho)X_+(\la,z,\rho)\,.
\end{align}
The symbol $w(\la,z,\rho)$ in the left-hand side is to remind us that whenever $\cM(w)$ is rewritten as shown on the right-hand side, $w$ must always be substituted using its relation 
(\ref{r-alg}) with $\la(=\la_{(+)})$.

\subsubsection*{Riemann-Hilbert problem : $M(z,\rho)$ from a given $\cM(w)$}

Based on the relation (\ref{fac-m}) between $M(z,\rho)$ and $\cM(w)$, the method of generating a new solution from a given seed solution via a dressing transformation (\ref{dressing-v}) can be reformulated as follows.
Starting from a seed solution with $M_{\text{seed}}(z,\rho)$, the corresponding monodromy matrix takes the form
\begin{align}
    \cM_{\text{seed}}(w(\la,z,\rho))=X_-^{(0)}(\la,z,\rho)M_{\text{seed}}(z,\rho)X_+^{(0)}(\la,z,\rho)\,,
\end{align}
Applying a dressing transformation generated by $g(w)\in G^{(\infty)}$ leads to a new monodromy matrix
\begin{align}
    \cM_{\text{seed}}(w)\mapsto \cM(w)=g(w)^{\natural}\cM_{\text{seed}}(w)g(w)\,.
\end{align}
To extract the corresponding $M(z,\rho)$, one must factorize $\cM(w)$ again into the form (\ref{fac-m}), which, however, is not an easy task.
The task finding such a factorization for a given meromorphic matrix function $\cM(w)$ is known as the matrix-valued Riemann-Hilbert problem.
The reconstructed matrix $M(z,\rho)$, obtained through the factorization process, has been shown to satisfy the equations of motion of the coset sigma model \cite{Camara:2017hez}, and thus yields a solution to the Einstein equations or the equations of motion of supergravity theories.
To summarize, the construction of gravitational solutions involves two steps: (i) specifying an appropriate monodromy matrix, and (ii) performing its factorization.

\subsubsection*{Candidate monodromy matrix $\cM(w)$ from a given $M(z,\rho)$}

A general prescription for determining the appropriate monodromy matrix corresponding to a specific black hole solution is not yet fully established.
Nevertheless, when a black hole solution is given and the corresponding $M(z,\rho)$ is known, a practical prescription has been proposed for constructing a candidate monodromy matrix $\cM(w)$ from $M(z,\rho)$ along the $z$-axis.
In regions where rods are present, $M(z,\rho)$ might exhibit potential divergences.
To avoid such divergences, a candidate monodromy matrix $\cM(w)$ is defined by taking the limit limit $\rho\to 0^+$ of $M(z,\rho)$ far from the finite length rods
\cite{Breitenlohner:1986um,Chakrabarty:2014ora}:
\begin{align}\label{sub-rule}
    \cM(w)=\lim_{\rho\to 0^+}M(z=w,\rho)\qquad \text{for}\quad z<-R\,,
\end{align}
 where $R$ is a sufficiently large positive number ensuring that $M(z=w,\rho)$ for $z<-R$ remains finite in the limit $\rho\to 0^+$.

\subsubsection{Factorization of monodromy matrix}

Here, we explicitly perform the factorization of a specific type of monodromy matrices by following \cite{Katsimpouri:2013wka}.
In general, solving the Riemann–Hilbert problem is a highly nontrivial task.
Fortunately, when $\cM(w)$ only contains simple poles, the associated Riemann-Hilbert problem simplifies to an algebraic problem solving a system of linear equations \cite{Nicolai:1991tt}.
Hence, we consider the following ansatz of the monodromy matrix, which consists of $N$ simple poles on the $w$-plane \cite{Katsimpouri:2012ky,Katsimpouri:2013wka,Katsimpouri:2014ara}\footnote{In general, the positions $w_j$ of simple poles can take complex values, but we assume them to be real. }
\begin{align}\label{monodromy-pole}
    \cM(w)=Y+\sum_{j=1}^{N}\frac{A_j}{w-w_j}\,,\qquad w_j\in \mathbb{R}\,.
\end{align}
Since $\cM(w)\in G^{(\infty)}$, the inverses must satisfy $\cM^{-1}(w)=\eta \cM(w)^{T}\eta$.
The residues $A_{j}$ and the constant part $Y$ in the spectral parameter space $w$ are $8\times 8$ matrices satisfying the conditions $A_j^{\natural}=A_j$ and $Y^{\natural}=Y$, as required by (\ref{m-con}).
The choice of the constant matrix $Y$ is fixed by requiring that the monodromy matrix (\ref{monodromy-pole}) obeys the relation (\ref{sub-rule}) and as $\rho\to 0^+$ and $z\to -\infty$ i.e. $w\to -\infty $ from Eq.~(\ref{r-alg}),  
\begin{align}\label{sM-asymp}
    \lim_{\substack{\rho\to 0^+\\z\to -\infty}}M(z,\rho)=Y\,.
\end{align}

Once a factorization of a given $\cM(w)$ is accomplished, the coset matrix $M(z,\rho)$ can be reconstructed from the asymptotic behavior of $\cM(w)$ in the limit $w\to -\infty$.
In the $\la$-plane, the limit $w\to -\infty$ corresponds to $\la=\la_{(+)}\to \infty$ for fixed $z$ and $\rho$, as follows from the relation (\ref{rpm-alg}). Therefore, by applying the limit to the factorized expression (\ref{fac-m}), we can express $M(z,\rho)$ as \cite{Katsimpouri:2012ky,Katsimpouri:2013wka,Katsimpouri:2014ara}
\begin{align}\label{M-formula}
    M(z,\rho)=Y\,X_+^{-1}(\infty,z,\rho)=Y\,(\eta\,X_+^{T}(\infty,z,\rho)\eta)\,,
\end{align}
where we used the boundary condition (\ref{Xpm-bc}) of $X_-(\la,z,\rho)$.

\subsubsection*{Determining $X_+(\la,z,\rho)$}

The remaining task is to construct explicitly the matrix-valued function $X_{+}(\la,z,\rho)$ associated with (\ref{monodromy-pole}). 
By following \cite{Katsimpouri:2013wka}, we assume that the matrices $A_k$ are rank-2 and are expressed in a factorized form using the constant eight-component vectors $a_j$ and $b_j$ \footnote{In the $SL(n,\mathbb{R})$ case of $D=n+2$ pure gravity considered in \cite{Katsimpouri:2012ky}, an ansatz was taken such that the residue $A_j$ has rank 1. For an intuitive explanation of why the residue matrices should be taken to have rank 2 in our set up, see for example Sec. 3.1 of \cite{Katsimpouri:2013wka}.}
\begin{align}\label{rank2-mat}
    A_j=\alpha_j(a_j\otimes a_j)\eta'-\beta_j((\eta b_j)\otimes (\eta b_j))\eta'\,,
\end{align}
where $\alpha_j$ and $\beta_{j}$ are constants.
The constant vectors $a_j$ and $b_j$ satisfy
\begin{align}\label{ab-cond}
    a_j^{T}\eta a_j=0\,,\qquad  b_j^{T}\eta b_j=0\,,\qquad a_j^{T}b_j=0
\end{align}
or equivalently $A_j\eta A_j^{T}=0$.
The conditions (\ref{ab-cond}) are required from the vanishing of double poles of $\cM(w)\cM^{-1}(w)=\cM(w)\eta \cM(w)^{T}\eta$.
In \cite{Katsimpouri:2013wka}, to simply the derivation of $X_+$, additional constraints were imposed on the vectors $a_j$ and $b_j$ as follows
\begin{align}\label{ab-assume2}
   a_k^{T}\eta\,a_l=0\,,\qquad  b_k^{T}\eta\,b_l=0\qquad \text{for}\quad k\neq l\,.
\end{align}
However, as we will see in the following discussion, this assumption does not hold for the multi-black string solutions with bubbles. Therefore, we here proceed without imposing the condition (\ref{ab-assume2}).

Since the factorization of $\cM(w)$ is performed in the $\la$-plane, we rewrite each pole of (\ref{monodromy-pole}) as a pair of simple poles in the $\la$-plane corresponding to the so-called soliton and anti-soliton:
\begin{align}\label{w-la-map}
    \frac{1}{w-w_j}=\nu_j\left( \frac{\la_j}{\la-\la_j}+\frac{1}{1+\la \la_j}\right)\,,\qquad 
     \nu_j=-\frac{2}{\rho\left(\la_j+\la_j^{-1}\right)}\,,
\end{align}
where we used the relation (\ref{rpm-alg}).
The poles $\la=\la_j$ and $\la=\bar{\la}_j=-1/\la_j$ describe the positions of solition and anti-solition, respectively, and these are given by
\begin{align}
    \la_j=\frac{1}{\rho}\left((z-w_j)+\sqrt{(z-w_j)^2+\rho^2}\right)=-\frac{1}{\bar{\la}_{j}}\,.
\end{align}
The matrix functions $X_{+}(\la,z,\rho)$ and $X_{-}(\la,z,\rho)$ must contain these simple poles at $\la=\la_j$ or $-1/\la_{j}$ to ensure consistency with the presence of simple poles of $\cM(w)$ on the $w$-plane.
Following \cite{Katsimpouri:2013wka}, we consider the following ansatz for the matrix-valued function $X_+(\la,z,\rho)$ such that it consists only of simple poles at $\la=\bar{\la}_{j}=-1/\la_{j}$ for all $j$ :
\begin{align}\label{xp-ex}
    X_+(\la,z,\rho)=1-\sum_{j=1}^{N}\frac{\la C_j}{1+\la \la_j}\,,
\end{align}
where each residue $C_j$ is given by
\begin{align}
    C_j=(c_j\otimes a_j)\eta'-\left((\eta d_j)\otimes (\eta b_j)\right)\eta'\,.
\end{align}
Since $X_+(\la)\eta \cM^{T}\eta$ must have no simple poles at $\la=-1/\la_{j}$, the vectors $a_{j},b_{j},c_{j},d_j$ are not independent and satisfy the equations \cite{Katsimpouri:2013wka}
\begin{align}
    \eta'a_i&=\frac{\gamma_i}{\la_i}d_{i}+\sum_{j\neq i}^{N}\frac{(a_i^{T}b_j)}{\la_i-\la_j}d_j
    -\sum_{j\neq i}^{N}\frac{(a_j^{T}\eta a_i)}{\la_i-\la_j}\eta c_j\,,\\
    \eta'b_i&=\frac{\gamma_i}{\la_i}c_{i}+\sum_{j\neq i}^{N}\frac{(a_j^{T}b_i)}{\la_j-\la_i}c_{j}+\sum_{j\neq i}^{N}\frac{(b^{T}_j\eta b_i)}{\la_i-\la_j}\eta d_j\,,
\end{align}
or equivalently
\begin{align}\label{abcd-eq2}
    \eta'a&=d\,\Gamma^{(0)T}-(\eta c) \Gamma^{(a)T}\,,\qquad \eta'b=c\Gamma^{(0)}+(\eta d)\,\Gamma^{(b)T}\,,
\end{align}
where the $8\times N$ matrices $a,b,c,d$ are
\begin{align}
    a&=(a_1,\dots,a_N)\,,\quad  b=(b_1,\dots,b_N)\,,\quad c=(c_1,\dots,c_N)\,,\quad  d=(d_1,\dots,d_N)\,.
\end{align}
The $N\times N$ matrices $\Gamma^{(0)},\Gamma^{(a)}$ and $\Gamma^{(b)}$ are defined as
\begin{align}
\Gamma_{kl}^{(0)}&=
\begin{cases}
    \frac{\gamma_k}{\la_k}\qquad &\text{for}\qquad k=l\\
    \frac{a_k^{T}b_l}{\la_k-\la_l}\qquad &\text{for}\qquad k\neq l
\end{cases}\,,\label{gamma0}\\
\Gamma^{(a)}_{kl}&=\begin{cases}
    0\qquad &\text{for}\qquad k=l\\
    \frac{(a_k^{T}\eta a_l)}{\la_k-\la_l}\qquad &\text{for}\qquad k\neq l
\end{cases}
\,,\label{gammaa}\\
\Gamma^{(b)}_{kl}&=\begin{cases}
    0\qquad &\text{for}\qquad k=l\\
    \frac{(b_k^{T}\eta b_l)}{\la_k-\la_l}\qquad &\text{for}\qquad k\neq l
\end{cases}\,,\label{gammab}
\end{align}
and the functions $\gamma_i=\gamma_{i}(z,\rho)$ in $\Gamma^{(0)}$ can be obtained by solving the equations 
\begin{align}
\begin{split}
\cA_i\eta \eta'a_i&=\nu_i\beta_i \gamma_i (\eta b_i)\,,\qquad
    (\eta b_i)\eta'\eta \cA_i^{T}=(\nu_i\alpha_i \gamma_{i} )a_i\,,\\
    \cA_i&=\left(\cM(w(\la,z,\rho))-\frac{\nu_kA_k}{1+\la \la_k}\right)\biggl\lvert_{\la=-1/\la_k}\,,\label{gamma-def}
\end{split}
\end{align}
which ensure the absence of simple poles in $\cM(w)\eta \cM^{T}(w)\eta$ at $\la=-1/\la_k$.
By solving the equations (\ref{abcd-eq2}), we obtain the more general expressions for the vectors $c_k$ and $d_k$ than in Ref.~\cite{Katsimpouri:2013wka}, given by
\begin{align}
    c&=\left(\eta' b-\eta \eta' a(\Gamma^{(0)T})^{-1}\Gamma^{(b)T}\right)(\Gamma^{(0)})^{-1}
    \left(1+\Gamma^{(a)T}(\Gamma^{(0)T})^{-1}\Gamma^{(b)T}(\Gamma^{(0)})^{-1}\right)^{-1}\,,\label{c-vec}\\
    d&=\left(\eta' a+\eta \eta' b(\Gamma^{(0)})^{-1}\Gamma^{(a)T}\right)(\Gamma^{(0)T})^{-1}
    \left(1+\Gamma^{(b)T}(\Gamma^{(0)})^{-1}\Gamma^{(a)T}(\Gamma^{(0)T})^{-1}\right)^{-1}\label{d-vec}\,.
\end{align}
If we impose the constraint (\ref{ab-assume2}) i.e. $\Gamma^{(a)}=\Gamma^{(b)}=0$, the expressions (\ref{c-vec}) and (\ref{d-vec}) reduce to the result in \cite{Katsimpouri:2013wka}
\begin{align}\label{cd-ab-sim}
     c=\eta' b(\Gamma^{(0)})^{-1}\,,\qquad 
     d=\eta' a (\Gamma^{(0)T})^{-1}\,.
\end{align}

\subsubsection{Conformal factor }

Finally, we present a closed expression for the conformal factor $e^{2\nu}$.
In the special case $\Gamma^{(a)}=\Gamma^{(b)}=0$, a simplified formula has been provided in \cite{Katsimpouri:2013wka} (see also \cite{Katsimpouri:2012ky}) and it is given by
\begin{align}\label{conf-formula0}
    e^{2\nu}=k_{\rm BM}\cdot \prod_{j=1}^{N}(\la_j\nu_j)\,{\rm det}\,\Gamma^{(0)}\,.
\end{align}
The overall constant $k_{\rm BM}$ is determined from the asymptotic condition of $e^{2\nu}$.
In principle, the formula (\ref{conf-formula0}) can be extended to more general case $(\Gamma^{(a)},\Gamma^{(b)})\neq (0,0)$ by following the derivation of (\ref{conf-formula0}) presented in \cite{Katsimpouri:2012ky,Katsimpouri:2013wka}. However, this approach becomes highly cumbersome when attempting to derive a general expression for $e^{2\nu}$ for the multi-neutral black string with an arbitrary number of bubbles, as considered in the following subsection.
Instead, in this work, we follow the approach described in \cite{Korotkin:1994au,Korotkin:1996vi}, emphasizing the analytic structure of the function $X_+(\la,z,\rho)$, and extend it to our current setup to obtain a formula for the conformal factor.

Let us begin with Eq.~(\ref{conf-eom}) for the conformal factor, rewritten in differential form on the 2D $x^+x^-$-plane  as follows\footnote{We used the relation $P=\frac{1}{2}(V^{\natural})^{-1}(dMM^{-1})V^{\natural}$.}
\begin{align}\label{con-eom2}
   d(\ln e^{2\nu})=-\frac{i\rho}{2}\Tr(P_{+}^2)dx^++\frac{i\rho}{2}\Tr(P_{-}^2)dx^-\,,
\end{align}
and recast it in terms of the matrix function $X_{+}$ or equivalently $\cV$.
The relation between the current $P_{\pm}$ and $\cV$ can be obtained from the fact that the on-shell Lax pair admits two expressions (\ref{BM-lax}) and (\ref{lax-pure}), and by evaluating the residues of both expressions at $\la=\pm i$.
The residue of the latter expression (\ref{lax-pure})  can be evaluated by rewriting it as
\begin{align}\label{lax-pure-2}
    \cL_{\pm}
    &=\partial_{\pm}\cV(\la,z,\rho)\cV(\la,z,\rho)^{-1}\biggl\lvert_{\la:\text{fixed}}\pm i \frac{\la}{\rho}\frac{i\pm \la}{i\mp \la}\partial_{\la}\cV(\la,z,\rho)\cV(\la,z,\rho)^{-1}\,,
\end{align}
using the identity satisfied by the spectral parameter $\la$\,:
\begin{align}
    \partial_{\pm}\la=\pm i \frac{\la}{\rho}\frac{i\pm \la}{i\mp \la}\,.
\end{align}
Taking into account the assumption $\cV$ does not have poles at $\la=\pm i$, 
evaluating the residues of (\ref{BM-lax}) and (\ref{lax-pure-2}) leads to the relations
\begin{align}\label{lax-res}
   P_{\pm}=-\frac{1}{\rho}\,\left(\frac{\partial}{\partial \la}\cV(\la,z,\rho)\cV(\la,z,\rho)^{-1}\right)\biggl\lvert_{\la=\pm i}\,.
\end{align}
By substituting (\ref{lax-res}) into (\ref{con-eom2}) and using the definition (\ref{xp-def}) of $X_+$, we can rewrite (\ref{con-eom2}) as
\begin{align}\label{con-eom3}
   d\left(\ln e^{2\nu}\right)&=-\frac{i}{2\rho}\,
    \Tr\left( \frac{\partial}{\partial \la}X_+(\la,z,\rho)X_+^{-1}(\la,z,\rho)\right)^2\biggl\lvert_{\la=+ i}dx^+\no\\
    &\quad +\frac{i}{2\rho}\,
    \Tr\left( \frac{\partial}{\partial \la}X_+(\la,z,\rho)X_+^{-1}(\la,z,\rho)\right)^2\biggl\lvert_{\la=-i}dx^-\,.
\end{align}
To proceed further, it is useful to specify the analytic properties of the function $X_+$. Following \cite{Korotkin:1996vi}, we impose the condition that the right-invariant current for $X_+(\la,z,\rho)$ along the $\la$ direction consists of only simple poles at $\la=\bar{\la}_j=-1/\la_j$ i.e.
\begin{align}\label{rXp-ass}
     \frac{\partial}{\partial \la}X_+(\la,z,\rho)X_+^{-1}(\la,z,\rho)=\sum_{j=1}^{N}\frac{B_j}{\la-\bar{\la}_j}\,,
\end{align}
where the matrices $B_j\,(j=1,\dots,N)$ are independent of $\la, z,\rho$\footnote{Requiring the matrices $B_j$ to be Weyl-Papapetrou coordinate-independent is a somewhat strong constraint. In fact, the closed expression (\ref{conf-formula-a}) for the conformal factor can still be obtained, as long as $\Tr(B_iB_j)$ in (\ref{dtau-bb}) are coordinate-independent.}.
This assumption indicates that the matrix meromorphic function $X_{+}$ has no essential singularities, and that the eigenvalues of $B_j$ specify the degree of ramification of $X_+$ at $\la=\bar{\la}_j$. According to \cite{Korotkin:1994au,Korotkin:1996vi}, $X_+$ satisfying (\ref{rXp-ass}) describes an isomonodromic solution of the BM linear system.
We further require that the function $X_+(\la,z,\rho)$ is regular at $\la=\infty$, which leads to the constraint
\begin{align}\label{sum-b}
    B_{\infty}=\sum_{j=1}^{N}B_j=0\,.
\end{align}
We then define a closed form specified by a new scalar function $\tau(z,\rho;\{w_j\})$ as \footnote{The scalar function $\tau(z,\rho;\{w_j\})$ is closely related to the $\tau$-function associated with Schlesinger equations, which describe isomonodromic solutions of the BM linear system. For more details, see \cite{Korotkin:1994au,Korotkin:1996vi}.}
\begin{align}\label{dtau-bb}
    d\left(\ln \tau\right)=\frac{1}{2}\sum_{p<q}^{N}\Tr(B_pB_q)d(\ln(\bar{\la}_{p}-\bar{\la}_{q}))\,,
\end{align}
and under the above assumptions, the equation (\ref{con-eom3}) becomes a simple expression
\begin{align}\label{d-conf}
   d(\ln e^{2\nu})
   &=   d\left(\ln \tau\right)+\frac{1}{4}\sum_{j=1}^{N}\Tr(B_j^2)\,d\left(\ln\left(\frac{\partial \bar{\la}_j}{\partial w_j} \right)\right)
   \,,
\end{align}
where we used the identity
\begin{align}
    \sum_{p<q}^{N}\Tr(B_pB_q)=\frac{1}{2}\Tr(B_{\infty}^2)-\frac{1}{2}\sum_{j=1}^{N}\Tr(B_j^2)=-\frac{1}{2}\sum_{j=1}^{N}\Tr(B_j^2)\,.
\end{align}
Finally, integrating out (\ref{d-conf}) yields a closed expression for the conformal factor $ e^{2\nu}$ given by
\begin{align}
    e^{2\nu}&=K_{\rm BM}\,\cdot \prod_{j=1}^{N}\left(\frac{\partial \bar{\la}_j}{\partial w_j}\right)^{\frac{1}{4}\Tr(B_j^2)}\,\tau(z,\rho;\{w_q\})\no\\
    &=K_{\rm BM}\,\cdot \prod_{j=1}^{N}(\bar{\la}_j \bar{\nu}_j)^{\frac{1}{4}\Tr(B_j^2)}\prod_{p<q}^{N}(\bar{\la}_{p}-\bar{\la}_{q})^{\frac{1}{2}\Tr(B_pB_q)}\,,\label{conf-formula-a}
\end{align}
where $K_{\rm BM}$ is an integration constant, and $\bar{\nu}_j$ is defined as $\nu_j$ by replacing $\la_j$ with $\bar{\la}_j$ in (\ref{w-la-map}).
It is noted that the expression (\ref{conf-formula-a}) is invariant under the exchange $\bar{\la}_j\leftrightarrow \la_j$.
This can easily be seen by using the constraint (\ref{sum-b}) imposed on the matrices $B_j$ and the relations
\begin{align}
    \frac{\partial \bar{\la}_j}{\partial w_j}=\frac{1}{\la_j^2}\frac{\partial \la_j}{\partial w_j}\,,\qquad \bar{\la}_{i}-\bar{\la}_{j}=\frac{1}{\la_i\la_j}(\la_{i}-\la_{j})\,.
\end{align}
Hence, the conformal factor $e^{2\nu}$ can also take the expression
\begin{align}
    e^{2\nu}
    &=K_{\rm BM}\,\cdot \prod_{j=1}^{N}(\la_j \nu_j)^{\frac{1}{4}\Tr(B_j^2)}\prod_{p<q}^{N}(\la_{p}-\la_{q})^{\frac{1}{2}\Tr(B_pB_q)}\,.\label{conf-formula}
\end{align}
The formula (\ref{conf-formula}) can be regarded as an extension of (\ref{conf-formula0}) in the sense that it is applicable even for the case of $\Gamma^{(a)}\neq 0\,, \Gamma^{(b)}\neq 0$.
Furthermore, while we imposed the constraint (\ref{sum-b}) on the constant matrices $B_j$, it is easy to generalize (\ref{conf-formula}) to the case where this condition is relaxed.

\subsection{Multi-neutral black strings from solving Riemann-Hilbert problem}\label{sec:mnst-fac}

In this subsection, we explicitly write down the monodromy matrix $\cM(w)$ corresponding to the multi-neutral black strings and then derive the gravitational solution by performing the factorization of $\cM(w)$.

\subsubsection{Asymptotic matrix $Y$ for multi-neutral black string}

To this end, we must first determine the constant matrix $Y$, which encodes the asymptotic behavior of the multi-neutral black string. 
Since the solution (\ref{mbstring}) asymptotically approaches $R^{1,4}\times S^1\times T^4$, we take the 10D flat spacetime as the seed solution. 
In general, the matrix $M(z,\rho)$ could contain divergent components in the asymptotic limit (\ref{sM-asymp}), depending on the choice of coordinates.
To avoid this subtlety, we follow \cite{Giusto:2007fx} and take the coordinate system using the angular coordinates $(\phi,\psi)$ introduced in (\ref{u1-fib-coord}):
\begin{align}
    ds^2_{\text{seed}}&=-dt^2+dy^2+\frac{1}{Z_0}(d\psi+H_0 d\phi)^2+Z_0\left(e^{2\nu}(d\rho^2+dz^2)+\rho^2 d\phi^2\right)+ds_{T^4}^2\,,\label{6dflat-met}
\end{align}
where the scalar function $Z_0\,, H_0$ and the conformal factor $e^{2\nu}$ are given by
\begin{align}
\begin{split}
    Z_0&=\frac{1}{\sqrt{\rho^2+z^2}} \,,\quad H_0=\frac{z}{\sqrt{\rho^2+z^2}}\,d\phi\,,\quad  e^{2\nu}=1\,.
\end{split}
\end{align}
The relation between the spherical coordinates $(r,\theta)$ and the Weyl-Papapetrou coordinate $(\rho,z)$ are obtained by taking a limit $l\to 0$ of (\ref{weyl-def}).
An advantage of choosing the coordinate system (\ref{6dflat-met}) is that the gauge invariant element $M_{\text{seed}}(z,\rho)$ approaches a constant matrix as $r\to \infty$.
Indeed, when the matrix $M_{\text{seed}}(z,\rho)$ corresponding to the metric (\ref{6dflat-met}) is computed by following the definitions (\ref{iwasawa-rep}) and (\ref{ginv-M}), it takes the form
\begin{align}
    M_{\text{seed}}(z,\rho)=
        \begin{pmatrix}
        1&0&0&0&0&0&0&0\\
        0&1&0&0&0&0&0&0\\
        0&0&\frac{1}{\sqrt{\rho^2+z^2}}&0&0&0&0&1\\
        0&0&0&\frac{1}{\sqrt{\rho^2+z^2}}&0&0&-1&0\\
        0&0&0&0&1&0&0&0\\
        0&0&0&0&0&1&0&0\\
        0&0&0&1&0&0&0&0\\
        0&0&-1&0&0&0&0&0
    \end{pmatrix}\,,
\end{align}
and is found to approach a constant matrix in the asymptotic region
\begin{align}
    \lim_{\substack{\rho\to 0^+\\z\to -\infty}}M_{\text{seed}}(z,\rho)=Y_{\text{flat}}+\cO\left(\frac{1}{r^2}\right)\,,
\end{align}
where the asymptotic matrix $Y_{\text{flat}}\in \mathfrak{so}(4,4)$ is
\begin{align}
    Y_{\text{flat}}=
    \begin{pmatrix}
        1&0&0&0&0&0&0&0\\
        0&1&0&0&0&0&0&0\\
        0&0&0&0&0&0&0&1\\
        0&0&0&0&0&0&-1&0\\
        0&0&0&0&1&0&0&0\\
        0&0&0&0&0&1&0&0\\
        0&0&0&1&0&0&0&0\\
        0&0&-1&0&0&0&0&0
    \end{pmatrix}
    \,,\qquad 
    Y_{\text{flat}}^{\natural}=Y_{\text{flat}}
    \,.
\end{align}

\subsubsection{Neutral black string}

\begin{figure}
\begin{center}
\begin{tikzpicture}[scale=0.65]
\node[font=\small ] at (-1.5,4.5) {$(1,0,0,0,0,0,0,0)$};
\node[font=\small ] at (-10,4) {$t$};
\node[font=\small ] at (-10,3) {$y$};
\node[font=\small ] at (-10,2) {$\varphi_1$};
\node[font=\small ] at (-10,1) {$\varphi_2$};
\node[font=\small ] at (-10,0) {$z$};
\node[font=\small ] at (-5,-0.5) {$0$};
\node[font=\small ] at (2,-0.5) {$\frac{l^2}{4}$};
\draw[gray,line width = 0.8] (-9,4) -- (6,4);
\draw[black,line width = 5] (-5,4) -- (2,4);
\draw[gray,line width = 0.8] (-9,3) -- (6,3);
\draw[gray,line width = 0.8] (-9,2) -- (6,2);
\draw[black,line width = 5,dashed ] (-9,2) -- (-8.5,2);
\draw[black,line width = 5] (-8.4,2) -- (-5,2);
\draw[black,dashed ] (-5,0) -- (-5,4);
\draw[black,dashed ] (2,0) -- (2,4);
\draw[gray,line width = 0.8] (-9,1) -- (6,1);
\draw[black,line width = 5] (2,1) -- (5.4,1);
\draw[black,line width = 5,dashed ] (5.5,1) -- (6,1);
\draw[->,black,line width = 1] (-9,0) -- (6,0);
\end{tikzpicture}
\caption{The rod structure of black string solution }\label{rod-btz-able2}
\end{center}
\end{figure}

Let us first consider the Riemann-Hilbert problem for the simplest case, the black string solution (\ref{d1d5p-e}) in the neutral limit.
This calculation has already been performed in \cite{Katsimpouri:2014ara}, but before considering the multi-black string case, we present the simplest example as a demonstration of solving the Riemann-Hilbert problem.

Since the neutral black string has the rod structure given in Fig.\,\ref{rod-btz-able2}, we perform a dressing transformation $g(w)$ such that the deformed monodromy matrix $\cM_{\rm String}(w)$ has two simple poles in the $w$-plane
\begin{align}
    \cM_{\rm String}(w)=Y_{\text{flat}}+\frac{A_1}{w}+\frac{A_2}{w-\frac{l^2}{4}}\,.
\end{align}
Each pole represents to the contact point on the $z$-axis between two distinct rods.
The residue matrices $A_1$ and $A_2$ are given by
\begin{align}
    A_1&=    
    \begin{pmatrix}
        -\frac{l^2}{4}&0&0&0&0&0&0&0\\
        0&0&0&0&0&0&0&0\\
        0&0&-1&0&0&0&0&-\frac{l^2}{8}\\
        0&0&0&0&0&0&0&0\\
        0&0&0&0&0&0&0&0\\
        0&0&0&0&0&0&0&0\\
        0&0&0&0&0&0&0&0\\
        0&0&\frac{l^2}{8} &0&0&0&0&\frac{l^4}{64}
    \end{pmatrix}\,,\qquad A_2=
        \begin{pmatrix}
        0&0&0&0&0&0&0&0\\
        0&0&0&0&0&0&0&0\\
        0&0&0&0&0&0&0&0\\
        0&0&0&-1&0&0&-\frac{l^2}{8}&0\\
        0&0&0&0&\frac{l^2}{4}&0&0&0\\
        0&0&0&0&0&1&0&0\\
        0&0&0&\frac{l^2}{8}&0&0&\frac{l^4}{64}&0\\
        0&0&0&0&0&0&0&0
    \end{pmatrix}\,,
\end{align}
and we find that these two matrices are of rank 2.  
The residue matrices $A_j$ can be expressed by the equation (\ref{rank2-mat}) from the perspective of two vectors
\begin{align}
\begin{split}
    a_1&=(1,0,0,0,0,0,0,0)^{T}\,,\quad
    a_2=\left(0,0,0,1,0,0,-\frac{l^2}{8},0\right)^{T}\,,\\
    b_1&=\left(0,0,0,\frac{l^2}{8},0,0,-1,0\right)^{T}\,,\quad
    b_2=\left(-\frac{l^2}{4},0,0,0,0,0,0,0\right)^{T}\,,
\end{split}
\end{align}
and the constants 
\begin{align}
    \alpha_1=\frac{l^2}{4}\,,\quad \alpha_2=-1\,,\quad \beta_1=-1\,,\quad \beta_2=\frac{4}{l^2}\,.
\end{align}
We can observe that the $2\times 2$ matrix $\Gamma^{(0)}$ takes a symmetric form
\begin{align}\label{bs-gamma}
    \Gamma^{(0)}=              -\frac{l^2}{4}\frac{1}{\la_{1,2}} \begin{pmatrix}
        0&1\\
       1&0
    \end{pmatrix}\,,
\end{align}
and two $2\times 2$ matrices $\Gamma^{(a)}$ and $\Gamma^{(b)}$ vanish.
The residues $C_{j}$ in $X_+(z,\rho,\la)$ are given by
\begin{align}
    C_1=
               \begin{pmatrix}
        \la_{12}&0&0&0&0&0&0&0\\
        0&0&0&0&0&0&0&0\\
        0&0&\frac{1}{2}\la_{1,2}&0&0&0&0&\frac{l^2}{16}\la_{1,2}\\
        0&0&0&0&0&0&0&0\\
        0&0&0&0&0&0&0&0\\
        0&0&0&0&0&0&0&0\\
        0&0&0&0&0&0&0&0\\
        0&0&\frac{4}{l^2}\la_{1,2}&0&0&0&0&\frac{1}{2}\la_{1,2}\\
    \end{pmatrix}\,,\quad 
C_2=
    \begin{pmatrix}
        0&0&0&0&0&0&0&0\\
        0&0&0&0&0&0&0&0\\
        0&0&0&0&0&0&0&0\\
        0&0&0&-\frac{1}{2}\la_{1,2}&0&0&-\frac{l^2}{16}\la_{1,2}&0\\
        0&0&0&0&-\la_{12}&0&0&0\\
        0&0&0&0&0&0&0&0\\
        0&0&0&-\frac{4}{l^2}\la_{1,2}&0&0&-\frac{1}{2}\la_{1,2}&0\\
        0&0&0&0&0&0&0&0\\
    \end{pmatrix}\,,
\end{align}
where we introduced the notation $\la_{1,2}=\la_1-\la_2$.
It can be shown that the monodromy matrix $\cM_{\rm String}(w)$ is factorized as
\begin{align}\label{nstring-cM-M}
    \cM_{\rm String}(w(\la,z,\rho))=X_-(\la,z,\rho)M_{\rm String}(z,\rho)X_+(\la,z,\rho)\,.
\end{align}
From the formula (\ref{M-formula}), the gauge invariant matrix $M_{\rm String}(z,\rho)$ can be computed, and the associated 16 scalar fields are given by
\begin{align}
\begin{split}
 e^{2U}&=\frac{\hat{r}^2}{4}\left(1-\frac{2m}{\hat{r}^2}\right)^{1/2}\,,\quad
    x^I=0\,,\\ 
    y^1&=\left(1-\frac{2m}{\hat{r}^2}\right)^{1/2}\,,\quad y^2=\left(1-\frac{2m}{\hat{r}^2}\right)^{1/2}\,,\quad  y^3=\left(1-\frac{2m}{\hat{r}^2}\right)^{1/2}\,,\\
    \tilde{\zeta}_\Lambda&=0\,,\quad
    \zeta^\Lambda=0\,,\quad
    \sigma=2T_0\,.
\end{split}
\end{align}
where we set $l^2=2m$ and $\hat{r}^2=r^2+2m$. The set of the scalar fields describes the black string solution (\ref{d1d5p-e}) in the neutral limit $\delta_1=\delta_5=0$.

Finally, we derive the conformal factor $e^{2\nu}$. Since $\Gamma^{(a)}=\Gamma^{(b)}=0$, we can use the simplified formula (\ref{conf-formula0}). 
Substituting (\ref{bs-gamma}) into (\ref{conf-formula0}) leads to the conformal factor 
\begin{align}
    e^{2\nu}=k_{\rm BM}\cdot \prod_{j=1}^{2}(\la_j\nu_j)\,{\rm det}\,\Gamma^{(0)}=-k_{\rm BM}\left(\frac{l^2}{2}\right)^2\frac{\la_1^2\la_2^2}{\rho^2(1+\la_1^2)(1+\la_2^2)\la_{1,2}^2}\,.
\end{align}
This expression is equal to the conformal factor for the black string up to the overall factor:
\begin{align}\label{bs-conf}
    e^{2\nu}=\frac{r^2(r^2+l^2)}{(r^2+l^2\cos^2\theta)(r^2+l^2\sin^2\theta)}\,,
\end{align}
which is the simplest case of (\ref{conf-mbtz}) i.e. $G=1$.
The overall factor $k_{\rm BM}$ is determined as $k_{\rm BM}=-1$ from the asymptotic structure of (\ref{bs-conf}).
The same result can be obtained by using the formula (\ref{conf-formula}), and this can be explicitly seen in the multi-neutral black string case.

\subsubsection{Multi-neutral black string}

\begin{figure}
\begin{center}
\begin{tikzpicture}[scale=0.65]
\node[font=\small ] at (-10,4) {$t$};
\node[font=\small ] at (-10,3) {$y$};
\node[font=\small ] at (-10,2) {$\varphi_1$};
\node[font=\small ] at (-10,1) {$\varphi_2$};
\node[font=\small ] at (-10,0) {$z$};
\node[font=\small ] at (-7,-0.5) {$0$};
\node[font=\small ] at (-5,-0.5) {$\frac{l^2_1}{4}$};
\node[font=\small ] at (-3,-0.5) {$\frac{l^2_1+l_2^2}{4}$};
\node[font=\small ] at (-1,-0.5) {$\frac{l^2_1+l_2^2+l_3^2}{4}$};
\node[font=\small ] at (2,-0.5) {$\cdots\cdots$};
\node[font=\small ] at (5,-0.5) {$\frac{l^2-l_7^2}{4}$};
\node[font=\small ] at (7,-0.5) {$\frac{l^2}{4}$};
\draw[black,dashed ] (-7,0) -- (-7,4);
\draw[black,dashed ] (-5,0) -- (-5,4);
\draw[black,dashed ] (-3,0) -- (-3,4);
\draw[black,dashed ] (-1,0) -- (-1,4);
\draw[black,dashed ] (1,0) -- (1,4);
\draw[black,dashed ] (3,0) -- (3,4);
\draw[black,dashed ] (5,0) -- (5,4);
\draw[black,dashed ] (7,0) -- (7,4);
\draw[gray,line width = 0.8] (-9,4) -- (9,4);
\draw[black,line width = 5] (-7,4) -- (-5,4);
\draw[black,line width = 5] (-3,4) -- (-1,4);
\draw[black,line width = 5] (1,4) -- (3,4);
\draw[black,line width = 5] (5,4) -- (7,4);
\draw[gray,line width = 0.8] (-9,3) -- (9,3);
\draw[black,line width = 5] (-5,3) -- (-3,3);
\draw[black,line width = 5] (-1,3) -- (1,3);
\draw[black,line width = 5] (3,3) -- (5,3);
\draw[gray,line width = 0.8] (-9,2) -- (9,2);
\draw[black,line width = 5,dashed ] (-9,2) -- (-8.5,2);
\draw[black,line width = 5] (-8.4,2) -- (-7,2);
\draw[gray,line width = 0.8] (-9,1) -- (9,1);
\draw[black,line width = 5] (7,1) -- (8.4,1);
\draw[black,line width = 5,dashed ] (8.5,1) -- (9,1);
\draw[->,black,line width = 1] (-9,0) -- (9,0);
\end{tikzpicture}
\caption{The rod structure of the multi-black string solution with $n=7$. The sets of $U_t$ and $U_y$ for this case are $U_t=\{1,3,5,7\}$ and $U_y=\{2,4,6\}$.}\label{rod-mbstring-able}
\end{center}
\end{figure}
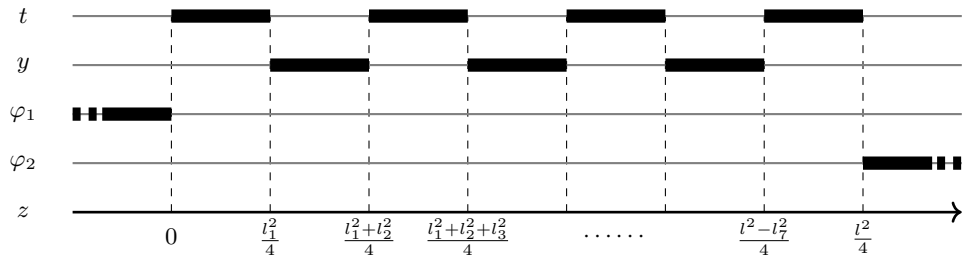

Next we consider a derivation of the multi-black string solution (\ref{mbstring}) with $n_b$ bubbles whose the associated rod structure is illustrated in Fig.\,\ref{rod-mbstring-able}. We apply a dressing transformation to flat spacetime such that the transformed monodromy matrix $\cM_{\rm mString}(w)$ has $N=n+1=2n_b+2$ simple poles in the $w$-plane
\begin{align}\label{mnst-mono}
    \cM_{\rm mString}(w)=Y_{\text{flat}}+\sum_{j=1}^{N}\frac{A_j}{w-w_j}\,,
\end{align}
where the positions of poles are
\begin{align}
    w_1=0\,,\quad w_2=\frac{l_1^2}{4}\,,\dots\,,\quad w_i=\sum_{j=1}^{i-1}\frac{l_j^2}{4}\,,\dots\,,\quad  w_{N}=\frac{l^2}{4}\,.
\end{align}
Although there is currently no systematic prescription for determining the residue matrices $A_j$ from a given rod structure, candidate monodromy matrices can be constructed using (\ref{sub-rule}) from the gauge invariant coset matrix associated with the neutral multi-black string:
\begin{align}
    M_{\rm mString}(z,\rho)=
        \begin{pmatrix}
        f&0&0&0&0&0&0&0\\
        0&1&0&0&0&0&0&0\\
        0&0&\frac{4f}{\hat{r}^2}W_b&0&0&0&0&\frac{m-\hat{r}^2}{2m-\hat{r}^2}W_b\\
        0&0&0&\frac{4}{\hat{r}^2}W_b^{-1}&0&0&-(1-\frac{m}{\hat{r}^2})W_b^{-1}&0\\
        0&0&0&0&f^{-1}&0&0&0\\
        0&0&0&0&0&1&0&0\\
        0&0&0&(1-\frac{m}{\hat{r}^2})W_b^{-1}&0&0&-\frac{m^2}{4\hat{r}^2}W_b^{-1}&0\\
        0&0&-\frac{m-\hat{r}^2}{2m-\hat{r}^2}W_b&0&0&0&0&-\frac{m^2}{4\hat{r}^2}fW_b\\
    \end{pmatrix}\,,\label{nmbs-m}
\end{align}
where the warp factor $W_b$ for the bubble part is defined in (\ref{bubble-part}) and the function $f$ is
\begin{align}
  f=\frac{1}{1-\frac{2m}{\hat{r}^2}}=1+\frac{2m}{r^2}\,.
\end{align}
The corresponding 16 scalar fields are given by
\begin{align}
\begin{split}\label{nmblack-string}
 e^{2U}&=\frac{\hat{r}^2}{4}\left(1-\frac{2m}{\hat{r}^2}\right)^{1/2}\,,\quad
    x^I=0\,,\\ 
    y^1&=f^{-1}W_b^{-1}\,,\quad y^2=\hat{Z}_1^{-1}f^{-1}\,,\quad  y^3=\hat{Z}_5^{-1}f^{-1}\,,\\
    \tilde{\zeta}_\Lambda&=0\,,\quad
    \zeta^\Lambda=0\,,\quad
    \sigma=2T_0\,.
\end{split}
\end{align}
By using (\ref{sub-rule}), we can find a candidate monodaromy matrix 
\tiny
\begin{align}
        \cM_{\rm mString}(w)&=
        \begin{pmatrix}
        1-\frac{l^2}{4w}&0&0&0&0&0&0&0\\
        0&1&0&0&0&0&0&0\\
        0&0&-\frac{1}{w}\cW_b(w)&0&0&0&0&\left(1-\frac{l^2}{8w}\right)\cW_b(w)\\
        0&0&0&-\frac{1}{w-w_{N}}\cW_b(w)^{-1}&0&0&-\left(1+\frac{l^2}{8}\frac{1}{w-w_{N}}\right)\cW_b(w)^{-1}&0\\
        0&0&0&0&1+\frac{l^2}{4}\frac{1}{w-w_{N}}&0&0&0\\
        0&0&0&0&0&1&0&0\\
        0&0&0&\left(1+\frac{l^2}{8}\frac{1}{w-w_{N}}\right)\cW_b(w)^{-1}&0&0&\frac{l^4}{64}\frac{1}{w-w_{N}}\cW_b(w)^{-1}&0\\
        0&0&-\left(1-\frac{l^2}{8w}\right)\cW_b(w)&0&0&0&0&\frac{l^4}{64w}\cW_b(w)\\
    \end{pmatrix}\label{mono-mat}
    \,,
\end{align}
\normalsize
where we set $l^2=2m$ and the function $\cW_b(w)$ is obtained by taking a limit of $\rho\to 0^+$ of the function $W_b(w)$ for $z<-R\,(R\gg w_{N})$ :
\begin{align}
       \cW_{b}(z)=\lim_{\rho\to 0^+}W_b&=\prod_{i\in U_y}\left(\frac{z-w_{i}}{z-w_{i+1}}\right)\,.
\end{align}
The matrices $A_k$ in (\ref{mnst-mono}) can be computed by taking the residue of each pole of (\ref{mono-mat}), and these matrices are found to be rank 2. The explicit expressions of $A_j$ and the corresponding eight-components vectors $a_{j}, b_{j}$ are presented in Appendix \ref{sec:A-mat}.
As shown in Appendix \ref{sec:vanih-gamma}, we can see $\gamma_i=0$ for every $i$ i.e.
\begin{align}
    \cA_i\eta \eta'a_i&=0\,,\qquad
    (\eta b_i)^{T}\eta'\eta \cA_i^{T}=0\,,
\end{align}
where $\cA_i$ are defined in (\ref{gamma-def}).
The non-zero components of the $N\times N$ matrices $\Gamma^{(0)},\Gamma^{(a)}, \Gamma^{(b)}$ from the definitions (\ref{gamma0}), (\ref{gammaa}) and (\ref{gammab}) are
\begin{align}
\begin{split}
    \Gamma^{(0)}_{1,N}&=\frac{l^2}{4\la_{1,N}}=\Gamma^{(0)}_{N,1}\,,\qquad 
    \Gamma^{(0)}_{2i,1}=-\frac{w_{2i}}{4\la_{2i,1}}\qquad 
  \Gamma^{(0)}_{N,2i+1}=-\frac{l^2}{8\la_{N,2i+1}}\,,\\
        \Gamma^{(0)}_{2i,2i+1}&=\frac{l^2-8w_{2i}}{8\la_{2i,2i+1}} \,,\qquad \Gamma^{(0)}_{2i+1,2i}=-\frac{8(l^2-8w_{2i+1})}{l^4\la_{2i+1,2i}}\,,
\end{split}
\end{align}
and
\begin{align}
\begin{split}
    \Gamma^{(a)}_{2i,2i+1}&=\frac{l^4-(l^2-8w_{2i})(l^2-8w_{2i+1})}{l^4\la_{2i,2i+1}}=-\Gamma^{(a)}_{2i+1,2i}\,,\\
    \Gamma^{(a)}_{2i+1,N}&=\frac{2(l^2-4w_{2i+1})}{l^2\la_{2i+1,N}}=-\Gamma^{(a)}_{N,2i+1}\,,
\end{split}
\end{align}
and
\begin{align}
\begin{split}
    \Gamma^{(b)}_{2i,2i+1}&=\frac{1}{\la_{2i,2i+1}}=-\Gamma^{(b)}_{2i+1,2i}\,,\qquad
    \Gamma^{(b)}_{1,2i}=\frac{1}{\la_{1,2i}}=-\Gamma^{(b)}_{2i,1}\,.
\end{split}
\end{align}
Here, we introduced the notation $\la_{i,j}=\la_i-\la_j$. 
Unlike the black string case discussed in the previous section, the matrices $\Gamma^{(a)}$ and $\Gamma^{(b)}$ are non-vanishing, and hence the simplified formula (\ref{cd-ab-sim}) for the vectors $c_j$ and $d_j$ cannot be used.

In principle, one can obtain an explicit expression for $X_+$ for arbitrary $N$ by solving the equations (\ref{c-vec}) and (\ref{d-vec}) for vectors $c_j$ and $d_j$ with the above matrices.
However, since carrying out this computation explicitly is somewhat cumbersome, we begin by deriving the explicit expressions of the matrices $C_j$ for small $N$. Based on these results, we then infer the general forms of $C_j$ for arbitrary $N$.
The resulting expressions for $C_j$ are given as follows:
\begin{align}
\begin{split}
    C_1&=
        \begin{pmatrix}
         \la_{1,N}&0&0&0&0&0&0&0\\
        0&0&0&0&0&0&0&0\\
        0&0&\frac{1}{2} \la_{1,N}\cF_{b}(\la_1)&0&0&0&0&\frac{l^2}{16}\la_{1,N}\cF_{b}(\la_1)\\
        0&0&0&0&0&0&0&0\\
        0&0&0&0&0&0&0&0\\
        0&0&0&0&0&1&0&0\\
        0&0&0&0&0&0&0&0\\
        0&0&\frac{4}{l^2}\la_{1,N}\cF_{b}(\la_1)&0&0&0&0&\frac{1}{2}\la_{1,N} \cF_{b}(\la_1)\\
    \end{pmatrix}\,,\label{C1-mst}\\
    C_{2j}&=
       \begin{pmatrix}
       0&0&0&0&0&0&0&0\\
        0&0&0&0&0&0&0&0\\
        0&0&0&0&0&0&0&0\\
        0&0&0&\frac{1}{2}\frac{\la_{N}+\la_1-2\la_{2j}}{\la_{N,2j}}\widetilde{\cF}_b(\la_{2j})^{-1}&0&0&-\frac{l^2}{16}\frac{\la_{N,1}}{\la_{N,2j}}\widetilde{\cF}_b(\la_{2j})^{-1}&0\\
        0&0&0&0&0&0&0&0\\
        0&0&0&0&0&0&0&0\\
        0&0&0&-\frac{4}{l^2}\frac{\la_{N,1}}{\la_{N,2j}}\widetilde{\cF}_b(\la_{2j})^{-1}&0&0&\frac{1}{2}\frac{\la_{N}+\la_1-2\la_{2j}}{\la_{N,2j}} \widetilde{\cF}_b(\la_{2j})^{-1}&0\\
        0&0&0&0&0&0&0&0\\
    \end{pmatrix}\,,\\
    C_{2j+1}&=
   \begin{pmatrix}
      0&0&0&0&0&0&0&0\\
        0&0&0&0&0&0&0&0\\
        0&0&\frac{1}{2}\frac{\la_{N}+\la_{1}-2\la_{2j+1}}{\la_{1,2j+1}}\widetilde{\cF}_b(\la_{2j+1})&0&0&0&0&\frac{l^2}{16}\frac{\la_{N,1}}{\la_{1,2j+1}}\widetilde{\cF}_b(\la_{2j+1})\\
        0&0&0&0&0&0&0&0\\
        0&0&0&0&0&0&0&0\\
        0&0&0&0&0&0&0&0\\
        0&0&0&0&0&0&0&0\\
        0&0&\frac{4}{l^2}\frac{\la_{N,1}}{\la_{1,2j+1}} \widetilde{\cF}_b(\la_{2j+1})&0&0&0&0&\frac{1}{2}\frac{\la_{N}+\la_{1}-2\la_{2j+1}}{\la_{1,2j+1}}\widetilde{\cF}_b(\la_{2j+1})\\
    \end{pmatrix}\,,\\
        C_{N}&=
        \begin{pmatrix}
        0&0&0&0&0&0&0&0\\
        0&0&0&0&0&0&0&0\\
        0&0&0&0&0&0&0&0\\
        0&0&0&-\frac{1}{2}\la_{1,N}\cF_{b}(\la_{N})^{-1}&0&0&-\frac{l^2}{16}\la_{1,N}\cF_{b}(\la_{N})^{-1}&0\\
        0&0&0&0&-\la_{1,N}&0&0&0\\
        0&0&0&0&0&0&0&0\\
        0&0&0&-\frac{4}{l^2}\la_{1,N}\cF_{b}(\la_{N})^{-1}&0&0&-\frac{1}{2}\la_{1,N}\cF_{b}(\la_{N})^{-1}&0\\
        0&0&0&0&0&0&0&0\\
    \end{pmatrix}\,.
\end{split}
\end{align}
Here, we introduced a new scalar function $\cF_{b}(\la)$ defined as
\begin{align}
    \cF_{b}(\la)=\prod_{i=1}^{n_b}\frac{\la-\la_{2i}}{\la-\la_{2i+1}}\,.
\end{align}
The values at $\la=\la_1,\la_{N}$ and the residues at $\la=\la_{2j}$ and $\la=\la_{2j+1}$ are given by
\begin{align}
\begin{split}\label{cF-value}
    \cF_{b}(\la_1)&=\prod_{i=1}^{n_b}\frac{\la_{1,2i}}{\la_{1,2i+1}}\,,\qquad\qquad\qquad 
    \cF_{b}(\la_{N})=\prod_{i=1}^{n_b}\frac{\la_{N,2i}}{\la_{N,2i+1}}\,,\\
    \widetilde{\cF}_b(\la_{2j})^{-1}&=\underset{\la=\la_{2j}}{\rm res}\cF_b(\la)^{-1}=-\la_{2j+1,2j}\prod_{\substack{i=1\\i\neq j}}^{n_b}\frac{\la_{2j,2i+1}}{\la_{2j,2i}}\,,\\ 
   \widetilde{\cF}_b(\la_{2j+1})&=\underset{\la=\la_{2j+1}}{\rm res}\cF_b(\la)=\la_{2j+1,2j}\prod_{\substack{i=1\\i\neq j}}^{n_b}\frac{\la_{2j+1,2i}}{\la_{2j+1,2i+1}}\,.
\end{split}
\end{align}
As we will see in appendix \ref{subsec:fac-mono}, the function $\cF_{b}(\la)$ could be regarded as a counterpart in the $\la$-plane of $W_b(w)$ which describes the bubbles.
In that appendix, we give a proof that the matrices $X_{\pm}$, constructed from $C_{j}$ given in (\ref{C1-mst}), enable the factorization of the monodromy matrix (\ref{mono-mat}) for arbitrary $N$, and reproduce the gauge invariant element (\ref{nmbs-m}) corresponding to the neutral black string solution.
In this way, the monodromy matrix (\ref{mnst-mono}) describes the multi-neutral black string solution.

\subsubsection*{Conformal factor}

As explained in Sec.\,\ref{sec:mbst}, the conformal factor $e^{2\nu}$ for the multi-black string coincides with that of the multi-BTZ black hole, and is independent of the RR charges. Accordingly, the multi-neutral black string also has the same conformal factor
\begin{align}\label{mnst-conf}
\begin{split}
    e^{2\nu}&=G(r,\theta)\frac{r^2(r^2+l^2)}{(r^2+l^2\cos^2\theta)(r^2+l^2\sin^2\theta)}\,,\\
    G&=\prod_{\substack{i\in U_y\\j\in U_t}}\frac{\left((r_i^2+l_i^2)\cos^2\theta_i+r^2_j\sin^2\theta_j\right)
    \left(r_i^2\cos^2\theta_i+(r_j^2+l_j^2)\sin^2\theta_j\right)}{\left((r_i^2+l_i^2)\cos^2\theta_i+(r_j^2+l_j^2)\sin^2\theta_j\right)\left(r_i^2\cos^2\theta_i+r_j^2\sin^2\theta_j\right)}\,.
\end{split}
\end{align}
In the remainder of this Section, we derive the above expression by using the formula (\ref{conf-formula}).

To this end, it is convenient to rewrite (\ref{mnst-conf}) in terms of the positions $\la_j$ of solitions.
From the definition (\ref{ri-cosi}) of the local spherical coordinates, we obtain useful identities
\begin{align}
\begin{split}
    r^2&=l^2\frac{\la_{N}}{\la_{1,N}}\,,\qquad\qquad r^2+l^2=l^2\frac{\la_1}{\la_{1,N}}\,,\\
    \frac{1}{r^2+l^2\sin^2\theta}&=\frac{1}{2\rho}\frac{\la_{N}}{\la_{N}^2+1}\,,\qquad \frac{1}{r^2+l^2\cos^2\theta}=\frac{1}{2\rho}\frac{\la_1}{\la_1^2+1}\,,
\end{split}
\end{align}
and
\begin{align}
\begin{split}
    r_i^2 \cos^2\theta_i&=2\rho\,\la_{i+1}\,,\qquad   r_i^2 \sin^2\theta_i=2\rho\,\la_i^{-1}\,,\\
     (r_i^2+l_i^2)\cos^2\theta_i&=2\rho\,\la_i\,,\qquad 
    (r_i^2+l_i^2)\sin^2\theta_i=2\rho \,\la_{i+1}^{-1}\,.
\end{split}
\end{align}
The conformal factor (\ref{mnst-conf}) can be rewritten as
\begin{align}\label{conf-ex-mst}
e^{2\nu}&=\prod_{j=1}^{n_b}\frac{1}{1+\la_{2j}^2}\frac{1}{1+\la_{2j+1}^2}\frac{(1+\la_1\la_{2j})(1+\la_{2j+1}\la_{N})}{(1+\la_1\la_{2j+1})(1+\la_{2j}\la_{N})}\no\\
    &\quad\times \prod_{p<q}^{n_b}\frac{1}{(1+\la_{2p+1}\la_{2q+1})^2(1+\la_{2p}\la_{2q})^2}
    \prod_{p,q=1}^{n_b}(1+\la_{2p}\la_{2q+1})^2\no\\
    &\quad\times \left(\frac{l^4}{4}\frac{1}{\rho^2}\frac{\la_1^2}{1+\la_1^2}\frac{\la_{N}^2}{1+\la_{N}^2}\frac{1}{\la_{1,N}^2}\right)\no\\
   &\propto \frac{1}{\rho^{N}}\prod_{j=1}^{n_b}\frac{\la_{2j}^2}{1+\la_{2j}^2}\frac{\la_{2j+1}^2}{1+\la_{2j+1}^2}\frac{\la_{1,2j+1}}{\la_{1,2j}}\frac{\la_{2j,N}}{\la_{2j+1,N}}\no\\
    &\quad\times \prod_{p<q}^{n_b}\la_{2p+1,2q+1}^2\la_{2p,2q}^2\prod_{p,q=1}^{n_b}\frac{1}{\la_{2p,2q+1}^2}\left(\frac{\la_1^2}{1+\la_1^2}\frac{\la_{N}^2}{1+\la_{N}^2}\frac{1}{\la_{1,N}^2}\right)\,,
\end{align}
where we used 
\begin{align}
    1+\la_i\la_j=-\frac{2(w_i-w_j)}{\rho}\frac{\la_i\la_j}{\la_{i,j}}\,.
\end{align}

Now, we derive the conformal factor (\ref{conf-ex-mst}) by using the formula (\ref{conf-formula}).
To apply this formula, it is necessary to show that the right-invariant current for $X_+(\la,z,\rho)$ in the $\la$-direction takes the form (\ref{rXp-ass}) which has only simple poles in the $\la$-plane.
For the $n_b=1,2$ cases, we have explicitly verified that $X_+(\la,z,\rho)$ characterized by $C_j$ given in (\ref{C1-mst}) takes the desired form (\ref{rXp-ass}). Based on the explicit expressions of $B_j$ for a few $n_b$, we expect that the residue matrices $B_j\,(j=1,\dots,N)$ for any $N$ are given by
\begin{align}
\begin{split}
    B_1&=
    \begin{pmatrix}
        -1&0&0&0&0&0&0&0\\
        0&0&0&0&0&0&0&0\\
        0&0&-\frac{1}{2}&0&0&0&0&-\frac{w_{N}}{4}\\
        0&0&0&\frac{1}{2}&0&0&\frac{w_{N}}{4}&0\\
        0&0&0&0&1&0&0&0\\
        0&0&0&0&0&0&0&0\\
        0&0&0&\frac{1}{w_{N}}&0&0&\frac{1}{2}&0\\
        0&0&-\frac{1}{w_{N}}&0&0&0&0&-\frac{1}{2}
    \end{pmatrix}
    =-B_{N}\,,\\
        B_{2j+1}&=
    \begin{pmatrix}
        0&0&0&0&0&0&0&0\\
        0&0&0&0&0&0&0&0\\
        0&0&-1&0&0&0&0&0\\
        0&0&0&1&0&0&0&0\\
        0&0&0&0&0&0&0&0\\
        0&0&0&0&0&0&0&0\\
        0&0&0&0&0&0&1&0\\
        0&0&0&0&0&0&0&-1
    \end{pmatrix}
=-B_{2j}
    \,.\label{B-def}
\end{split}
\end{align}
The tracelessness of the matrices $B_j$ follows from the fact that the right-invariant current $\partial_{\la}X_+X_{+}^{-1}$ takes a value in $\mathfrak{so}(4,4)$.
Although we do not provide an explicit derivation of $B_j$, we proceed using these expressions of $B_j$ to derive the conformal factor (\ref{conf-ex-mst}).
From (\ref{B-def}), the matrices $B_j$ satisfy
\begin{align}
  B_{\infty}=\sum_{i=1}^{N}B_i=0\,,\qquad  \Tr(B_j^2)=4\qquad \text{for}\,\, j=1,\dots, N\,,
\end{align}
and
\begin{align}
\begin{split}
    \Tr(B_1B_{N})&=-4\,,\qquad \Tr(B_1B_{2j})=-2\,,\qquad \Tr(B_1B_{2j+1})=2\,,\\
    \Tr(B_{2i}B_{2j+1})&=-4\,,\qquad \Tr(B_{2i}B_{2j})=4\,,\qquad \Tr(B_{2i+1}B_{2j+1})=4\,,\\
   \Tr(B_{2j}B_{N})&=2 \,,\qquad \Tr(B_{2j+1}B_{N})=-2\,,
\end{split}
\end{align}
where $i,j=1,\dots,n_b$.
By using the formula (\ref{conf-formula}) along with the above data for the matrices $B_j$, we can obtain
\begin{align}
    e^{2\nu}
    &=K_{\rm BM}\,\cdot \prod_{j=1}^{N}(\la_j \nu_j)\prod_{p<q}\la_{p,q}^{\frac{1}{2}\Tr(B_pB_q)}\no\\
    &=K_{\rm BM}\,\frac{(-2)^{N}}{\rho^{N}}\prod_{j=1}^{n_b}\frac{\la_{2j}^2}{1+\la_{2j}^2}\frac{\la_{2j+1}^2}{1+\la_{2j+1}^2}\frac{\la_{1,2j+1}}{\la_{1,2j}}\frac{\la_{2j,N}}{\la_{2j+1,N}}\no\\
    &\quad\times \prod_{p<q}^{n_b}\la_{2p+1,2q+1}^2\la_{2p,2q}^2\prod_{p,q=1}^{n_b}\frac{1}{\la_{2p,2q+1}^2}\left(\frac{\la_1^2}{1+\la_1^2}\frac{\la_{N}^2}{1+\la_{N}^2}\frac{1}{\la_{1,N}^2}\right)\,.
\end{align}
This precisely matches the conformal factor (\ref{conf-ex-mst}) for the multi-neutral black string up to overall constant.

\section{multi-BTZ black hole from global $SO(4,4)$ transformations}\label{sec:harrison}

In this section, we demonstrate that the multi-BTZ black hole solution~(\ref{btzs-sol}) can be obtained from the multi-neutral black string solution~(\ref{nmblack-string}) by applying global $SO(4,4)$ transformations 
\begin{align}\label{mono-trans-btz-st}
    \cM_{\rm mBTZ}(w)=g^{\natural}\cM_{\rm mString}(w)g\qquad \text{for}\quad g\in SO(4,4)\,.
\end{align}
Under any global $SO(4,4)$ transformation, the factorized form of the monodromy matrix is preserved. Thus, the transformation~(\ref{mono-trans-btz-st}) is naturally inherited as a transformation of the gauge-invariant element:
\begin{align}\label{ginv-btz-bs}
    M_{\rm mBTZ}(z,\rho)=g^{\natural}M_{\rm mString}(z,\rho)g\,.
\end{align}
This shows that the two classical solutions belong to the same orbit within the symmetric coset space~(\ref{so44-coset-iib}).
In the case of a single BTZ black hole, the relation~(\ref{ginv-btz-bs}) was established in \cite{Cvetic:2011dn,Cvetic:2013cja,Sahay:2013xda} by using a sequence of global $SO(4,4)$ transformations, referred to as ``Harrison transformations". 
As we will show, the same set of global $SO(4,4)$ transformations used in \cite{Cvetic:2013cja, Sahay:2013xda} also ensures that the multi-BTZ black holes are related to multi-black strings through the relations ~(\ref{mono-trans-btz-st}) or (\ref{ginv-btz-bs}), while preserving the rod structure. This result indicates that the multi-BTZ black holes are subtracted geometries of the multi-black strings, extending the results of \cite{Cvetic:2011dn, Cvetic:2013cja, Sahay:2013xda}.

\subsection{Global $SO(4,4)$ transformations}

Let us first present the explicit expression of the global $SO(4,4)$ element $g$ that ensures the relations (\ref{mono-trans-btz-st}) and (\ref{ginv-btz-bs}).
The global $SO(4,4)$ transformation can be decomposed into three distinct transformations : (i) charging $g_c$, (ii) subtraction $g_{\text{sub}}$, (iii) scaling $g_{\text{scale}}$, 
\begin{align}\label{so44-total}
    g=g_{c}\cdot g_{\text{sub}}\cdot g_{\text{scale}}\in SO(4,4)\,,
\end{align}
where $g_c\,, g_{\text{sub}} \,, g_{\text{scale}}\in SO(4,4)$ are given by
\begin{align}
    g_{c}&=\exp[\beta_2(E_2+F_2)+\beta_3(E_3+F_3)]\,,\\
    g_{\text{sub}}&=\exp[d_2F_2+d_3F_3]\,,\\
    g_{\text{scaling}}&=\exp[-c_2H_2-c_3H_3]\,.
\end{align}
The real parameters $\beta_k, d_k, c_k$ are chosen as
\begin{align}
   \beta_2&=-\delta_1\,,\qquad \beta_3=-\delta_5\,,\\
   d_2&=1\,,\qquad d_3=1\,,\\
   c_2&=-\delta_1+\frac{1}{2}\log\left(\frac{2m}{Q_1}\right)\,,\qquad
    c_3=-\delta_5+\frac{1}{2}\log\left(\frac{2m}{Q_5}\right)\,,
\end{align}
where it should be noted from Eq.~(\ref{conf-eom}) that  the conformal factor $e^{2\nu}$ is invariant under global $SO(4,4)$ transformations. 
More detailed explanation of each transformation will be provided later. We briefly outline the overview of these transformations.

More specifically, the first transformation, $g_c$, adds D1 and D5 charges to the multi-neutral black string while preserving its asymptotic structure, leading to the multi-charged black string solution. 
The subsequent transformation, $g_{\text{sub}}$, changes the asymptotic structure of the multi-black string (or equivalently the constant matrix $Y_{\text{flat}}$), resulting in the multi-BTZ black hole, which asymptotically approaches \ads spacetime. 
The final transformation, $g_{\text{scaling}}$, scales the overall factor of the scalar fields $x^I$ and $y^I$, allowing us to adjust the magnitude of the D1 and D5 charges. 
By tracing the path of these three transformations in the symmetric coset space, we find that only four scalar fields, ${x^{2}, x^{3}, y^{2}, y^{3}}$, undergo changes. For clarity, we present the scalar fields ${x^{2}, x^{3}, y^{2}, y^{3}}$ corresponding to these three gravitational solutions:
\paragraph{Multi-neutral black string}
\begin{align}
\begin{split}\label{xy-mnst}
    x^{2}&=0\,,\qquad x^{3}=0\,,\\
    y^2&=\hat{Z}_1^{-1}\left(1-\frac{2m}{\hat{r}^2}\right)^{1/2}\,,\qquad  y^3=\hat{Z}_5^{-1}\left(1-\frac{2m}{\hat{r}^2}\right)^{1/2}\,.
\end{split}
\end{align}
\paragraph{Multi-(charged) black string}
\begin{align}
\begin{split}\label{d1d5-scasol}
 x^2&=\frac{2ms_1c_1}{\hat{r}^2\hat{Z}_1}\,,\quad x^3=\frac{2ms_5c_5}{\hat{r}^2\hat{Z}_5}\,,\\
 y^2&=\hat{Z}_1^{-1}\left(1-\frac{2m}{\hat{r}^2}\right)^{1/2}\,,\quad  y^3=\hat{Z}_5^{-1}\left(1-\frac{2m}{\hat{r}^2}\right)^{1/2}\,.
\end{split}
\end{align}
\paragraph{Multi-BTZ black hole (\ref{mbtz-pot})}
\begin{align}
\begin{split}
x^2&=\frac{r^2+\frac{l^2}{2}}{Q_1}\,,\quad x^3=\frac{r^2+\frac{l^2}{2}}{Q_5}\,,\\ 
 y^2&=\frac{r^2}{Q_1}\sqrt{1+\frac{l^2}{r^2}}\,,\quad  y^3=\frac{r^2}{Q_5}\sqrt{1+\frac{l^2}{r^2}}\,.
\end{split}
\end{align}
It is easy to see that the global $SO(4,4)$ transformation~(\ref{so44-total}) acts on only four scalar fields, ${x^{2}, x^{3}, y^{2}, y^{3}}$. This is due to the vanishing of $\zeta^{\Lambda}$ and $\tilde{\zeta}_{\Lambda}$ in both solutions, which allows the coset representative $V \in SO(4,4)$ to be expressed as the product of representatives from four coset spaces $SL(2, \mathbb{R}) / SO(2)$:
\begin{align}
    V\lvert_{(\ref{d1d5-scasol-gen})}=\underbrace{\left(e^{-U\,H_0}e^{-\frac{1}{2}\sigma E_0}\right)}_{\frac{SL(2,\mathbb{R})_0}{SO(2)_0}}\cdot\prod_{I=1}^{3}\underbrace{\left(e^{-\frac{1}{2}(\log y^I)H_I}\cdot e^{-x^IE_{I}}\right)}_{\frac{SL(2,\mathbb{R})_I}{SO(2)_I}}\,,
\end{align}
where each $\mathfrak{sl}(2,\mathbb{R})_{\Lambda}$ ($\Lambda=0,1,2,3$) is formed by the generators  
\begin{align}
    \mathfrak{so}(4,4)\supset \bigoplus_{\Lambda=0}^{3}\mathfrak{sl}(2,\mathbb{R})_{\Lambda}= \bigoplus_{\Lambda=0}^{3}\text{span}_{\mathbb{R}}\{ E_{\Lambda},F_{\Lambda},H_{\Lambda}\}\,.
\end{align}
Since the generators of different sets commute with each other, (\ref{so44-total}) transforms the scalar fields $\{x^{2},x^{3},y^{2},y^{3}\}$.
Below, we give brief comments on some properties of the above three $SO(4,4)$ transformations, and for more details, see \cite{Cvetic:2011dn,Cvetic:2013cja,Sahay:2013xda}.

\subsubsection{(i) Charging transformation $g_{c}$}

The first transformation, generated by $g_c$, known as charging transformation, adds D1 and D5 charges to the multi-neutral black string with (\ref{xy-mnst}), and then we obtain the multi-black string solution, characterized by the scalar fields~(\ref{d1d5-scasol}).
The group element $g_c$ is defined such that it preserves the constant matrix $Y_{\text{flat}}$:
\begin{align}\label{charge_con}
    g_{c}^{\natural}\,Y_{\text{flat}}\,g_{c}=Y_{\text{flat}}\,.
\end{align}
The global $SO(4,4)$ transformations satisfying the condition~(\ref{charge_con}) can be expressed in terms of elements of the subgroup $H=SO(2,2)\times SO(2,2)$ \cite{Katsimpouri:2014ara}\footnote{The transformation generated by $E_1+F_1$ with a real parameter $\beta_1$ also satisfy the condition. However, its effect on the gravitational solution can be removed by performing a coordinate transformation on $(t,y)$\cite{Cvetic:2013cja}. Hence, for simplicity, we set $\beta_1=0$.}, and this  transformation preserves the asymptotic structure $\rm R^{1,4}\times \rm S^1\times  \rm T^4$  of the multi-black string.

\subsubsection{(ii) Subtraction $g_{\text{sub}}$}

Next, we discuss the transformation generated by
\begin{align}\label{sub-g}
    g_{\text{sub}}=\exp[d_2F_2+d_3F_3]\,, 
\end{align}
which changes the asymptotic structure of the gravitational solution.
For general real deformation parameters $d_2$ and $d_3$, the action of $g_{\text{sub}}$ on the constant matrix $Y_{\text{flat}}$ is given by
\begin{align}
    g_{\text{sub}}^{\natural}\,Y_{\text{flat}}\,g_{\text{sub}}=        \begin{pmatrix}
        (1-d_2^2)(1-d_3^2)&(1-d_2^2)d_3&0&0&-d_2d_3&(1-d_3^2)d_2&0&0\\
        -(1-d_2^2)d_3&(1-d_2^2)&0&0&-d_2& -d_2d_3&0&0\\
        0&0&0&0&0&0&0&1\\
        0&0&0&0&0&0&-1&0\\
        -d_2d_3&d_2&0&0&1&d_3&0&0\\
        -(1-d_3^2)d_2&-d_2d_3&0&0&-d_3&1-d_3^2&0&0\\
        0&0&0&1&0&0&0&0\\
        0&0&-1&0&0&0&0&0\\
    \end{pmatrix}\,. \label{tY-fa}
\end{align}
The transformed asymptotic matrix (\ref{tY-fa}) indicates that setting the real parameters $d_2$ and $d_3$ to $\pm 1$ significantly changes the asymptotic structure of the original black string.
Indeed, by extracting the four scalar fields $x^{2,3}$ and $y^{2,3}$ from the transformed gauge-invariant element $\widetilde{M}(z,\rho)=g_{\text{sub}}^{\natural}\,M_{\rm mString}\,g_{\text{sub}}$, we can find that at the extremal values $d_2=d_3=\pm1$, they take the form 
\begin{align}
\begin{split}\label{xy-btz-h}
x^2&=\frac{\hat{r}^2e^{2\delta_1}}{2m}\left(1-\frac{2me^{-\delta_1}c_1}{\hat{r}^2}\right)=\frac{r^2+(1-e^{-2\delta_1})m }{Q'_1}\,,\\ 
x^3&=\frac{\hat{r}^2e^{2\delta_5}}{2m}\left(1-\frac{2me^{-\delta_5}c_5}{\hat{r}^2}\right)=\frac{r^2+(1-e^{-2\delta_5})m }{Q'_5}\,,\\ 
y^2&=\frac{\hat{r}^2e^{2\delta_1}}{2m}\left(1-\frac{2m}{\hat{r}^2}\right)^{1/2}=\frac{r^2}{Q_1'}\sqrt{1+\frac{l^2}{r^2}}\,,\\
y^3&=\frac{\hat{r}^2e^{2\delta_5}}{2m}\left(1-\frac{l^2}{\hat{r}^2}\right)^{1/2}=\frac{r^2}{Q_5'}\sqrt{1+\frac{l^2}{r^2}}\,.
\end{split}
\end{align}
Since the constant terms in $x^I$ are gauge dependent,
the scalar fields (\ref{xy-btz-h}) correspond to those of the multi-BTZ black hole carrying D1- and D5-brane charges
\begin{align}
    Q_1'=2m e^{-2\delta_1}\,,\qquad Q_5'=2m e^{-2\delta_5}\,.
\end{align}
The asymptotic constant matrix (\ref{tY-fa}) becomes
\begin{align}
    \widetilde{Y}_{AdS_3\times S_3\times T^4}&=\lim_{d_{2,3}\to 1}g_{\text{sub}}^{\natural}\,Y_{\text{flat}}\,g_{\text{sub}}\no\\
    &=        \begin{pmatrix}
        0&0&0&0&-1&0&0&0\\
        0&0&0&0&-1& -1&0&0\\
        0&0&0&0&0&0&0&1\\
        0&0&0&0&0&0&-1&0\\
        -1&1&0&0&1&1&0&0\\
       0&-1&0&0&-1&0&0&0\\
        0&0&0&1&0&0&0&0\\
        0&0&-1&0&0&0&0&0\\
    \end{pmatrix}\,. 
\end{align}
Thus, the global $SO(4,4)$ transformation (\ref{sub-g}) realizes an interpolation between the multi-black string and the multi-BTZ black hole by varying real parameters $d_2$ and $d_3$ from 0 to $\pm 1$.
This transformation, which interpolates between an asymptotically $R^{1,4}\times S^{1}\times T^4$ and an \ads spacetime, was first discussed in the relationship between a single black string and a single BTZ black hole \cite{Cvetic:2011dn,Cvetic:2013cja,Sahay:2013xda}.
In that context, the BTZ black hole is often referred to as the ``subtracted geometry" of the black string, a concept introduced in \cite{Cvetic:2011dn} to better understand the internal structure of the black string.
In this sense, the multi-BTZ black hole can also be regarded as the ``subtracted geometry" of the multi-black string.

\subsubsection{(iii) Scaling transformation $g_{\text{scale}}$}

We already obtain a multi-BTZ black hole from the second transformation, but the D1 and D5 charges don’t match the multi-black string.
If one wishes to preserve these charges during the $SO(4,4)$ Harrison transformations, this can be done by considering a scaling transformation generated by $g_{\text{scaling}}$, which is also an element of $SO(4,4)$.
Under this transformation, the scalar fields $x^I$ and $y^I$ are transformed as
\begin{align}
    x^I\to e^{2c_I}x^I\,,\qquad y^I\to e^{2c_I}y^I\,,
\end{align}
and by taking three parameters $c_I$ as
\begin{align}
   c_1=0\,,\qquad c_2=\frac{1}{2}\log\left(\frac{Q_1'}{Q_1}\right)\,,\qquad
    c_3=\frac{1}{2}\log\left(\frac{Q_5'}{Q_5}\right)\,,
\end{align} 
we can obtain the scalar fields $x^I\,, y^I$ corresponding to the multi-BTZ black hole with $Q_1$ and $Q_5$ charges defined in (\ref{flat-q1q5}).

The variation of the scalar fields $y^{2,3}$ affects the asymptotic structure of the warp factors $Z_{1,5}$ at large $r$, the constant matrix $\widetilde{Y}_{AdS_3\times S_3\times T^4}$ is also transformed as
\begin{align}
    Y_{AdS_3\times S_3\times T^4}&=g_{\text{scaling}}^{\natural}\,   \widetilde{Y}_{AdS_3\times S_3\times T^4}\,g_{\text{scaling}}\no\\
    &=        \begin{pmatrix}
        0&0&0&0&-1&0&0&0\\
        0&0&0&0&-\frac{Q_5'}{Q_5}& -1&0&0\\
        0&0&0&0&0&0&0&1\\
        0&0&0&0&0&0&-1&0\\
        -1&\frac{Q_5'}{Q_5}&0&0&\frac{Q_1'Q_5'}{Q_1Q_5}&\frac{Q_1'}{Q_1}&0&0\\
       0&-1&0&0&-\frac{Q_1'}{Q_1}&0&0&0\\
        0&0&0&1&0&0&0&0\\
        0&0&-1&0&0&0&0&0\\
    \end{pmatrix}\,. 
\end{align}
We note that the transformed asymptotic matrix $Y_{AdS_3\times S_3\times T^4}$ only depends on the real parameters $\delta_1$ and $\delta_5$.

To summarize the above discussion, we have shown that by applying a series of $SO(4,4)$ transformations to the monodromy matrix corresponding to the multi-neutral black string, we can construct the monodromy matrix associated with a multi-BTZ black hole. 
Its general form can be schematically represented as follows
\begin{align}\label{ads3-monodromy}
     \cM(w)=Y_{AdS_3\times S_3\times T^4}+\sum_{j=1}^{N}\frac{A_{j}}{w-w_{j}}\,,
\end{align}
where the constant matrix $Y_{AdS_3\times S_3\times T^4}$ encodes the asymptotics structure of the corresponding gravitational solution. 
Since the $SO(4,4)$ Harrison transformations preserve the rank of the residue matrices $A_j$, these matrices remain  $8\times 8$ matrices with rank-2.
If we consider more general rank-2 residue matrices $A_j$, it is expected that new asymptotically \ads black hole solutions can be generated. 
We leave the exploration of this direction for future work.

\section{Conclusion and discussion}

In this paper, we have formulated axisymmetric solutions described by the D1-D5-KKm system within the framework of the integrable BM linear system \cite{Breitenlohner:1986um}, and developed a procedure for constructing black hole solutions with asymptotically \ads geometry based on this integrable system. 
By employing the BM approach and the $SO(4,4)$ transformation, we have successfully derived the recently constructed multi-BTZ black hole solutions \cite{Bah:2022pdn}, which are non-BPS black hole solutions with asymptotically \ads geometry.
Following the procedure developed in previous works \cite{Breitenlohner:1986um, Katsimpouri:2012ky, Chakrabarty:2014ora, Katsimpouri:2013wka, Katsimpouri:2014ara}, we adapt an approach in which black hole solutions are constructed as soliton solutions through the factorization of the monodromy matrix associated with the BM linear system.

Our approach begins by solving the Riemann-Hilbert problem for the neutral multi-black string rather than directly tackling the multi-BTZ black hole case due to the simplicity of the  monodromy matrices.
We have then generated the multi-BTZ black hole solution by applying a series of $SO(4,4)$-transformations, including charging, subtraction and scaling,  to this seed solution. 
As a result, we have found that the monodromy matrix corresponding to the multi-BTZ black holes consists of simple poles with rank-2 residue matrices, along with a constant matrix encoding the asymptotic AdS structure, similar to the asymptotically flat black hole solutions considered in \cite{Katsimpouri:2013wka}.
In this study, we have found that, compared to the single black string case, factorizing the monodromy matrix for the multi-neutral black string requires relaxing a specific assumption made in \cite{Katsimpouri:2013wka}. 
Nevertheless, we successfully perform an explicit factorization of the monodromy matrix, even in the presence of an arbitrary number of bubbles.
Furthermore, for the computation of the conformal factor $e^{2\nu}$, we have extended the procedure developed in \cite{Korotkin:1994au, Korotkin:1996vi}, which heavily relies on the analytic properties of the monodromy matrix, within our framework. 
This extension allows us to obtain a closed formula for the conformal factor that remains valid in more general settings. As a result, we successfully have obtained the conformal factor for the multi-BTZ black holes explicitly.
These results show that, as in the asymptotically flat cases, the approach of constructing black hole solutions via solving the Riemann-Hilbert problem remains a very powerful solution-generating technique, even for asymptotically AdS black hole solutions.

One future direction of this work is the construction of a broad class of non-BPS, non-extremal black hole solutions with asymptotically AdS geometry. This can be achieved by appropriately modifying the rank-2 residue matrices in the monodromy matrix (Eq. \ref{ads3-monodromy}). Constructing such more general AdS black hole solutions presents a natural direction for future research. In this regard, a recent work \cite{Chakraborty:2025ger} has suggested that non-BPS microstate geometries can also be constructed using solution-generating techniques associated with coset model symmetry, the Ernst formalism, and the inverse scattering method. By applying the techniques developed in this study, it may be possible to construct non-BPS microstate geometries not only in asymptotically flat space but also in asymptotically AdS spacetimes. Additionally, exploring a microscopic description of non-BPS AdS black hole solutions would be an important avenue for further investigation.

\section*{Acknowledgements}

J.Y.\ thank Ryotaku Suzuki for useful discussions.
S.T.\ was supported by JSPS KAKENHI Grant Number 21K03560.

\newpage

\appendix

\section*{Appendix}

\section{Dimensional reduction of type IIB supergravity}\label{sec:dim_red}

In this appendix, we will give a small review of a dimensional reduction of type IIB supergravity from 10D to 4D.

\subsection{10D type IIB supergravity}

Here we present the action and the equations of motion for 10D type IIB supergravity in the string frame.
The bosonic part of the type IIB supergravity action in the string frame is given by
\begin{align}\label{IIBaction}
    S_{\rm 10D}&=\int d^{10}x\,\sqrt{-\hat{g}_{10}}\left(e^{-2\Phi_{10}}(\hat{R}_{10}+4\partial_{\mu}\Phi_{10}\partial^{\mu}\Phi_{10}-\frac{1}{2}|\hat{H}_3|^2)-\frac{1}{2}\sum_{p=1,3,5}|\hat{F}_p|^2\right)\no\\
    &\qquad \qquad +\frac{1}{2}\int F_5\wedge F_3\wedge B_2\,,
\end{align}
where $H_3=dB_2$ is the field strength of the NS-NS 2-form field $B_2$ and $F_1\,, F_3\,, F_5$ are the RR field strengths
\begin{align}
    F_n=dC_{n-1}+H_3\wedge C_{n-3}\,.
\end{align}
We use the notation $|\hat{F}_p|^2=\frac{1}{p!}\hat{g}^{k_1l_1}\dots \hat{g}^{k_pl_p}F_{k_1\dots k_{p}}F_{l_1\dots l_{p}}$, and $\hat{g}_{mn}(m,n=0,1,\dots,9)$ expresses the ten-dimensional metric in the string frame.
The equations of motion are given by 
\begin{align}
\begin{split}\label{iib-eom}
    &\hat{R}_{10}+4\nabla^2\Phi_{10}-4\nabla_m\Phi_{10}\nabla^m\Phi_{10}-\frac{1}{2}|\hat{H}_3|^2=0\,,\\
    &\hat{R}_{mn}-\frac{1}{4}H_{mkl}H_{n}{}^{kl}+2\nabla_m\nabla_n\Phi_{10}=T_{mn}\,,\\
    &\frac{1}{2}\nabla^kH_{kmn}+\frac{1}{2}F^{k}F_{kmn}+\frac{1}{12}F_{mnklp}F^{klp}-\partial_k\Phi_{10} H^{k}{}_{mn}=0\,,\\
    &\nabla^{m}\cF_{m}-\partial^{m}\cF_{m}-\frac{1}{6}H^{mnk}\cF_{mnk}=0\,,\\
    &\nabla^{k}\cF_{kmn}-\partial^{k}\Phi_{10} \cF_{kmn}-\frac{1}{6}H^{kpq}\cF_{kpqmn}=0\,,\\
    &\nabla^{k}\cF_{kmnpq}-\partial^{k}\Phi_{10} \cF_{kmnpq}+\frac{1}{36}\epsilon_{mnpqrstuvw}H^{rst}\cF^{uvw}=0\,,
\end{split}
\end{align}
where $\cF_p=e^{\Phi_{10}}F_{p}$ and $T^{mn}$ is the matter contribution from the RR fields defined as
\begin{align}
    T_{mn}=e^{2\Phi_{10}}\left(|\hat{F}_{1}|^2+\frac{1}{4}F_{mkl}F_{n}{}^{kl}+\frac{1}{4\times 4!}F_{mklsr}F_{n}^{klsr}-\frac{1}{4}\hat{g}_{mn}\left(|\hat{F}_{1}|^2+|\hat{F}_3|^2\right)\right)\,.
\end{align}
Imposing the self-duality of the RR 5-form leads to $|\hat{F}_5|^2=0$.

\subsection{Dimensional reduction from 10D to 3D }

We first consider a dimensional reduction of the ten dimensional solution (\ref{d1d5-bg}) over $S^1(t)\times S^1(y)\times T^4$. The resulting Euclidean 4D supergracity is the $\cN=2$ Euclidean STU model.

\paragraph{10D to 6D} To see this, let us first consider a dimensional reduction of the type IIB supergravity over $T^4$ with the ansatz
\begin{align}
\begin{split}
    ds^2_{10}&=ds_{6}^2+e^{\frac{1}{\sqrt{2}}\Phi}\sum_{a=1}^{4}(dy^a)^2\,,\qquad
     e^{2\Phi_{10}}=e^{\sqrt{2}\Phi}\,,\\
    C_2&=C_2^{(\rm 6D)}\,,\qquad
    C_0=C_4=B_2=0\,.
\end{split}
\end{align}
Through this reduction, the type IIB supergravity action (\ref{IIBaction}) reduces to the bosonic part of 6D supergravity in the Einstein frame \cite{Duff:1995sm}
\begin{align}\label{6dsugra-action}
    S_{\rm 6D}&=\int d^6x\sqrt{-g_6}\biggl(R_6-\frac{1}{2}\partial_{\mu} \Phi\partial^{\mu}\Phi -\frac{1}{2}e^{\sqrt{2}\Phi}|F_3|^2\biggr)\,,
\end{align}
where we introduced the notation $|F_p|^2=\frac{1}{p!}g_6^{k_1l_1}\dots g_6^{k_pl_p}F_{k_1\dots k_{p}}F_{l_1\dots l_{p}}$.
Here, the dilaton $\Phi$ is renormalized such that the kinetic term has a canonical form.

\paragraph{6D to 5D} 
Next, we perform the standard Kaluza-Klein reduction along the $y$ direction with the ansatz
\begin{align}
\begin{split}\label{6d-ansatz}
    ds_6^2&=e^{-\sqrt{\frac{3}{2}}\Psi}(dy+A^1)^2+e^{\frac{1}{\sqrt{6}}\Psi}ds_5^2\,,\\
   F_3&=F_3^{(\rm 5D)}+dA^2\wedge (dy+A^1)\,,\\
     F_3^{(\rm 5D)}&=dC_2^{(\rm 5D)}-dA^2\wedge A^1\,.
\end{split}
\end{align}
In the five dimension, the RR 2-form potential $C_2^{(\rm 5D)}$ can be mapped to a gauge field $A^3$ using the Hodge duality $\star_{5}$ for the 5D metric, and the field strength $F_3^{(\rm 5D)}$ can be rewritten as 
\begin{align}
   F_3^{(\rm 5D)}=- (h^3)^{-2}\star_5dA^3-dA^2\wedge A^1\,.
\end{align}
This duality transformation reveals a triality in the vector fields $A^I(I=1,2,3)$.
For the scalar fields $\Phi$ and $\Psi$, we introduce three scalar fields $h^I(I=1,2,3)$ that satisfy the relation
\begin{align}
    h^1h^2h^3=1
\end{align}
in a way that makes the triality manifest and embed them in \footnote{In this choice, the 6D metric (\ref{6d-ansatz}) is expressed as $ds_6^2=(h^1)^{-\frac{3}{2}}(dy+A^1)^2+(h^1)^{\frac{1}{2}}ds_5^2$.}
\begin{align}
    h^1=e^{\sqrt{\frac{2}{3}}\Psi}\,,\qquad h^2=e^{-\frac{1}{\sqrt{6}}\Psi-\frac{1}{\sqrt{2}}\Phi}\,,\qquad h^3=e^{-\frac{1}{\sqrt{6}}\Psi+\frac{1}{\sqrt{2}}\Phi}\,.
\end{align}
The five-dimensional $U(1)^3$ supergravity action, in which triality is manifest, is written down in
\begin{align}\label{5d_sugra_action}
    S_{\rm 5D}&=\int d^5x\sqrt{-g_5}\left(R_5-\frac{1}{2}G_{IJ}\partial_{\mu}h^I\partial^{\mu}h^J-\frac{1}{2}G_{IJ}F_{\mu\nu}^IF^{J\mu\nu}\right)\no\\
    &\qquad\qquad-\frac{1}{6}\int |\epsilon_{IJK}|\,F^I\wedge F^J\wedge A^K\,, 
\end{align}
where $\epsilon_{IJK}$ is an anti-symmetric tensor normalized as $\epsilon_{123}=1$, and the metric $G_{IJ}$ of the moduli space of the scalar fields $h^I$ is $G_{IJ}=\text{diag}\left((h^1)^{-2},(h^2)^{-2},(h^3)^{-2}\right)$.
For the details of a derivation of the 5D action (\ref{5d_sugra_action}) via the reduction, see for example appendix A in \cite{Virmani:2012kw}.

Applying the above dimensional reduction to the ansatz (\ref{d1d5-bg}) of the D1-D5-KKm system, 
the fields $\{g_{\mu\nu}^{\rm 5D}, h^I,A^I\}$ in the five-dimensional $U(1)^3$ supergravity take the following form:
\begin{align}
\begin{split}
    ds_{5}^2&=\frac{-dt^2}{(W Z_1Z_5)^{\frac{2}{3}}}+(WZ_1Z_5)^{\frac{1}{3}}\biggl[\frac{1}{Z_0}(d\psi+H_0 d\phi)^2+Z_0\left(e^{2\nu}(d\rho^2+dz^2)+\rho^2 d\phi^2\right)\biggr]\,,\\
h^1&=W^{-1}\left(WZ_1Z_5\right)^{\frac{1}{3}}\,,\quad h^2=Z_1^{-1}\left(WZ_1Z_5\right)^{\frac{1}{3}}\,,\quad h^3=Z_5^{-1}\left(WZ_1Z_5\right)^{\frac{1}{3}}\,,\\
A^1&=0\,,\quad
        A^2=-T_1dt\,,\quad
        A^3=-T_5dt\,.
\end{split}
\end{align}
Here, the gauge field $A^3$ is obtained from the relation
\begin{align}
   dA^3= (h^3)^{2}\star_5dC_2^{(5D)}=\frac{1}{\rho Z_5^2 }(\star_2dH_5)\wedge dt=-dT_5\wedge dt\,,
\end{align}
where we used $\star_5^2=-1$ and (\ref{TH-dual}).

\paragraph{5D to 4D} 
We then consider the Kaluza-Klein reduction along the time direction as in \cite{Sahay:2013xda},
\begin{align}\label{5d-d1d5p-e-gen}
\begin{split}
       ds_5^2&=-f^2(dt+\check{A}^0)^2+f^{-1}ds_4^2\,,\\
    A^{I}&=\chi^I(dt+\check{A}^0)+\check{A}^I\,.
\end{split}
\end{align}
We collectively denote the set of the KK gauge field $\check{A}^0$ and the gauge fields $\check{A}^I$ as $\check{A}^{\Lambda}$ with the corresponding field strengths $\check{F}^{\Lambda}=d\check{A}^{\Lambda}$.
The resulting four-dimensional $\cN=2$ supergravity theory is the Euclidean STU model with three vector multiplets and a superpotential 
\begin{align}
    F=-\frac{X^1X^2X^3}{X^0}\,.
\end{align}
Here, $X^{\Lambda}=(X^0,X^1,X^2,X^3)$ are complex scalar fields and play a role a role of holomorphic coordinates on the special-K\"ahler manifold
\begin{align}
    \cM_V=\left(\frac{SU(2,\mathbb{R})}{U(1)}\right)^3\,.
\end{align}
By fixing a gauge $X^0=1$, the bosonic action of the Euclidean STU model \cite{Ceresole:1995jg} takes the form \cite{Cortes:2003zd,Cortes:2005uq,Cortes:2009cs,Gutowski:2012yb}
\begin{align}\label{4dstu}
    S_{\rm 4D}=\int d^4x\sqrt{g_4}\left(R_4-2G_{I\bar{J}}\partial_{\mu}X^I\partial^{\mu}\bar{X}^J\right) +\frac{1}{2}\int \check{F}^{\Lambda}\wedge \check{G}_{\Lambda}\,,
\end{align}
where three complex scalar fields $X^I(I=1,2,3)$ are parameterized by
\begin{align}
    X^I=x^I+e\,y^I=-\chi^I+e\,fh^I\,.
\end{align}
Here, $e$ denotes a split-complex unit which satisfies the usual conjugation law $\bar{e}=-e$, but satisfying $e^2=+1$ rather than $-1$.
The K\"ahler metric $G_{I\bar{J}}$ is obtained by taking the second derivatives of the K\"ahler potential
\begin{align}
    G_{I\bar{J}}=\frac{\partial^2}{\partial X^{I}\partial \bar{X}^{J}}\cK\,,\qquad \cK=-\log\left[-e(\bar{X}^{\Lambda}F_{\Lambda} - \bar{F}_{\Lambda}X^{\Lambda})\right]\,,
\end{align}
where $F_{\Lambda}=\partial_{\Lambda}F$ is the derivative of the superpotential $F$ with respect to $X^{\Lambda}$.
The two forms $\check{G}_{\Lambda}$ are defined as
\begin{align}
    \check{G}_{\Lambda}=({\rm Re}\,N)_{\Lambda \Sigma}\check{F}^{\Sigma}_2+({\rm Im}\,N)_{\Lambda\Sigma}\star_{4}\check{F}^{\Sigma}\,,
\end{align}
where the complex symmetric matrix $N_{\Lambda \Sigma}$ is constructed from the prepotential $F(X)$ as
\begin{align}
    N_{\Lambda \Sigma}=\bar{F}_{\Lambda \Sigma}+2e\frac{({\rm Im}\,F\cdot X)_{\Lambda}({\rm Im}\,F\cdot X)_{\Sigma}}{X\cdot {\rm Im}F \cdot X}\,,
\end{align}
where $F_{\Lambda\Sigma}=\partial_{\Lambda}\partial_{\Sigma}F$ and ${\rm Im}\,F_{\Lambda\Sigma}$ denotes its imaginary part.

For the case of the ansatz (\ref{d1d5-bg}), the warp factor $f$ is given by
\begin{align}
    f=(y^1y^2y^3)^{\frac{1}{3}}=\frac{1}{(WZ_1Z_5)^{\frac{1}{3}}}\,.
\end{align}
The fields $\{g_{\mu\nu}^{(\rm 4D)}, x^I,y^I,\check{A}^{\Lambda}\}$ in the Euclidean STU model are
\begin{align}
\begin{split}
    ds_4^2&=\frac{1}{Z_0}(d\psi+H_0 d\phi)^2+Z_0\left(e^{2\nu}(d\rho^2+dz^2)+\rho^2 d\phi^2\right)\,,\\
    x^1&=0\,,\quad x^2=T_1\,,\quad x^3=T_5\,,\\ 
    y^1&=W^{-1}\,,\quad y^2=Z_1^{-1}\,,\quad  y^3=Z_5^{-1}\,,\quad
    \check{A}^\Lambda=0\,.
\end{split}
\end{align}
Since the explicit expressions of $G_{I\bar{J}}, N_{\Lambda \Sigma}$, and related quantities are not necessary for our analysis, we do not write down them here.

\subsubsection*{4D to 3D}

Finally, we perform a reduction of the four-dimensional fields to three-dimensional ones by taking the ansatz 
\begin{align}
\begin{split}\label{4d-3d-m-iib}
    ds_4^2&=e^{2U}(d\psi+\omega_3)^2+e^{-2U}ds_3^2\,,\\
     \check{A}^{\Lambda}&=\zeta^{\Lambda}(d\psi+\omega_3)+\hat{A}^{\Lambda}\,.
\end{split}
\end{align}
The field strengths $\hat{F}_2^{\Lambda}$ and $\hat{F}_2$ for the one-form fields $\hat{A}^{\Lambda}$ and $\omega_3$ are defined by
\begin{align}
   \hat{F}_2^{\Lambda}=d\hat{A}^{\Lambda} \,,\qquad \hat{F}_2=d\omega_3\,.
\end{align}
In order to access the integrable structure of the solution (\ref{d1d5-bg}), we utilize the fact that in three-dimensional space, the field strengths $\hat{F}_2^{\Lambda}$ and $\hat{F}_2$ can be dualized to scalar fields $\tilde{\zeta}_{\Lambda}$ and $\sigma$ using the Hodge duality as follows \cite{Sahay:2013xda}
\begin{align}
    -d\tilde{\zeta}_{\Lambda}&=e^{2U}({\rm Im}\,N)_{\Lambda \Sigma}\star_{3}(\hat{F}_2^{\Sigma}+\zeta^{\Sigma}\hat{F}_2)+({\rm Re}\,N)_{\Lambda\Sigma}d\zeta^{\Sigma}\,,\label{dual-tzeta-eq}\\
   -d\sigma&= -2e^{4U}\star_{3}\hat{F}_2+\tilde{\zeta}_{\Lambda}d\zeta^{\Lambda}-\zeta^{\Lambda}d\tilde{\zeta}_{\Lambda} \,,\label{dual-sig-eq}
\end{align}
where $\star_3$ is the Hodge star operator with respect to the three-dimensional base space $ds_3^2$.

We now give the 16 scalar fields $\{U,x^I,y^I, \tilde{\zeta}_\Lambda,\zeta^\Lambda,\sigma\}$ corresponding to the ten-dimensional ansatz (\ref{d1d5-bg}).
By comparing it with (\ref{4d-3d-m-iib}), we obtain 
\begin{align}
    e^{2U}=\frac{1}{Z_0}\,,\qquad \omega_3=H_0\,d\phi\,,\qquad \zeta^{\Lambda}=0\,.
\end{align}
The first Hodge dual relation (\ref{dual-tzeta-eq}) leads to $\tilde{\zeta}_{\Lambda}=0$, and the second equation (\ref{dual-sig-eq}) is reduced to
\begin{align}
    \frac{1}{2}d\sigma=\frac{1}{Z_0^2}\star_3d\omega_3=-\frac{1}{\rho Z_0^2}\star_2dH_0=dT_0\,,
\end{align}
where $\star_2$ is the Hodge dual operator for the 2D flat Euclidean space $(\rho,z)$.
In this way, the complete set of 16 scalar fields is given by
\begin{align}
\begin{split}\label{d1d5-scasol-gen}
 e^{2U}&=Z_0^{-1}\,,\quad
    x^1=0\,,\quad x^2=T_1\,,\quad x^3=T_5\,,\\ 
    y^1&=W^{-1}\,,\quad y^2=Z_1^{-1}\,,\quad  y^3=Z_5^{-1}\,,\\
    \tilde{\zeta}_\Lambda&=0\,,\quad
    \zeta^\Lambda=0\,,\quad
    \sigma=2T_0\,.
\end{split}
\end{align}

\section{Matrix realization of $\mathfrak{so}(4,4)$}\label{sec:so44_rep}

Here, we give the matrix realization of $\mathfrak{so}(4,4)$ presented in \cite{Virmani:2012kw}.
Let $E_{ij}$ be the standard basis of matrices, where the $(i,j)$-th element is 1 and all other elements are 0.
The generators of the semi-simple Lie algebra $\mathfrak{so}(4,4)$ are expressed as
\begin{align}
\begin{split}
    H_0&=E_{33}+E_{44}-E_{77}-E_{88}\,,\qquad H_1=E_{33}-E_{44}-E_{77}+E_{88}\,,\\
    H_2&=E_{11}+E_{22}-E_{55}-E_{66}\,,\qquad H_3=E_{11}-E_{22}-E_{55}+E_{66}\,,
\end{split}
\end{align}
and
\begin{align}
\begin{split}
    E_0&=E_{47}-E_{38}\,,\quad E_{1}=E_{87}-E_{34}\,,\\
    E_{2}&=E_{25}-E_{16}\,,\quad E_{3}=E_{65}-E_{12}\,,\\
    F_{0}&=E_{74}-E_{83}\,,\quad F_{1}=E_{78}-E_{43}\,,\\
    F_{2}&=E_{52}-E_{61}\,,\quad F_{3}=E_{56}-E_{21}\,,\\
    E_{q_0}&=E_{41}-E_{58}\,,\quad E_{q_1}=E_{57}-E_{31}\,,\\
    E_{q_2}&=E_{46}-E_{28}\,,\quad E_{q_3}=E_{42}-E_{68}\,,\\
    F_{q_0}&=E_{14}-E_{85}\,,\quad F_{q_1}=E_{75}-E_{13}\,,\\
    F_{q_2}&=E_{64}-E_{82}\,,\quad F_{q_3}=E_{24}-E_{86}\,,\\
    E_{p_0}&=E_{17}-E_{35}\,,\quad E_{p_1}=E_{18}-E_{45}\,,\\
    E_{p_2}&=E_{67}-E_{32}\,,\quad E_{p_3}=E_{27}-E_{36}\,,\\
    F_{p_0}&=E_{71}-E_{53}\,,\quad F_{p_1}=E_{81}-E_{54}\,,\\
    F_{p_2}&=E_{76}-E_{23}\,,\quad F_{p_3}=E_{72}-E_{63}\,.
\end{split}
\end{align}
The set $\{H_0,H_1,H_2,H_3\}$ forms the Cartan subagebra of $\mathfrak{so}(4,4)$.

\section{Details of monodromy matrix for multi-neutral black string}

In this appendix, we give the explicit expressions of the residue matrices $A_j$ of the monodromy matrix for the multi-neutral black string and present a proof of its factorization.  

\subsection{Expressions of residue matrices}\label{sec:A-mat}

The explicit expressions of the residue matrices $A_j$ corresponding to the multi-neutral black string are given by
\begin{align}
    A_1&=
        \begin{pmatrix}
        -\frac{l^2}{4}&0&0&0&0&0&0&0\\
        0&0&0&0&0&0&0&0\\
        0&0&-\cW_b(0)&0&0&0&0&-\frac{l^2}{8} \cW_b(0)\\
        0&0&0&0&0&0&0&0\\
        0&0&0&0&0&0&0&0\\
        0&0&0&0&0&1&0&0\\
        0&0&0&0&0&0&0&0\\
        0&0&\frac{l^2}{8} \cW_b(0)&0&0&0&0&\frac{l^4}{64} \cW_b(0)\\
    \end{pmatrix}\,,\\
    A_{2j}&=
       \begin{pmatrix}
       0&0&0&0&0&0&0&0\\
        0&0&0&0&0&0&0&0\\
        0&0&0&0&0&0&0&0\\
        0&0&0&\frac{1}{w_{N}-w_{2j}}\widetilde{\cW}_b(w_{2j})^{-1}&0&0&-\frac{(l^2-8w_{2j})}{8(w_{N}-w_{2j})}\widetilde{\cW}_b(w_{2j})^{-1}&0\\
        0&0&0&0&0&0&0&0\\
        0&0&0&0&0&0&0&0\\
        0&0&0&\frac{l^2-8w_{2j}}{8(w_{N}-w_{2j})}\widetilde{\cW}_b(w_{2j})^{-1}&0&0&-\frac{l^4}{64(w_{N}-w_{2j})}\widetilde{\cW}_b(w_{2j})^{-1}&0\\
        0&0&0&0&0&0&0&0\\
    \end{pmatrix}\,,\\
    A_{2j+1}&=
   \begin{pmatrix}
      0&0&0&0&0&0&0&0\\
        0&0&0&0&0&0&0&0\\
        0&0&-\frac{1}{w_{2j+1}}\widetilde{\cW}_b(w_{2j+1})&0&0&0&0&-\frac{l^2-8w_{2j+1}}{8w_{2j+1}}\widetilde{\cW}_b(w_{2j+1})\\
        0&0&0&0&0&0&0&0\\
        0&0&0&0&0&0&0&0\\
        0&0&0&0&0&0&0&0\\
        0&0&0&0&0&0&0&0\\
        0&0&\frac{l^2-8w_{2j+1}}{8w_{2j+1}}\widetilde{\cW}_b(w_{2j+1})&0&0&0&0&\frac{l^4}{64w_{2j+1}}\widetilde{\cW}_b(w_{2j+1})\\
    \end{pmatrix}\,,\\
        A_{N}&=
        \begin{pmatrix}
        0&0&0&0&0&0&0&0\\
        0&0&0&0&0&0&0&0\\
        0&0&0&0&0&0&0&0\\
        0&0&0&-\cW_b(w_{N})^{-1}&0&0&-\frac{l^2}{8}\cW_b(w_N)^{-1}&0\\
        0&0&0&0&\frac{l^2}{4}&0&0&0\\
        0&0&0&0&0&0&0&0\\
        0&0&0&\frac{l^2}{8}\cW_b(w_{N})^{-1}&0&0&\frac{l^4}{64}\cW_b(w_{N})^{-1}&0\\
        0&0&0&0&0&0&0&0\\
    \end{pmatrix}\,,
\end{align}
where $\cW_b(w_1)$ and $\cW_b(w_N)$ take the forms
\begin{align}
    \cW_b(w_{1})&=\prod_{i=1}^{n_b}\frac{w_{2i}}{w_{2i+1}}\,,\qquad \cW_b(w_{N})=\prod_{i=1}^{n_b}\frac{w_{N}-w_{2i}}{w_{N}-w_{2i+1}}\,,
\end{align}
and the residues of $\cW_b(w)$ and its inverse are
\begin{align}
    \widetilde{\cW}_b(w_{2j})^{-1}&=\underset{w=w_{2j}}{\rm res}\cW_b(w)^{-1}=-(w_{2j+1}-w_{2j})\prod_{\substack{i=1\\i\neq j}}^{n_b}\frac{w_{2j}-w_{2i+1}}{w_{2j}-w_{2i}}\,,\\
    \widetilde{\cW}_b(w_{2j+1})&=\underset{w=w_{2j+1}}{\rm res}\cW_b(w)=(w_{2j+1}-w_{2j})\prod_{\substack{i=1\\i\neq j}}^{n_b}\frac{w_{2j+1}-w_{2i}}{w_{2j+1}-w_{2i+1}}\,.
\end{align}
These matrices $A_j$ are found to be of rank 2, and can be written in terms of eight-component vectors $a_j,b_j$ and the constants $\alpha_j,\beta_j$ as shown in (\ref{rank2-mat}).
We take $a_j$ and $b_j$ as
\begin{align}
\begin{split}
    a_1&=(1,0,0,0,0,0,0,0)^{T}\,,\\
    a_{2j}&=\left(0,0,0,1,0,0,0,\frac{l^2-8w_{2j}}{8}\right)^{T}\,,\\
    a_{2j+1}&=\left(0,0,-\frac{8}{l^4}(l^2-8w_{2j+1}),0,0,0,0,1\right)^{T}\,,\\
    a_N&=\left(0,0,0,1,0,0,-\frac{l^2}{8},0\right)^{T}\,,
\end{split}
\end{align}
and
\begin{align}
\begin{split}
    b_1&=\left(0,0,0,-\frac{l^2}{8},0,0,1,0\right)^{T}\,,\\
    b_{2j}&=(0,0,1,0,0,0,0,0)^{T}\,,\\
    b_{2j+1}&=(0,0,0,0,0,0,1,0)^{T}\,,\\
    b_N&=\frac{l^2}{4}(1,0,0,0,0,0,0,0)^{T}\,.
\end{split}
\end{align}
With this choice, the constants $\alpha_j$ and $\beta_j$ become
\begin{align}
\begin{split}
    \alpha_1&=\frac{l^2}{4}\,,\quad \alpha_{2j}=\frac{1}{w_{N}-w_{2j}}\widetilde{\cW}_b(w_{2j})^{-1}\,,\\
    \alpha_{2j+1}&=\frac{l^4}{64 w_{2j+1}}\widetilde{\cW}_b(w_{2j+1})\,,\quad \alpha_{N}=-\frac{1}{\cW_b(w_{N})}\,,\\
    \beta_1&=-\cW_b(w_1)\,,\quad \beta_{2j}=-w_{2j}\widetilde{\cW}_b(w_{2j})^{-1}\,,\\
    \beta_{2j+1}&=-\frac{64  (w_{N}-w_{2j+1})}{l^4}\widetilde{\cW}_b(w_{2j+1})\,,\quad \beta_{N}=\frac{4}{l^2}\,.
\end{split}
\end{align}

\subsection{Vanishing of $\gamma_j$}\label{sec:vanih-gamma}

Here, we will show that the constants $\gamma_j$ defined in (\ref{gamma-def}) vanish for all $j$\,($j=1,\dots,N$).

\subsubsection{$\gamma_1$ case}

Let us start by computing the constant $\gamma_1$ associated with the first rod. This can be obtained by substituting the factorized monodromy matrix (\ref{fac-m}) into $\cA_1\eta \eta'a_1$ in (\ref{gamma-def}).
The non-trivial contributions to this quantity are 
\begin{align}
      A_{N} \eta \eta' a_{1}&=w_{N} Y_{\text{flat}}\eta \eta' a_{1}\,,\\
        Y_{\text{flat}}\eta \eta' a_{1}&=\left(0,0,0,0,-1,0,0,0\right)^{T}\,,
\end{align}
and then $\cA_1\eta \eta'a_1$ is computed as
\begin{align}
    \cA_1\eta \eta'a_1&=\left(Y_{\text{flat}}\eta \eta' a_{1}+ \frac{1}{w_1-w_{N}}A_{N}\eta \eta'a_1\right)\no\\
    &=\left(0,0,0,0,0,0,0,0\right)^{T}\,,
\end{align}
where we used $w_1=0$.
Hence, the constant $\gamma_1$ vanishes.

\subsubsection{$\gamma_{N}$ case}

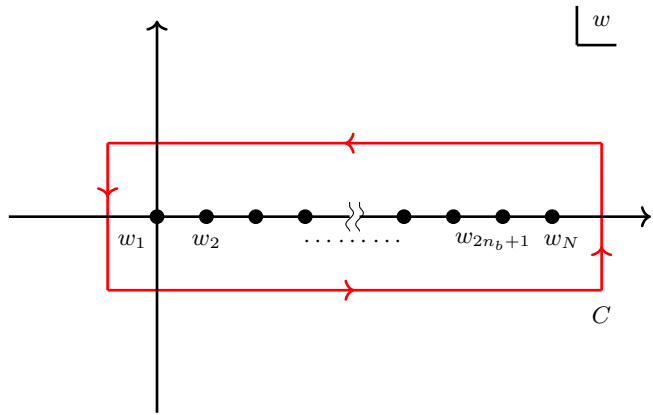
\begin{figure}
\begin{center}
\begin{tikzpicture}[scale=0.65]
\node[font=\small ] at (9,4) {$w$};
\node[font=\small ] at (9,-2) {$C$};
\node[font=\small ] at (-0.5,-0.5) {$w_1$};
\node[font=\small ] at (1,-0.5) {$w_{2}$};
\node[font=\small ] at (3.3,-0.5) {$\cdots$};
\node[font=\small ] at (4.0,-0.5) {$\cdots$};
\node[font=\small ] at (4.7,-0.5) {$\cdots$};
\node[font=\small ] at (6.8,-0.5) {$w_{2n_b+1}$};
\node[font=\small ] at (8.2,-0.5) {$w_{N}$};
\fill[black] (0,0) circle (0.15);
\fill[black] (1,0) circle (0.15);
\fill[black] (2,0) circle (0.15);
\fill[black] (3,0) circle (0.15);
\fill[black] (5,0) circle (0.15);
\fill[black] (6,0) circle (0.15);
\fill[black] (7,0) circle (0.15);
\fill[black] (8,0) circle (0.15);
\draw[-,black,line width = 1] (8.5,4.3) -- (8.5,3.5);
\draw[-,black,line width = 1] (8.5,3.5) -- (9.3,3.5);
\draw[decorate,decoration={snake, amplitude=0.5mm, segment length=3mm}] (3.9,-0.3) -- (3.9,0.3);
\draw[decorate,decoration={snake, amplitude=0.5mm, segment length=3mm}] (4.1,-0.3) -- (4.1,0.3);
\draw[-,postaction={decorate},
        decoration={
          markings,
          mark=at position 0.7 with {\arrow{<}}
        },red,line width = 1] (-1,-1.5) -- (-1,1.5);
\draw[-,postaction={decorate},
        decoration={
          markings,
          mark=at position 0.3 with {\arrow{>}}
        },red,line width = 1] (9,-1.5) -- (9,1.5);
\draw[-,postaction={decorate},
        decoration={
          markings,
          mark=at position 0.5 with {\arrow{<}}
        },red,line width = 1] (-1,1.5) -- (9,1.5);
\draw[-,postaction={decorate},
        decoration={
          markings,
          mark=at position 0.5 with {\arrow{>}}
        },red,line width = 1] (-1,-1.5) -- (9,-1.5);
\draw[->,black,line width = 1] (0,-4) -- (0,4);
\draw[-,black,line width = 1] (-3,0) -- (3.9,0);
\draw[->,black,line width = 1] (4.1,0) -- (10,0);
\end{tikzpicture}
\caption{The red rectangle represents the closed curve $C$ in the complex $w$-plane and encloses all the junction points $w_i(i=1,\dots,N=2n_b+2)$ of the rods describing the multi-neutral black string.}\label{fig:C-W}
\end{center}
\end{figure}

Next, we turn to the computation of $\gamma_{N}$. In this case, the non-vanishing terms in $\cA_N\eta \eta'a_N$ are listed as
\begin{align}
        A_{2j+1} \eta \eta' a_{N}&=-\frac{w_{N}-w_{2j+1}}{w_{2j+1}}\widetilde{\cW}_b(w_{2j+1}) Y_{\text{flat}}\eta \eta' a_{N}\,,\\ 
    A_1 \eta \eta' a_{N}&=-w_{N}\cW_b(0) Y_{\text{flat}}\eta \eta' a_{N}\,,\\
       Y_{\text{flat}}\eta \eta' a_{N}&=\left(0,0,1,0,0,0,0,-\frac{l^2}{8}\right)^{T}\,.
\end{align}
Using these expressions, we evaluate $\cA_N\eta \eta'a_N$ as
\begin{align}
     &\cA_N\eta \eta'a_N=\left(Y_{\text{flat}}+\frac{1}{w_{N}-w_{1}}A_{1}+\sum_{j=1}^{n_b}\frac{1}{w_{N}-w_{2j+1}}A_{2j+1}\right)\eta \eta'a_{N}\no\\
     &=\left(1-\cW_b(0)-\sum_{j=1}^{n_b}\frac{ \widetilde{\cW}_b(w_{2j+1})}{w_{2j+1}}\right) Y_{\text{flat}}\eta \eta' a_{N}=\left(0,0,0,0,0,0,0,0\right)^{T}\,.
\end{align}
The final equality follows from the identity 
\begin{align}\label{wb-sum1}
    \cW_b(0)+\sum_{j=1}^{n_b}\frac{ \widetilde{\cW}_b(w_{2j+1})}{w_{2j+1}}=\frac{1}{2\pi i}\int_{C}\frac{\cW_{b}(w)}{w}dw=\frac{1}{2\pi i}\int_{C_0}\frac{\cW_{b}(\xi^{-1})}{\xi}d\xi=1\,,
\end{align}
where $\xi=1/w$. The closed contour $C$ encloses all the poles of the integrand and is illustrated in Fig.\,\ref{fig:C-W}. This contour can be smoothly deformed into the contour $C_0$, which is the unit circle centered at $\xi=0$.
In this way, the constant $\gamma_{N}$ is zero.

\subsubsection{$\gamma_{2j}$ case}

The non-vanishing contributions to $\cA_{2j}\eta \eta'a_{2j}$ are given by
\begin{align}
        A_1 \eta \eta' a_{2j}&=-w_{2j}\cW_b(0)Y_{\text{flat}}\eta \eta' a_{2j}+\frac{w_{N,2j}w_{2j}\cW_b(0)}{k}
        \left(0,0,0,0,0,0,0,1\right)^{T}\,,\\
    A_{2i+1} \eta \eta' a_{2j}&=\frac{\widetilde{\cW}_b(w_{2i+1})w_{2i+1,2j}}{w_{2i+1}}Y_{\text{flat}}\eta \eta' a_{2j}+\frac{w_{N,2j}w_{2j}\widetilde{\cW}_b(w_{2i+1})}{kw_{2i+1}}\left(0,0,0,0,0,0,0,1\right)^{T}\,,\\
    Y_{\text{flat}}\eta \eta' a_{2j}&=\left(0,0,1,0,0,0,0,\frac{l^2-8w_{2j}}{8}\right)^{T}\,.
\end{align}
Thus, $\cA_{2j}\eta \eta'a_{2j}$ becomes
\begin{align}\label{ga2j-eq}
     &\cA_{2j}\eta \eta'a_{2j}=\left(Y_{\text{flat}}+\frac{1}{w_{2j}-w_{1}}A_{1}+\sum_{i=1}^{n_b}\frac{1}{w_{2j}-w_{2i+1}}A_{2i+1}\right)\eta \eta'a_{2j}\no\\
     &=\left(1-\cW_b(0)-\sum_{j=1}^{n_b}\frac{ \widetilde{\cW}_b(w_{2j+1})}{w_{2j+1}}\right) Y_{\text{flat}}\eta \eta' a_{2j}\no\\
     &\quad +(w_{N}-w_{2j})\left(\cW_b(0)-\sum_{i=1}^{n_b}\frac{w_{2j}\widetilde{\cW}_b(w_{2i+1})}{w_{2i+1}(w_{2i+1}-w_{2j})} \right)\left(0,0,0,0,0,0,0,1\right)^{T}\,.
\end{align}
The first line in (\ref{ga2j-eq}) vanishes due to the identity (\ref{wb-sum1}), and the expression in parentheses on the second line can be rewritten in the integral form
\begin{align}
   -\cW_b(0)+ \sum_{i=1}^{n_b}\frac{w_{2j}\widetilde{\cW}_b(w_{2i+1})}{w_{2i+1}(w_{2i+1}-w_{2j})} =\frac{1}{2\pi i}\int_{C}\frac{w_{2j}}{w(w-w_{2j})}\cW_{b}(w)dw\,,
\end{align}
where we used $\cW_b(w_{2j})=0$. Since the integrand behaves like $\cO(w^{-2})$ at large $|w|$, the right hand side vanishes. As result, we obtain $\gamma_{2j}=0$ for $j=1,2,\dots, n_b$.

\subsubsection{$\gamma_{2j+1}$ case}

Finally, let us evaluate the constants $\gamma_{2j+1}$. The non-vanishing terms in $\cA_{2j+1} \eta \eta' a_{2j+1}$ are
\begin{align}
    A_{2i} \eta \eta' a_{2j+1}&=\frac{w_{2j+1,2i}}{w_{N,2i}\widetilde{\cW}_b(w_{2i})} Y_{\text{flat}}\eta \eta' a_{2j+1} +\frac{64\,w_{N,2j+1}w_{2j+1}}{l^4w_{N,2i}\widetilde{\cW}_b(w_{2i})}\left(0,0,0,1,0,0,0,0\right)^{T}\,,\\
    A_{N} \eta \eta' a_{2j+1}&=\frac{w_{N,2j+1}}{\tilde{W}_b(w_{N})}\left(0,0,0,-\frac{8}{l^2},0,0,1,0\right)^{T}\,,\\
    Y_{\text{flat}}\eta \eta' a_{2j+1}&=\left(0,0,0,-\frac{8(l^2-8w_{2j+1})}{l^4},0,0,1,0\right)^{T}\,.
\end{align}
Then, $\cA_{2j+1} \eta \eta' a_{2j+1}$ are evaluated as
\begin{align}
     &\cA_{2j+1}\eta \eta'a_{2j+1}=\left(Y_{\text{flat}}+\sum_{i=1}^{n_b}\frac{1}{w_{2j+1}-w_{2i}}A_{2i}+\frac{1}{w_{2j+1}-w_{N}}A_{N}\right)\eta \eta'a_{2j+1}\no\\
     &=Y_{\text{flat}}\eta \eta'a_{2j+1}\no\\
     &\quad+\frac{l^2}{l^2-8w_{2j+1}}\left(-\frac{1}{\cW_b(w_{N})}+\sum_{i=1}^{n_b} \left(\frac{2w_{2i}(w_{N}-w_{2j+1})}{w_{N}(w_{2i}-w_{2j+1})}-1\right)\frac{1}{(w_{N}-w_{2i})\widetilde{\cW}_b(w_{2i})}
     \right) Y_{\text{flat}}\eta \eta' a_{2j+1}\no\\
     &\quad +\frac{8w_{2j+1}}{l^2-8w_{2j+1}}\left(\frac{1}{\cW_b(w_{N})}-\sum_{i=1}^{n_b}\frac{w_{N}-w_{2j+1}}{(w_{2i}-w_{2j+1})(w_{N}-w_{2i})\widetilde{\cW}_b(w_{2i})} \right)\left(0,0,0,0,0,0,1,0\right)^{T}\,.
\end{align}
We can demonstrate that this vector also vanishes by rewriting the expressions in parentheses on each line as complex contour integrals and explicitly evaluating the integrals as follows:
\begin{align}
    &-\frac{1}{\cW_b(w_{N})}+\sum_{i=1}^{n_b} \left(\frac{2w_{2i}(w_{N}-w_{2j+1})}{w_{N}(w_{2i}-w_{2j+1})}-1\right)\frac{1}{(w_{N}-w_{2i})\widetilde{\cW}_b(w_{2i})}\no\\
    &=  \frac{1}{2\pi i}\int_{C}\left(\frac{2w(w_{N}-w_{2j+1})}{w_{N}(w-w_{2j+1})}-1\right)\frac{1}{(w_{N}-w)\cW_b(w)}dw\no\\
     &=\frac{1}{2\pi i}\int_{C_0}\left(\frac{2(w_{N}-w_{2j+1})}{w_{N}(1-\xi w_{2j+1})}-1\right)\frac{1}{(\xi w_{N}-1)\cW_b(1/\xi)}\frac{d\xi}{\xi}\no\\
     &=-\frac{w_{N}-2w_{2j+1}}{w_{N}}=-\frac{l^2-8w_{2j+1}}{l^2}\,,
\end{align}
and
\begin{align}
    &\frac{1}{2\pi i}\int_{C}\frac{w_{N}-w_{2j+1}}{(w-w_{2j+1})(w_{N}-w)}\cW_b(w)^{-1}dw\no\\ 
    &=-\frac{1}{\cW_b(w_{N})}+\sum_{i=1}^{n_b}\frac{w_{N}-w_{2j+1}}{(w_{2i}-w_{2j+1})(w_{N}-w_{2i})\widetilde{\cW}_b(w_{2i})}=0\,. 
\end{align}
In this way, we find that the constants $\gamma_{2j+1}$ for $j=1,2,\dots, n_b$ vanish.

\subsection{Factorization of monodromy matrix}\label{subsec:fac-mono}

We now explicitly show that the monodromy matrix $\cM(w)$ given in (\ref{mnst-mono}) for the multi-neutral black string can be factorized into the expression $X_-M(z,\rho)X_+$, which is denoted by $\widetilde{\cM}(\la,z,\rho)$. Here, the matrix $X_+$ is defined by taking the matrices $C_j$ given in (\ref{C1-mst}). Before doing this, we begin with making the following few preparations:
\begin{itemize}
    \item Since $\cM(w)$ and $\widetilde{\cM}(\la,z,\rho)$ are described in terms of different sets of variables, it is convenient to rewrite both matrices in terms of the spectral parameter $\la$ and the soliton positions $\la_j$ (Sec.\,\ref{sec:rewriteM}). 
    \item By using the analytic structure in $\la$ of $\cM(w(\la,z,\rho))$ and $\widetilde{\cM}(\la,z,\rho)$, together with their symmetric structure, we can verify that it is enough to examine the constant terms and the residues at the simple poles $\la=-1/\la_j (j=1,\dots,N)$ of certain components of the two matrix functions (Sec.\,\ref{sec:sym_M}).
\end{itemize}
After that, we explicitly compute the constant terms and the residues of specific components of $\cM(w)$ and $\widetilde{\cM}(\la,z,\rho)$ in order to demonstrate their equivalence.

\subsubsection{Useful expressions of $\cM(w(\la,z,\rho))$ and $\widetilde{\cM}(\la,z,\rho)$ }\label{sec:rewriteM}

Let us begin by rewriting  $\cM(w(\la,z,\rho))$ and $\widetilde{\cM}(\la,z,\rho)$ in terms of the spectral parameter $\la$ and the soliton positions $\la_{j}$, such that the Weyl-Papapetrou coordinates $(z,\rho)$ do not appear explicitly in the resulting expressions.

We start with the case of the factorized monodromy matrix $\widetilde{\cM}(\la,z,\rho)$. In this case, it is enough to rewrite the coset matrix $M_{\rm mString}(z,\rho)$ given in  (\ref{nmbs-m}).
By using definition of the soliton positions $\la_j$, we can rewrite $M_{\rm mString}(z,\rho)$ as
{\footnotesize
\begin{align}
    M(z,\rho)=
     \begin{pmatrix}
        f&0&0&0&0&0&0&0\\
        0&1&0&0&0&0&0&0\\
        0&0&-\frac{4}{l^2}\left(1-\frac{\la_1}{\la_{N}}\right)W_b&0&0&0&0&\frac{1}{2}\left(1+\frac{\la_1}{\la_{N}}\right)W_b\\
        0&0&0&\frac{4}{l^2}\left(1-\frac{\la_{N}}{\la_{1}}\right)W_b^{-1}&0&0&-\frac{1}{2}\left(1+\frac{\la_{N}}{\la_{1}}\right)W_b^{-1}&0\\
        0&0&0&0&f^{-1}&0&0&0\\
        0&0&0&0&0&1&0&0\\
        0&0&0&\frac{1}{2}\left(1+\frac{\la_{N}}{\la_{1}}\right)W_b^{-1}&0&0&-\frac{l^2}{16}\left(1-\frac{\la_{N}}{\la_{1}}\right)W_b^{-1}&0\\
        0&0&-\frac{1}{2}\left(1+\frac{\la_1}{\la_{N}}\right)W_b&0&0&0&0&\frac{l^2}{16}\left(1-\frac{\la_{1}}{\la_{N}}\right)W_b\\
    \end{pmatrix}
    \,,\label{M-larep}
\end{align}}
where we replaced $ M_{\rm mString}(z,\rho)$ with $M(z,\rho)$, and the scalar functions $f$ and $W_b$ are expressed as
\begin{align}
    f=\frac{\la_1}{\la_{N}}\,,\qquad W_b=\prod_{j=1}^{n_b}\frac{\la_{2j+1}}{\la_{2j}}\,.
\end{align}
Several components of $M(z,\rho)$ are found to be equal to each other up to constant factors as follows:
\begin{align}\label{M-rel1}
    M_{88}=-\frac{l^4}{64}M_{33}\,,\quad M_{77}=-\frac{l^4}{64}M_{44}\,,
    \quad M_{38}=-M_{83}\,,\quad M_{74}=-M_{47}\,,
\end{align}
Furthermore, from the expression (\ref{M-larep}) of $M(z,\rho)$, we find that the exchange of the solition positions $\la_{1}\leftrightarrow \la_{N}$ and $\la_{2j}\leftrightarrow \la_{2j+1}$ leads to the identities
\begin{align}\label{M-rel2}
    M_{55}&=M_{11}\Bigl\lvert_{\la_1\leftrightarrow \la_{N}}\,,\quad
    M_{44}=-M_{33}\Bigl\lvert_{\substack{\la_1\leftrightarrow \la_{N}\\ \la_{2j+1}\leftrightarrow \la_{2j}}}\,,\quad
    M_{47}=M_{83}\Bigl\lvert_{\substack{\la_1\leftrightarrow \la_{N}\\ \la_{2j+1}\leftrightarrow \la_{2j}}}\,.
\end{align}
These properties of $M(z,\rho)$ reflect the symmetric configuration of the rods as shown in Fig.~\ref{rod-mbstring-able}.

Next, we rewrite $\cM(w)$ given in (\ref{mono-mat}) in terms of $\la$ and $\la_j$ by using the relation (\ref{w-la-map}) or equivalently
\begin{align}\label{w-wj-la}
    w-w_j=\frac{1}{\nu_j}\frac{(\la-\la_j)(1+\la \la_j)}{\la(1+\la_j^2)}\,.
\end{align}
As mentioned at the beginning of section \ref{subsec:fac-mono}, in the later discussion it is enough to focus only on the residues of the simple poles of $\cM(w)$ at $w=w_1,w_2,\dots, w_{N}$. These residues can be expressed in terms of the scalar function $\cW_b(w)$, which characterizes the bubbles. In what follows, we rewrite relevant expressions involving $\cW_b(w)$ by employing the identity (\ref{w-wj-la}).
To this end, we first write down some useful identities derived from (\ref{w-wj-la}).
By setting $w=w_1=0$ or $w_j=w_1=0$ in (\ref{w-wj-la}), we can obtain  
\begin{align}\label{nu1-la}
    w_{i}\nu_{1}=\frac{\la_{i,1}}{\la_{i}}\frac{1+\la_1\la_{i}}{1+\la_1^2}\,,\qquad 
     w_i\nu_{i}=\frac{\la_{i,1}}{\la_{1}}\frac{1+\la_1\la_{i}}{1+\la_i^2}\,.
\end{align}
By combining (\ref{w-wj-la}) and (\ref{nu1-la}), we also find 
\begin{align}\label{nu1-la2}
    w_{2j+1}-w_{2j}&=w_{2j+1}\frac{\la_1}{\la_{2j}}\frac{\la_{2j+1,2j}}{\la_{2j+1,1}}\frac{1+\la_{2j+1}\la_{2j}}{1+\la_1\la_{2j+1}}\\
    &=  -(w_{2j}-w_{N})\frac{\la_N}{\la_{2j+1}}\frac{\la_{2j,2j+1}}{\la_{2j,N}}\frac{1+\la_{2j+1}\la_{2j}}{1+\la_N\la_{2j}}\,,\\
    (w_1-w_{N})\frac{\nu_{2j+1}}{\la_{2j+1}}&=\frac{\la_{1,N}}{\la_1\la_{N}}\frac{1+\la_1\la_{N}}{1+\la_{2j+1}^2}\,.\label{w1wn-2j1}
\end{align}
From these relations, the relevant expressions related to $\cW_{b}(w)$ can be rewritten as
\begin{align}
    \cW_{b}(w_{1})&=\prod_{j=1}^{n_b}\frac{w_{2j}}{w_{2j+1}}=\prod_{j=1}^{n_b}\frac{\la_{2j+1}}{\la_{2j}}\frac{\la_{1,2j}}{\la_{1,2j+1}}\frac{1+\la_1\la_{2j} }{1+\la_1 \la_{2j+1}}\,,\label{Ww1}\\
\cW_b(w_{N})&=\prod_{j=1}^{n_b}\frac{w_{N}-w_{2j}}{w_{N}-w_{2j+1}}
=\prod_{j=1}^{n_b}\frac{\la_{2j+1}}{\la_{2j}}\frac{\la_{N,2j}}{\la_{N,2j+1}}\frac{1+\la_{N}\la_{2j}}{1+\la_{N}\la_{2j+1}}\,,\label{WwN}\\
    \widetilde{\cW}_b(w_{2j})^{-1}&=-(w_{2j+1}-w_{2j})\prod_{\substack{i=1\\i\neq j}}^{n_b}\frac{w_{2j}-w_{2i+1}}{w_{2j}-w_{2i}}\no\\
    =(w_{2j}-&w_{N})\frac{\la_{N}}{\la_{2j+1}}\frac{\la_{2j,2j+1}}{\la_{2j,N}}\frac{1+\la_{2j+1}\la_{2j}}{1+\la_{2j}\la_{N}}\prod_{\substack{i=1\\i\neq j}}^{n_b}\frac{\la_{2i}}{\la_{2i+1}}\frac{\la_{2j,2i+1}}{\la_{2j,2i}}\frac{1+\la_{2j}\la_{2i+1}}{1+\la_{2i}\la_{2j}}\,,\label{W2j-2}\\
    \widetilde{\cW}_b(w_{2j+1})&=(w_{2j+1}-w_{2j})\prod_{\substack{i=1\\i\neq j}}^{n_b}\frac{w_{2j+1}-w_{2i}}{w_{2j+1}-w_{2i+1}}\no\\
    =w_{2j+1}&\frac{\la_1}{\la_{2j}}\frac{\la_{2j+1,2j}}{\la_{2j+1,1}}\frac{1+\la_{2j+1}\la_{2j}}{1+\la_1\la_{2j+1}}\prod_{\substack{i=1\\i\neq j}}^{n_b}\frac{\la_{2i+1}}{\la_{2i}}\frac{\la_{2j+1,2i}}{\la_{2j+1,2i+1}}\frac{1+\la_{2i}\la_{2j+1}}{1+\la_{2i+1}\la_{2j+1}}\,.
\end{align}
From the expressions $\cW_{b}(w)$, we can find that $\cM(w(\la,z,\rho))$ satisfies the relations similar to (\ref{M-rel1}) and (\ref{M-rel2}) i.e.
\begin{align}
\begin{split}\label{cM-rel}
    \cM_{88}&=-\frac{l^4}{64}\cM_{33}\,,\quad \cM_{77}=-\frac{l^4}{64}\cM_{44}\,,
    \quad \cM_{38}=-\cM_{83}\,,\quad \cM_{74}=-\cM_{47}\,,\\
    \cM_{55}&=\cM_{11}\Bigl\lvert_{\la_1\leftrightarrow \la_{N}}\,,\quad
    \cM_{44}=-\cM_{33}\Bigl\lvert_{\substack{\la_1\leftrightarrow \la_{N}\\ \la_{2j+1}\leftrightarrow \la_{2j}}}\,,\quad
    \cM_{47}=\cM_{83}\Bigl\lvert_{\substack{\la_1\leftrightarrow \la_{N}\\ \la_{2j+1}\leftrightarrow \la_{2j}}}\,.
\end{split}
\end{align}
For later discussion, we write down the relations between the two scalar functions $\cW_b(w)$ and $\cF_b(\la)$.
From the above expressions (\ref{Ww1}) and (\ref{WwN}), the scalar function $\cW_b(w)$ at $w=w_1,w_{N}$ are factorized in the following form: 
\begin{align}
    \cW_{b}(w_{1})&= W_b \cF_{b}(\la_1)\cF_{b}(-1/\la_1)\,,\label{w1-fac}\\
    \cW_{b}(w_{N})&= W_b \cF_{b}(\la_{N})\cF_{b}(-1/\la_{N})\,.
\end{align}
The residues of $\cW_b(w)$ at $w=w_{2j+1}$ are expressed as
\begin{align}\label{cw-2j1}
        \widetilde{\cW}_b(w_{2j+1})&
    =w_{2j+1}\frac{\la_1}{\la_{2j+1}}\frac{1}{\la_{2j+1,1}}\frac{1+\la_{2j+1}^2}{1+\la_1\la_{2j+1}}
    W_{b}
    \widetilde{\cF}_b(\la_{2j+1})\cF_b(-1/\la_{2j+1})\,.
\end{align}

\subsubsection{Symmetric structure of $\widetilde{\cM}(\la,z,\rho)$}\label{sec:sym_M}

The factorized monodromy matrix $\widetilde{\cM}(\la,z,\rho)=X_-M(z,\rho)X_+$ also satisfies the same relation as (\ref{cM-rel}).
To this end, we express each component of $\widetilde{\cM}$ in terms of $M$ and $X_{\pm}$. By denoting the non-vanishing components of $\widetilde{\cM}(\la,z,\rho)$ by
\begin{align}
     \widetilde{\cM}=X_-M(z,\rho)X_+
     =  \begin{pmatrix}
        \widetilde{\cM}_{11}&0&0&0&0&0&0&0\\
        0&1&0&0&0&0&0&0\\
        0&0&\widetilde{\cM}_{33}&0&0&0&0&\widetilde{\cM}_{38}\\
        0&0&0&\widetilde{\cM}_{44}&0&0&\widetilde{\cM}_{47}&0\\
        0&0&0&0&\widetilde{\cM}_{55}&0&0&0\\
        0&0&0&0&0&1&0&0\\
        0&0&0&\widetilde{\cM}_{74}&0&0&\widetilde{\cM}_{77}&0\\
        0&0&\widetilde{\cM}_{83}&0&0&0&0&\widetilde{\cM}_{88}\\
    \end{pmatrix}\,,
\end{align}
we can write each non-vanishing component as
\begin{align}
\begin{split}\label{tM-com}
    \widetilde{\cM}_{11}&=M_{11}X_{+,11}X_{-,11}\,,\\
    \widetilde{\cM}_{33}&=M_{33}\left(X_{-,33}X_{+,33}-\frac{l^4}{64}X_{-,38}X_{+,83}-\frac{l^2}{8}\frac{\la_{1}+\la_{N}}{\la_{1}-\la_{N}}(X_{-,38}X_{+,33}-X_{-,33}X_{+,83})\right)\,,\\
    \widetilde{\cM}_{44}&=M_{44}\left(X_{-,44}X_{+,44}-\frac{l^4}{64}X_{-,47}X_{+,74}+\frac{l^2}{8}\frac{\la_1+\la_{N}}{\la_1-\la_{N}}\left(X_{-,47}X_{+,44}-X_{-,44}X_{+,74}\right)\right)\,,\\
    \widetilde{\cM}_{55}&=M_{55}X_{-,55}X_{+,55}\,,\\
    \widetilde{\cM}_{77}&=M_{77}\left(X_{-,77}X_{+,77}-\frac{64}{l^4}X_{-,74}X_{+,47}
    -\frac{8}{l^2}\frac{\la_1+\la_{N}}{\la_1-\la_{N}}(X_{-,77}X_{+,47}-X_{-,74}X_{+,77})\right)\,,\\
    \widetilde{\cM}_{88}&=M_{88}\left(X_{-,88}X_{+,88}-\frac{64}{l^4}X_{-,83}X_{+,38}
    +\frac{8}{l^2}\frac{\la_1+\la_{N}}{\la_1-\la_{N}}(X_{-,88}X_{+,38}-X_{-,83}X_{+,88})\right)\,,\\
    \widetilde{\cM}_{38}&=M_{38}\left(X_{-,33}X_{+,88}-X_{-,38}X_{+,38}+\frac{\la_1-\la_{N}}{\la_1+\la_{N}}\left(\frac{8}{l^2}X_{-,33}X_{+,38}-\frac{l^2}{8}X_{-,38}X_{+,88}\right)\right)\,,\\
    \widetilde{\cM}_{83}&=M_{83}\left(X_{-,88}X_{+,33}-X_{-,83}X_{+,83}-\frac{\la_1-\la_{N}}{\la_1+\la_{N}}\left(\frac{8}{l^2}X_{-,83}X_{+,33}-\frac{l^2}{8}X_{-,88}X_{+,83}\right)\right)\,,\\
    \widetilde{\cM}_{47}&=M_{47}\left(X_{-,44}X_{+,77}-X_{-,47}X_{+,47}-\frac{\la_1-\la_{N}}{\la_1+\la_{N}}\left(\frac{8}{l^2}X_{-,44}X_{+,47}-\frac{l^2}{8}X_{-,47}X_{+,77}\right)\right)\,,\\
    \widetilde{\cM}_{74}&=M_{74}\left(X_{-,77}X_{+,44}-X_{-,74}X_{+,74}+\frac{\la_1-\la_{N}}{\la_1+\la_{N}}\left(\frac{8}{l^2}X_{-,74}X_{+,44}-\frac{l^2}{8}X_{-,77}X_{+,74}\right)\right)\,.
\end{split}
\end{align}
From the explicit expression (\ref{C1-mst}) of $C_j$, the matrices $X_{\pm}$ are found to satisfy the following relations:
\begin{align}
\begin{split}
\label{Xm38-88}
  X_{\pm,88}&=X_{\pm,33}\,,\quad  X_{+,38}=\frac{l^4}{64}X_{+,83}\,,\quad  X_{-,38}=\frac{64}{l^4}X_{-,83}
 \,,\\
  X_{\pm,77}&=X_{\pm,44}\,,\quad  X_{+,47}=\frac{l^4}{64}X_{+,74}\,,\quad  X_{-,47}=\frac{64}{l^4}X_{-,74}
 \,,
\end{split}
\end{align}
and
\begin{align}
\begin{split}\label{Xmxp-rel2}
    X_{\pm,55}&= X_{\pm,11}\Bigl\lvert_{\la_1\leftrightarrow \la_{N}}\,,\qquad X_{\pm,44}=X_{\pm,33}\Bigl\lvert_{\substack{\la_1\leftrightarrow \la_{N}\\ \la_{2j+1}\leftrightarrow \la_{2j}}}\,,\\
    X_{+,74}&=X_{+,83}\Bigl\lvert_{\substack{\la_1\leftrightarrow \la_{N}\\ \la_{2j+1}\leftrightarrow \la_{2j}}}\,,\qquad 
    X_{-,47}=X_{-,38}\Bigl\lvert_{\substack{\la_1\leftrightarrow \la_{N}\\ \la_{2j+1}\leftrightarrow \la_{2j}}}\,.
\end{split}
\end{align}
By combining (\ref{M-rel1}), (\ref{M-rel2}), (\ref{tM-com}), (\ref{Xm38-88}) and (\ref{Xmxp-rel2}), we confirm that $\widetilde{\cM}$ indeed satisfies the same relation as (\ref{cM-rel}).
Accordingly, it is enough to verify the equivalence between $(11), (33)$ and $(83)$ components of $\cM$ and $\widetilde{\cM}$ for demonstrating the factorization of $\cM$. 

To see this, recall that the factorized monodromy matrix $\widetilde{\cM}(\la,z,\rho)$, by construction, consists of only simple poles in the $\la$-plane. In particular, the components $\widetilde{\cM}_{11}, \widetilde{\cM}_{33}$ and $\widetilde{\cM}_{83}$ have simple poles at $\la=\la_1,-1/\la_1, \la_{2i+1},-1/\la_{2i+1} (i=1,2,\dots,n_b)$.
Moreover, each component of $\widetilde{\cM}(\la,z,\rho)$ is invariant under the exchange $\la \leftrightarrow -1/\la$.
Therefore, to show the equivalence of the corresponding components, it suffices to verify that both the residue at the simple poles at $\la=-1/\la_1, -1/\la_{2i+1} (i=1,2,\dots,n_b)$ and the constant term  coincide.
In the following, these residues and constant terms of each relevant component of $\widetilde{\cM}(\la,z,\rho)$ are computed explicitly and shown to match those of the corresponding components in $\cM(w(\la,z,\rho))$.

\subsubsection{$\cM_{11}$ and $\widetilde{\cM}_{11}$}

The equivalence of the (11) components of $\cM$ and $\widetilde{\cM}$ can be readily verified.
From the expression (\ref{C1-mst}) for the matrix $C_1$, the (11) component of $\widetilde{\cM}$ can be rewritten as
\begin{align}
      \widetilde{\cM}_{11}&=M_{11}X_{+,11}X_{-,11}
      =\frac{\la_1}{\la_{N}}\left(1-\frac{\la \la_{1,N}}{1+\la \la_1}\right)\left(1+\frac{\la_{1,N}}{\la - \la_1}\right)\no\\
      &=1-\frac{\la_{N,1}(1+\la_1\la_{N})}{\la_{N}(1+\la_1^2)}\left(\frac{\la_1}{\la-\la_1}+\frac{1}{1+\la \la_1}\right)\,.
\end{align}
By using the relation (\ref{w-la-map}), it is straightforward to show that $\widetilde{\cM}_{11}$ reduces to
\begin{align}
   \widetilde{\cM}_{11}=1-\frac{w_{N}}{w}\,,
\end{align}
which is equal to $\cM_{11}$.

\subsubsection{$\cM_{33}$ and $\widetilde{\cM}_{33}$}

 Next, we will show the equivalence between $\widetilde{\cM}_{33}(\la,z,\rho)$ and 
 \begin{align}
     \cM_{33}(w)=-\frac{1}{w}\cW_{b}(w)\,.
 \end{align}
As noted above, this is achieved by comparing their residues at $\la=-1/\la_1$ and $\la=-1/\la_{2i+1}$ as well as their constant terms.

\subsubsection*{Residue at $\la=-1/\la_1$}

The residues of $X_{+,33}$ and $X_{+,83}$ at $\la=-1/\la_1$ which $\widetilde{\cM}_{33}$ consists of are given by
\begin{align}
   \underset{\la=-1/\la_1}{\rm res}X_{+,33}=\frac{1}{\la_1^2}(C_1)_{33}\,, \qquad   \underset{\la=-1/\la_1}{\rm res}X_{+,83}=\frac{1}{\la_1^2}(C_1)_{83}=\frac{8}{l^2}\frac{1}{\la_1^2}(C_1)_{33}\,,
\end{align}
and thus the residue of $\widetilde{\cM}_{33}$ presented in (\ref{tM-com}) are computed as
\begin{align}\label{resm33-1}
   \underset{\la=-1/\la_1}{\rm res}\widetilde{\cM}_{33}
   &=
  \frac{M_{33}\cF_{b}(\la_1)}{\la_1}\biggl(X_{-,33}-\frac{l^2}{8} X_{-,38}\biggr)\biggl\lvert_{\la=-1/\la_1}\,,
\end{align}
where the $(33)$-component of $M$ is 
\begin{align}
   M_{33}= -\frac{4}{l^2}\left(1-\frac{\la_1}{\la_{N}}\right)W_b\,.
\end{align}
The term inside the parentheses in (\ref{resm33-1}) is related to $\cF_b(\la)$, expressed as
\begin{align}\label{X33-83-rel}
     X_{-,33}-\frac{l^2}{8} X_{-,38}&=1+\frac{\la_1-\la_{N}}{\la-\la_1}\cF_b(\la_1)
     +\sum_{j=1}^{n_b}\frac{\la_{2j+1}-\la_{N}}{(\la_{2j+1}-\la_1)(\la-\la_{2j+1})}\widetilde{\cF}_b(\la_{2j+1})\no\\
&=\frac{\la-\la_{N}}{\la-\la_1}\cF_{b}(\la)\,.
\end{align}
By using (\ref{X33-83-rel}), (\ref{w1-fac}) and (\ref{nu1-la}), we can see that the residue of $\widetilde{\cM}_{33}$ at $\la=-1/\la_1$ is equal to the one of $\cM_{33}(w(\la,z,\rho))$ :
\begin{align}
   \underset{\la=-1/\la_1}{\rm res}\widetilde{\cM}_{33}
  &=-\cW_{b}(w_1)\frac{\nu_1}{\la_1}= \underset{\la=-1/\la_1}{\rm res}\left(\frac{-\cW_{b}(w(\la,z,\rho))}{w(\la,z,\rho)-w_1}\right)\no\\
  &=\underset{\la=-1/\la_1}{\rm res}\cM_{33}(w(\la,z,\rho))\,.
\end{align}

\subsubsection*{Residues at $\la=-1/\la_{2j+1}$}

Next, we examine the residues at $\la=-1/\la_{2j+1}$. The residues of $X_{+,33}$ and $X_{+,83}$ at this value are given by
\begin{align}
   \underset{\la=-1/\la_{2j+1}}{\rm res}X_{+,33}&=\frac{1}{\la_{2j+1}^2}(C_{2j+1})_{33}\,,\\
   \underset{\la=-1/\la_{2j+1}}{\rm res}X_{+,83}&=\frac{1}{\la_{2j+1}^2}(C_{2j+1})_{83}=\frac{8}{l^2}\frac{1}{\la_{2j+1}^2}\frac{\la_{N,1}}{\la_{N}+\la_1-2\la_{2j+1}}(C_{2j+1})_{33}\,.
\end{align}
The residue of $\widetilde{\cM}_{33}$ at $\la=-1/\la_{2j+1}$ takes the form
\begin{align}\label{resm33-la2j1}
   \underset{\la=-1/\la_{2j+1}}{\rm res}\widetilde{\cM}_{33}
  &=\frac{M_{33}(C_{2j+1})_{33}}{\la_{2j+1}^2}
  \biggl(\left(1-\frac{\la_{1,N}}{\la_{N}+\la_1-2\la_{2j+1}}\frac{\la_1+\la_{N}}{\la_1-\la_{N}}\right)X_{-,33}\no\\
  &\qquad-\left(\frac{\la_1+\la_{N}}{\la_1-\la_{N}}-\frac{\la_{1,N}}{\la_{N}+\la_1-2\la_{2j+1}}\right)\frac{l^2}{8}X_{-,38}\biggr)\biggl\lvert_{\la=-1/\la_{2j+1}}
  \,.
\end{align}
The expression inside the parentheses satisfies the following identity:
\begin{align}\label{x33-x38-2}
    &\left(1-\frac{\la_{1,N}}{\la_{N}+\la_1-2\la_{2j+1}}\frac{\la_1+\la_{N}}{\la_1-\la_{N}}\right)X_{-,33}
    -\left(\frac{\la_1+\la_{N}}{\la_1-\la_{N}}-\frac{\la_{1,N}}{\la_{N}+\la_1-2\la_{2j+1}}\right)\frac{l^2}{8}X_{-,38}\no\\
    &=-\frac{2}{\la_{N}+\la_1-2\la_{2j+1}}
    \biggl[\la_{2j+1}+\frac{\la_1(\la_{2j+1}-\la_{N})}{\la-\la_1}\cF_{b}(\la_1)\no\\
    &\qquad\qquad\qquad\qquad\qquad\qquad+\sum_{j=1}^{n_b}\frac{(\la_{2j+1}^2-\la_1\la_{N})}{(\la-\la_{2j+1})(\la_{2j+1}-\la_1)}
    \widetilde{\cF}_b(\la_{2j+1})\biggr]\no\\
    &=-\frac{2(\la \la_{2j+1}-\la_1\la_{N})}{(\la_{N}+\la_1-2\la_{2j+1})(\la-\la_1)}
    \cF_b(\la)\,.
\end{align}
The second equality in (\ref{x33-x38-2}) follows by comparing residues on both sides.
Substituting the identity (\ref{x33-x38-2}) into (\ref{resm33-la2j1}) leads to
\begin{align}
   \underset{\la=-1/\la_{2j+1}}{\rm res}\widetilde{\cM}_{33}
    &=-\frac{1}{w_{N}}\frac{\la_{1,N}(1+\la_1\la_{N})}{\la_{N}\la_{2j+1}\la_{1,2j+1}(1+\la_1\la_{2j+1})}W_b \widetilde{\cF}_b(\la_{2j+1}) \cF_b(-1/\la_{2j+1})
\,.
\end{align}
By using (\ref{cw-2j1}) and (\ref{w1wn-2j1}), this can be rewritten as
\begin{align}
    \underset{\la=-1/\la_{2j+1}}{\rm res}\widetilde{\cM}_{33}&= \frac{1}{w_{N}w_{2j+1}}\frac{\la_{1,N}}{\la_1\la_{N}}
   \frac{1+\la_1\la_{N}}{1+\la_{2j+1}^2 }   \widetilde{\cW}_b(w_{2j+1})\no\\
   &=-\frac{1}{w_{2j+1}}\frac{\nu_{2j+1}}{\la_{2j+1}}\widetilde{\cW}_b(w_{2j+1})\no\\ 
   &=\underset{\la=-1/\la_{2j+1}}{\rm res}\left(\frac{-\cW_{b}(w(\la,z,\rho))}{w(\la,z,\rho)-w_1}\right)\no\\
  &=\underset{\la=-1/\la_{2j+1}}{\rm res}\cM_{33}(w(\la,z,\rho))\,.
\end{align}

\subsubsection*{Constant term}

The constant term of $\widetilde{\cM}_{33}$ can be obtained by taking a limit $\la\to \infty$, and it takes the form
\begin{align}
     \lim_{\la\to \infty}\widetilde{\cM}_{33}&=M_{33}\biggl(1-\frac{ (C_{1})_{3,3}}{\la_{1}}-\sum_{j=1}^{n_b}\frac{ (C_{2j+1})_{3,3}}{\la_{2j+1}}+\frac{l^2}{8}\frac{\la_{1}+\la_{N}}{\la_{1}-\la_{N}}\left(-\frac{ (C_{1})_{8,3}}{\la_{1}}-\sum_{j=1}^{n_b}\frac{ (C_{2j+1})_{8,3}}{\la_{2j+1}}\right)\biggr)\no\\
     &=M_{33}\left(1-\cF_{b}(\la_1)+\sum_{j=1}^{n_b}\frac{1}{\la_{1,2j+1}}\underset{\la=\la_{2j+1}}{\rm res}\cF_b(\la)\right)\,.
\end{align}
The sum involving $\cF_b(\la)$ can be combined into a complex integral
\begin{align}\label{f-int}
    \cF_{b}(\la_1)-\sum_{j=1}^{n_b}\frac{1}{\la_{1,2j+1}}\underset{\la=\la_{2j+1}}{\rm res}\cF_b(\la)=\oint_{C} \frac{\cF_{b}(\la)}{\la-\la_1}d\la\,,
\end{align}
where $C$ is a closed contour encircling the poles $\la=\la_1\,, \la_{2j+1} (j=1,\dots, n_b)$.
By making the change of the variable $\xi=1/\la$, the integral (\ref{f-int}) becomes
\begin{align}
        \cF_{b}(\la_1)-\sum_{j=1}^{n_b}\frac{\underset{\la=\la_{2j+1}}{\rm res}\cF_b(\la)}{\la_{1,2j+1}}=\oint_{C_{0}} \frac{\cF_{b}(1/\xi)}{1-\xi\la_1}\frac{d\xi}{\xi}=1\,,
\end{align}
where the closed contour $C_0$ encircles $\xi=0$.
Therefore, we obtain $\lim_{\la\to \infty}\widetilde{\cM}_{33}(\la,z,\rho)=0$ which is equal to the constant term of $\cM_{33}(w)$.

\begin{figure}
\begin{center}
\begin{tikzpicture}[scale=0.65]
\node[font=\small ] at (9,4) {$\la$};
\node[font=\small ] at (9,-2) {$C$};
\node[font=\small ] at (-0.5,-0.5) {$\la_1$};
\node[font=\small ] at (1,-0.5) {$\la_{2}$};
\node[font=\small ] at (3.3,-0.5) {$\cdots$};
\node[font=\small ] at (4.0,-0.5) {$\cdots$};
\node[font=\small ] at (4.7,-0.5) {$\cdots$};
\node[font=\small ] at (6.8,-0.5) {$\la_{2n_b+1}$};
\node[font=\small ] at (8.2,-0.5) {$\la_{N}$};
\fill[black] (0,0) circle (0.15);
\fill[black] (1,0) circle (0.15);
\fill[black] (2,0) circle (0.15);
\fill[black] (3,0) circle (0.15);
\fill[black] (5,0) circle (0.15);
\fill[black] (6,0) circle (0.15);
\fill[black] (7,0) circle (0.15);
\fill[black] (8,0) circle (0.15);
\draw[-,black,line width = 1] (8.5,4.3) -- (8.5,3.5);
\draw[-,black,line width = 1] (8.5,3.5) -- (9.3,3.5);
\draw[decorate,decoration={snake, amplitude=0.5mm, segment length=3mm}] (3.9,-0.3) -- (3.9,0.3);
\draw[decorate,decoration={snake, amplitude=0.5mm, segment length=3mm}] (4.1,-0.3) -- (4.1,0.3);
\draw[-,postaction={decorate},
        decoration={
          markings,
          mark=at position 0.7 with {\arrow{<}}
        },red,line width = 1] (-1,-1.5) -- (-1,1.5);
\draw[-,postaction={decorate},
        decoration={
          markings,
          mark=at position 0.3 with {\arrow{>}}
        },red,line width = 1] (9,-1.5) -- (9,1.5);
\draw[-,postaction={decorate},
        decoration={
          markings,
          mark=at position 0.5 with {\arrow{<}}
        },red,line width = 1] (-1,1.5) -- (9,1.5);
\draw[-,postaction={decorate},
        decoration={
          markings,
          mark=at position 0.5 with {\arrow{>}}
        },red,line width = 1] (-1,-1.5) -- (9,-1.5);
\draw[->,black,line width = 1] (0,-4) -- (0,4);
\draw[-,black,line width = 1] (-3,0) -- (3.9,0);
\draw[->,black,line width = 1] (4.1,0) -- (10,0);
\end{tikzpicture}
\caption{The red rectangle represents the closed curve $C$ and encloses all the junction points $\la_i(i=1,\dots,N)$ of the rods.}\label{fig:C-la}
\end{center}
\end{figure}
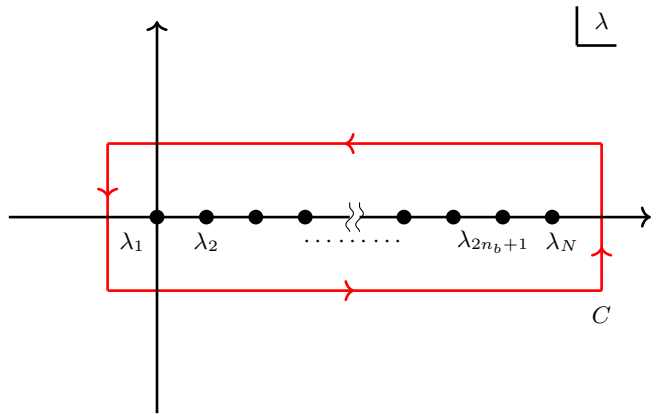

\subsubsection{$\cM_{83}$ and $\widetilde{\cM}_{83}$}

Finally, we will show the equivalence between $\widetilde{\cM}_{83}(\la,z,\rho)$ and 
 \begin{align}
     \cM_{83}(w)=-\left(1-\frac{w_{N}}{2w}\right)\cW_b(w)\,.
 \end{align}
We again compute their residues at $\la=-1/\la_1$ and $\la=-1/\la_{2i+1}$ as well as their constant terms.

\subsubsection*{Residue at $\la=-1/\la_1$}

The residue of $\widetilde{\cM}_{83}$ at $\la=-1/\la_1$ is evaluated as
\begin{align}\label{m83-res}
   \underset{\la=-1/\la_1}{\rm res}\widetilde{\cM}_{83}
     &=
  \frac{M_{83}\cF_{b}(\la_1)}{\la_1}\frac{\la_1-\la_{N}}{\la_1+\la_{N}}\biggl(X_{-,33}-\frac{l^2}{8} X_{-,38}\biggr)\biggl\lvert_{\la=-1/\la_1}\,,
\end{align}
where we used (\ref{Xm38-88}) and the $(83)$-component of $M$ is given by
\begin{align}
   M_{83}= -\frac{1}{2}\left(1+\frac{\la_1}{\la_{N}}\right)W_b\,.
\end{align}
By using (\ref{X33-83-rel}) and (\ref{nu1-la}), we can rewrite (\ref{m83-res}) as
\begin{align}
   \underset{\la=-1/\la_1}{\rm res}\widetilde{\cM}_{83}
  &=\frac{w_{N}}{2}\,\cW_{b}(w_1)\frac{\nu_1}{\la_1}\no\\
  &= \underset{\la=-1/\la_1}{\rm res}\left(-\left(1-\frac{w_{N}}{2(w(\la,z,\rho)-w_{1})}\right)\cW_{b}(w(\la,z,\rho))\right)\no\\
  &=\underset{\la=-1/\la_1}{\rm res}\cM_{33}(w(\la,z,\rho))\,.
\end{align}

\subsubsection*{Residues at $\la=-1/\la_{2j+1}$}

Next, we will see the residues at $\la=-1/\la_{2j+1}$. 
We can obtain
\begin{align}\label{m83-res2}
   \underset{\la=-1/\la_{2j+1}}{\rm res}\widetilde{\cM}_{83}
  &=\frac{M_{83}(C_{2j+1})_{33}}{\la_{2j+1}^2}
 \left(1-\frac{\la_{1,N}}{\la_{N}+\la_1-2\la_{2j+1}}\frac{\la_1-\la_{N}}{\la_1+\la_{N}}\right)X_{-,33}
 \,.
\end{align}
The expression inside the parentheses satisfies the following identity:
\begin{align}\label{x33-x38-3}
    &\left(1-\frac{\la_{1,N}}{\la_{N}+\la_1-2\la_{2j+1}}\frac{\la_1-\la_{N}}{\la_1+\la_{N}}\right)X_{-,33}
    -\left(\frac{\la_1-\la_{N}}{\la_1+\la_{N}}-\frac{\la_{1,N}}{\la_{N}+\la_1-2\la_{2j+1}}\right)\frac{l^2}{8}X_{-,38}\no\\
    &=\frac{2}{(\la_1+\la_{N})(\la_{N}+\la_1-2\la_{2j+1})}
    \biggl[(2\la_1\la_{N}-\la_{2j+1}(\la_1+\la_{N}))+\frac{\la_1\la_{N,1}\la_{2j+1,N}}{\la-\la_1}\cF_{b}(\la_1)\no\\
    &\qquad+\sum_{j=1}^{n_b}\frac{(\la_1+\la_{N})(\la_{2j+1}^2+\la_1\la_{N})-4\la_1\la_{N}\la_{2j+1}}{(\la-\la_{2j+1})(\la_{2j+1}-\la_1)}
    \widetilde{\cF}_b(\la_{2j+1})\biggr]\no\\
    &=\frac{2}{(\la_1+\la_{N})(\la_{N}+\la_1-2\la_{2j+1})}\left(\frac{\la(2\la_1\la_{N}-\la_{2j+1}(\la_1+\la_{N}))-\la_1\la_{N}(\la_{N}+\la_1-2\la_{2j+1})}{\la-\la_1}\right)
    \cF_b(\la)\,.
\end{align}
By using (\ref{cw-2j1}), (\ref{nu1-la}) and (\ref{w-wj-la}), we can rewrite the expression (\ref{m83-res2}) as
\begin{align}
    \underset{\la=-1/\la_{2j+1}}{\rm res}\widetilde{\cM}_{83}
  &=\frac{1}{2}\left(-2\frac{\la_{2j+1,N}}{\la_{2j+1}\la_{N}}\frac{1+\la_{2j+1}\la_{N}}{1+\la_{2j+1}^2}+\frac{\la_{1,N}}{\la_1\la_{N}}
   \frac{1+\la_1\la_{N}}{1+\la_{2j+1}^2 }\right)\frac{\widetilde{\cW}_b(w_{2j+1})}{w_{2j+1}}\no\\
  &=-\left(1-\frac{w_{N}}{2w_{2j+1}}\right)\frac{\nu_{2j+1}}{\la_{2j+1}}\widetilde{\cW}_b(w_{2j+1})\no\\
  &= \underset{\la=-1/\la_{2j+1}}{\rm res}\left(-\left(1-\frac{w_{N}}{2(w(\la,z,\rho)-w_{1})}\right)\cW_{b}(w(\la,z,\rho))\right)\no\\
  &=\underset{\la=-1/\la_{2j+1}}{\rm res}\cM_{83}(w(\la,z,\rho))\,.
\end{align}

\subsubsection*{Constant term}

Finally, we compute the constant terms of $\cM_{83}(w)$ and $\widetilde{\cM}_{83}(\la,z,\rho)$.
The constant term of $\cM_{83}(w)$ can be evaluated by taking the limit $w\to \infty$ :
\begin{align}
   \lim_{w\to \infty}\cM_{83}(w)= \lim_{w\to \infty}\left(-\left(1-\frac{l^2}{8w}\right)\cW_b(w)\right)= -1\,.
\end{align}
On the other hand, taking the limit $\la\to \infty$ leads to the constant term of $\widetilde{\cM}_{83}(\la,z,\rho)$ for fixed $z$ and $\rho$, and from the definition (\ref{tM-com}), we obtain
\begin{align}
       \lim_{\la\to \infty}\widetilde{\cM}_{83}
       &=M_{83}\left(1-\frac{\la_1-\la_{N}}{\la_1+\la_{N}}\cF_{b}(\la_1)
+\sum_{j=1}^{n_b}\frac{-2\la_1\la_{N}+\la_{2j+1}(\la_1+\la_{N})}{\la_{2j+1}\la_{1,2j+1}(\la_1+\la_{N})}
 \widetilde{\cF}_b(\la_{2j+1})\right)\,,
\end{align}
where $M_{83}$ is given by
\begin{align}
    M_{83}=-\frac{1}{2}\left(1+\frac{\la_1}{\la_{N}}\right)W_b
    =-\frac{\la_1+\la_{N}}{2\la_{N}}\cF_{b}(0)^{-1}\,.
\end{align}
The collection of terms involving the scalar function $\cF_b(\la)$ inside the parentheses can be simplified by rewriting it as a complex integral, as shown in
\begin{align}
&\frac{\la_1-\la_{N}}{\la_1+\la_{N}}\cF_{b}(\la_1)
-\sum_{j=1}^{n_b}\frac{-2\la_1\la_{N}+\la_{2j+1}(\la_1+\la_{N})}{\la_{2j+1}\la_{1,2j+1}(\la_1+\la_{N})}
 \widetilde{\cF}_b(\la_{2j+1})\no\\
    &=\frac{1}{\la_1+\la_{N}}\oint_{C}\frac{-2\la_1\la_{N}+\la(\la_1+\la_{N})}{\la-\la_1}\cF_b(\la)\frac{d\la}{\la}-\frac{2\la_{N}}{\la_1+\la_{N}}\cF_b(0)\no\\
    &=\frac{1}{\la_1+\la_{N}}\oint_{C_{0}}\frac{-2\la_1\la_{N}\xi+(\la_1+\la_{N})}{1-\xi\la_1}\cF_b(1/\xi)\frac{d\xi}{\xi}-\frac{2\la_{N}}{\la_1+\la_{N}}\cF_b(0)\no\\
    &=1-\frac{2\la_{N}}{\la_1+\la_{N}}\cF_b(0)\,,
\end{align}
where in the second equality, we made a change of variables by setting $\xi=1/\la$, and deformed the closed contour $C$ into $C_{0}$.
The closed contour $C$ encircles the simple poles $\la=0,\la_1$ and $\la_{2j+1}\,(j=1,\dots,n_b)$ of the integrand, whereas the closed contour $C_{0}$ surrounds only the point $\xi=0$.
Therefore, we obtain $\lim_{\la\to \infty}\widetilde{\cM}_{83}=-1$ which matches $\lim_{w\to \infty}\cM_{83}(w)$.

In summary, the (83) component of the monodromy matrix $\cM(w)$ coincides with the corresponding entry of the factrized form $\widetilde{\cM}_{83}(\la,z,\rho)$.

\end{document}